\documentclass[usenatbib,usegraphicx]{mn2e} 

\newcommand{\chandra}{\textit{Chandra}}
\newcommand{\xmm}{\textit{XMM-Newton}}

\newcommand{\zpup}{$\zeta$~Pup}

\newcommand{\taustar}{\ensuremath{\tau_{\ast}}}

\newcommand{\Ro}{\ensuremath{{R_{\mathrm o}}}}

\newcommand{\fcl}{\ensuremath{{f_{\mathrm cl}}}}

\newcommand{\Rstar}{\ensuremath{{R_{\ast}}}}

\newcommand{\Rsun}{\ensuremath{\mathrm {R_{\sun}}}}

\newcommand{\Msunyr}{\ensuremath{{\mathrm {M_{\sun}~{\mathrm yr^{-1}}}}}}

\newcommand{\kms}{km s$^{-1}$}
\newcommand{\vinf}{\ensuremath{v_{\infty}}}
\newcommand{\Teff}{\ensuremath{T_{\rm eff}}}
\newcommand{\Lx}{\ensuremath{L_{\rm X}}}
\newcommand{\Lbol}{\ensuremath{L_{\rm Bol}}}
\newcommand{\Hea}{${\rm He}{\alpha}$}
\newcommand{\Lya}{${\rm Ly}{\alpha}$}
\newcommand{\Lyb}{${\rm Ly}{\beta}$}
\newcommand{\Ha}{${\rm H}{\alpha}$}
\newcommand{\halpha}{${\rm H}{\alpha}$}

\newcommand{\Mdot}{\ensuremath{\dot{M}}}

\newcommand{\apj}{ApJ}
\newcommand{\apjs}{ApJS}

\newcommand{\aap}{A\&A}

\newcommand{\aapr}{A\&A~Rev.}
\newcommand{\aj}{AJ}
\newcommand{\mnras}{MNRAS}
\newcommand{\araa}{ARAA}
\newcommand{\pasp}{PASP}

\begin{document}

\title[Chandra mass-loss rates of O stars]{Measuring mass-loss rates
  and constraining shock physics using X-ray line profiles of O stars
  from the {\it Chandra} archive} 
\author[D. Cohen et al.]{David H.\ Cohen,$^{1}$\thanks{E-mail:
    cohen@astro.swarthmore.edu} Emma E.\ Wollman,$^{1,2}$ Maurice A.\
  Leutenegger,$^{3,4}$ \newauthor Jon O. Sundqvist,$^{5,6}$ Alex W.\
  Fullerton,$^{7}$ Janos Zsarg\'{o},$^{8}$ Stanley P.\ Owocki$^{5}$ \\  
 $^{1}$Department of Physics and Astronomy, Swarthmore College, Swarthmore, PA 19081, USA\\
  $^{2}$Department of Physics, California Institute of Technology, 1200 East California Blvd., Pasadena, CA 91125, USA \\
  $^{3}$NASA/Goddard Space Flight Center, Code 662, Greenbelt, MD 20771, USA \\
  $^{4}$CRESST and University of Maryland, Baltimore County, MD 21250, USA \\
  $^{5}$Bartol Research Institute, University of Delaware, Newark, DE 19716, USA \\
  $^{6}$Institut f\"ur Astronomie und Astrophysik der Universit\"at M\"unchen, Scheinerstr.\ 1, D-81679 M\"unchen, Germany \\
  $^{7}$Space Telescope Science Institute, 3700 San Martin Dr., Baltimore, MD 21218, USA \\
  $^{8}$Escuela Superior de Fisica y Matem\'{a}ticas, Instituto Polit\'{e}cnico Nacional, C.P.\ 07738, Mexico, D.F., Mexico 
}

\maketitle

\label{firstpage}

\begin{abstract}

\noindent
We quantitatively investigate the extent of wind absorption signatures
in the X-ray grating spectra of all non-magnetic, effectively single O
stars in the \chandra\/ archive via line profile fitting. Under the
usual assumption of a spherically symmetric wind with embedded shocks,
we confirm previous claims that some objects show little or no wind
absorption. However, many other objects do show asymmetric and blue
shifted line profiles, indicative of wind absorption. For these stars,
we are able to derive wind mass-loss rates from the ensemble of line
profiles, and find values lower by an average factor of 3 than those
predicted by current theoretical models, and consistent with \Ha\/ if
clumping factors of $f_{\rm cl} \approx 20$ are assumed. The same
profile fitting indicates an onset radius of X-rays typically at $r
\approx 1.5$ \Rstar, and terminal velocities for the X-ray emitting
wind component that are consistent with that of the bulk wind.  We
explore the likelihood that the stars in the sample that do not show
significant wind absorption signatures in their line profiles have at
least some X-ray emission that arises from colliding wind shocks with
a close binary companion. The one clear exception is $\zeta$ Oph, a
weak-wind star that appears to simply have a very low mass-loss rate.
We also reanalyse the results from the canonical O supergiant \zpup,
using a solar-metallicity wind opacity model and find $\Mdot = 1.8
\times 10^{-6}$ \Msunyr, consistent with recent multi-wavelength
determinations.

\end{abstract}


\section{Introduction} \label{sec:intro}

By losing mass at a rate of $\Mdot \sim 10^{-6}$ \Msunyr\/ via its
stellar wind, an O star can shed a significant portion of its mass
over the course of its lifetime \citep{Puls2008}. Not only can this
substantially reduce the mass of a core-collapse supernova progenitor,
but the wind transfers a significant amount of mass, momentum, and
energy to the surrounding interstellar medium (ISM). Thus, the wind
mass-loss rate is an important parameter in the study of both stellar
evolution and of the Galactic environment.  In recent years there has
been increased awareness of large systematic uncertainties in many
mass-loss rate diagnostics, primarily due to wind clumping, rendering
the actual mass-loss rates of O stars somewhat controversial (e.g.\
\citealt{fmp2006,Oskinova2007,Sundqvist2010}).

X-rays provide a potentially good clumping-insensitive mass-loss rate
diagnostic via the effect of wind attenuation on X-ray emission line
profile shapes. The characteristic line profile shape that provides
the diagnostic power arises because redshifted photons emitted from
the rear hemisphere of the wind are subject to more attenuation than
the blueshifted photons originating in the front hemisphere
(\citealt{MacFarlane1991,oc2001}; see fig. 2 of \citealt{Cohen2010}).
The degree of blue shift and asymmetry in these line profiles is then
directly proportional to the wind column density and thus to the
mass-loss rate.  By fitting a simple quantitative model \citep{oc2001}
to each emission line in a star's spectrum and then analysing the
ensemble of line optical depths, we can determine the star's mass-loss
rate \citep{Cohen2010,Cohen2011}. Complementary approaches that fit
all the lines simultaneously, along with fitting the broadband X-ray
properties, have also been employed recently
\citep{Herve2012,Herve2013}. While our approach does not use as many
different observational constraints it does have the advantage of
simplicity, which enables us to more easily explore the effects of
individual line properties, particularly those involving hot plasma
kinematics and absorption by the cold wind component, which are the
focus of this paper.

Because this X-ray absorption line profile diagnostic scales with the
column density rather than the square of the density, it avoids many
of the problems presented by traditional mass-loss rate diagnostics.
In particular, UV resonance line diagnostics are problematic due to
their sensitivity to ionization corrections which are highly uncertain
and are sensitive to clumping effects on density-squared recombination
\citep{Bouret2005}.  Further complications arise with UV lines from
optically thick clumping, including velocity-space clumping
\citep{Oskinova2007,Owocki2008,Sundqvist2010,Sundqvist2011}. For
direct density-squared diagnostics such as \Ha\/ and radio or IR
free-free emission, the mass-loss rate will be overestimated if
clumping is not accounted for. And even when clumping is accounted
for, there is a degeneracy between the mass-loss rate and the clumping
factor, as the quantity derived from these diagnostics is
$\Mdot\sqrt{\fcl}$ where the clumping factor, $f_{\rm cl} \equiv
\langle\rho^2\rangle/\langle\rho\rangle^2$. Using the X-ray absorption
diagnostic in conjunction with the density-squared emission
diagnostics can break this degeneracy and enable us to simultaneously
determine the mass-loss rate and the clumping factor.

Recent, more sophisticated application of the density-squared emission
diagnostics (\Ha, IR and radio free-free), assuming a radially
dependent clumping factor, has led to a downward revision of empirical
mass-loss rates of O stars \citep{Puls2006}.  These lowered mass-loss
rates provide a natural explanation for the initially surprising
discovery \citep{Cassinelli2001,Kahn2001} that X-ray line profiles are
not as asymmetric as traditional mass-loss rate estimates had implied.


\begin{table*}
\begin{center}
  \caption{Properties of programme stars}
\begin{tabular}{ccccccccc}
  \hline
  Star & Spectral Type & \Teff\/ & $R$ & log $g$ & \vinf\/ & MEG counts & HEG counts & exposure time
  \\
 & & (kK) & (\Rsun) & (cm s$^{-2}$) & (\kms) & & & (ksec) \\
  \hline
HD 93129A & O3 If* & 42.5$^a$ & 22.5$^a$ & 3.71$^a$ & 3200$^a$ & 2936 & 1258 & 137.7  \\
HD 93250 & O3.5 V & 46.0$^a$ & 15.9$^a$ & 3.95$^a$ & 3250$^b$ & 6169 & 2663 & 193.7 \\
9 Sgr & O4 V & 42.9$^c$ & 12.4$^c$ & 3.92$^c$ & 3100$^b$ & 4530 & 1365 & 145.8 \\
$\zeta$ Pup & O4 If & 40.0$^d$ & 18.9$^d$ & 3.63$^d$ & 2250$^b$ & 11018 & 2496 & 73.4 \\
HD 150136 & O5 III & 40.3$^c$ & 15.1$^c$ & 3.69$^c$ & 3400$^b$ & 8581 & 2889 & 90.3 \\
Cyg OB2 8A & O5.5 I & 38.2$^e$ & 25.6$^e$ & 3.56$^e$ & 2650$^e$ & 6575 & 1892 & 65.1 \\
HD 206267 & O6.5 V & 37.9$^c$ & 9.61$^c$  & 3.92$^c$ & 2900$^b$ & 1516 & 419 & 73.5 \\
15 Mon & O7 V & 37.5$^f$ & 9.9$^f$ & 3.84$^f$ & 2150$^b$ & 1621 & 393 & 99.8 \\
$\xi$ Per & O7.5 III & 35.0$^a$ & 14.0$^a$ & 3.50$^a$ & 2450$^b$ & 5603 & 1544 & 158.8 \\
$\tau$ CMa & O9 II & 31.6$^c$ & 17.6$^c$ & 3.41$^c$ & 2200$^b$ & 1300 & 311 & 87.1 \\
$\iota$ Ori & O9 III & 31.4$^f$ & 17.9$^f$ & 3.50$^f$ & 2350$^b$ & 4836 & 1028 & 49.9 \\
$\zeta$ Oph & O9 V & 32.0$^a$ & 8.9$^a$ & 3.65$^a$ & 1550$^b$ & 5911 & 1630 & 83.8 \\
$\delta$ Ori & O9.5 II & 30.6$^c$ & 17.7$^c$ & 3.38$^c$ & 2100$^b$ & 6144 & 1071 & 49.1 \\
$\zeta$ Ori & O9.7 Ib & 30.5$^c$ & 22.1$^c$ & 3.19$^c$ & 1850$^b$ & 9140 & 2003 & 59.6 \\
$\epsilon$ Ori & B0 Ia & 27.5$^g$ & 32.4$^g$ & 3.13$^g$ & 1600$^b$ & 6813 & 1474 & 91.7 \\
  \hline
\end{tabular}
{References: $^a$\citet{Repolust2004}; $^b$\citet{Haser1995}; $^c$\citet{Martins2005}; $^d$\citet{Najarro2011}; $^e$\citet{Mokiem2005}; $^f$\citet{Markova2004}; $^g$\citet{Searle2008} }

\label{tab:sample}
\end{center}
\end{table*}

While small-scale, optically thin clumping can reconcile the X-ray,
\Ha, IR, and radio data for these stars, there is no direct evidence
for optically thick clumping, or porosity, in the X-ray data
themselves\footnote{While optically thick clumping can affect UV
  resonance lines, the opacities of those lines are so large compared
  to X-ray continuum opacities that a given wind can easily have
  optically thick clumping in the UV but be very far from that regime
  in the X-ray.}
\citep{Cohen2008,Sundqvist2012,Herve2013,Leutenegger2013}.  Porosity
results from optically thick clumps, which can hide opacity in their
interiors, enhancing photon escape through the interclump channels.
While porosity has been proposed as an explanation for the
more-symmetric-than-expected observed X-ray line profiles
\citep{ofh2006}, very large porosity lengths are required in order for
porosity to have any effect on line profiles
\citep{oc2006,Sundqvist2012}, and levels of porosity consistent with
measured line profiles produce only modest (not more than about 25 per
cent) effects on derived mass-loss rates \citep{Leutenegger2013}. In
this paper, we derive mass-loss rates from the measured X-ray line
profiles under the assumption that porosity extreme enough to
significantly affect mass-loss rate determinations is not present.

The initial application of the X-ray line profile based mass-loss rate
diagnostic to the O supergiant \zpup\/ gave a mass-loss rate of $\Mdot
= 3.5 \times 10^{-6}$ \Msunyr\/ \citep{Cohen2010}. This represents a
factor of three reduction over the unclumped \Ha\/ value
\citep{Repolust2004,Puls2006}, and is consistent with the newer
analysis of \Ha, IR, and radio data which sets an upper limit of
$\Mdot < 4.2 \times 10^{-6}$ \Msunyr\/ when the effects of clumping
are accounted for \citep{Puls2006}. A similar reduction is found for
the very early O supergiant, HD 93129A, where the X-ray mass-loss rate
of $\Mdot = 6.8 \times 10^{-6}$ \Msunyr\/ is about a factor of 3.5
lower than inferred from unclumped \Ha\/ models, consistent with a
clumping factor $f_{\rm cl} = 3.5^2 \approx 12$ \citep{Cohen2011}.

The goal of this paper is to extend the X-ray line profile mass-loss
rate analysis to all the non-magnetic, effectively
single\footnote{Effectively single in the sense that there is no
  obvious wind-wind interaction-related X-ray emission.} O stars with
grating spectra in the \chandra\/ archive. It is already known that
some, especially later-type, O stars show no obvious wind attenuation
signatures \citep{Miller2002,Skinner2008,Naze2010,Huenemoerder2012},
and as one looks towards weaker winds in early B (V - III) stars, the
X-ray lines are not as broad as the wind velocities would suggest they
should be \citep{Cohen2008}.  Therefore, we have excluded from our
sample very late-O main sequence stars with relatively narrow lines,
but we do include late-O giants and supergiants, even when the
profiles appear unaffected by attenuation.  In these cases we want to
quantify the level of attenuation that may be hidden in the noise,
placing upper limits on their mass-loss rates.  Of course, it is
possible that the model assumptions break down for some of the stars
in the sample, not least of all if wind-wind interactions with a
binary companion are responsible for some of the X-ray emission, in
which case an intrinsically symmetric emission line profile may dilute
whatever attenuation signal is present.

An additional goal of this paper is to constrain wind-shock models of
X-ray production by extracting kinematic and spatial information about
the shock-heated plasma from the line profiles. The profiles are
Doppler broadened by the bulk motion of the hot plasma embedded in the
highly supersonic wind. Our quantitative line profile model allows us
to derive an onset radius of hot plasma and also, for the highest
signal-to-noise lines, the terminal velocity of the X-ray emitting
plasma.  We will use these quantities to test the predictions of
numerical simulations of wind-shock X-ray production.

The paper is organized as follows: In the next section we describe the
data and our sample of O stars taken from the \chandra\/ archive. In
\S3 we describe our data analysis and modeling methodology including
the line profile model, the line profile fitting procedure, and the
derivation of the mass-loss rate from an ensemble of line fits. In \S4
we present our results, including mass-loss rate determinations for
each star in our sample, and in \S5 we discuss the results for each
star in the sample and conclude with a discussion of the implications
of the line profile fitting results.

\section{The Programme Stars}
\label{sec:program_stars}

\subsection{Observations}
\label{subsec:observations}

All observations reported on in this paper were made with \chandra's
High Energy Transmission Grating Spectrometer (HETGS)
\citep{Canizares2005}. The HETGS has two grating arrays: the Medium
and High Energy Gratings (MEG and HEG). The MEG has a FWHM spectral
resolution of 0.023 \AA, while the HEG has a resolution of 0.012 \AA,
but lower sensitivity at the wavelengths of the lines we analyse in
this paper. The dispersion of the grating arrays onto the Advanced
CCD Imaging Spectrometer (ACIS) CCDs lead to bin sizes of 5 and 2.5
m\AA\/ for the MEG and HEG spectra, respectively. We used the standard
reduction procedure ({\sc ciao} 3.3 to 4.3) for most of the spectra,
but for Cyg OB 8A, which is in a crowded field, care had to be taken
to properly centroid the zeroth order spectrum of the target star,
which necessitated the use of a customized reduction procedure within
{\sc ciao}.

The observed spectra consist of a series of collisionally excited
emission lines superimposed on a primarily bremsstrahlung continuum.
The lines arise from high ionization states: most lines are from
helium-like or hydrogen-like ions from abundant (even atomic number)
elements O through Si, and the remainder come from iron L-shell
transitions, primarily in Fe\, {\sc xvii}, but also from higher
stages, especially for stars with hotter plasma temperature
distributions. \chandra\/ is sensitive in the wavelength range from
1.2 to 31 \AA\/ (0.4 to 10 keV).  However, the shortest-wavelength
line we are able to analyse in our sample stars is the Si\, {\sc xiv}
line at 6.182 \AA\/ and the longest is the O\, {\sc vii} line at
21.804 \AA.  The spectra vary in quality -- from 1611 to 13514 total
first-order MEG $+$ HEG counts -- and some suffer from significant
interstellar attenuation at longer wavelengths.  These two factors
determine the number of lines we are able to fit in each star.

\subsection{The sample}
\label{subsec:sample}

We selected every O and very early B star in the \chandra\/ archive as
of 2009 with a grating spectrum -- see {\sc xatlas}
\citep{Westbrook2008} -- that shows obviously wind-broadened emission
lines, aside from \zpup\/ and HD 93129A, which we have already
analysed \citep{Cohen2010,Cohen2011}.  We eliminated from our sample
those stars with known magnetic fields that are strong enough to
provide significant wind confinement \citep{Petit2013} (this includes
$\theta^1$ Ori C and $\tau$ Sco) and we also excluded obvious binary
colliding wind shock (CWS) X-ray sources (such as $\gamma^2$ Vel\/ and
$\eta$ Car) which are generally hard and variable. Some objects
remaining in the sample are possible CWS X-ray sources. They are
included because their spectra -- including their line profiles -- do
not obviously appear to deviate from the expectations of the embedded
wind shock (EWS) scenario, although we give special scrutiny to the
fitting results for these stars in \S\ref{sec:discussion}. It should
be noted that colliding wind binary systems can show non-thermal radio
emission without having significant CWS X-ray emission. We also
exclude main sequence stars and giants with spectral type O9.5 and
later, as these stars (including $\sigma$ Ori A and $\beta$ Cru) have
X-ray lines too narrow to be understood in the context of standard
embedded wind shocks. We ended up including one B star, the supergiant
$\epsilon$ Ori (B0 Ia), which has wind properties very similar to O
stars.  The sample stars and their important parameters are listed in
Table \ref{tab:sample}.  We also include HD 93129A and \zpup\/ in the
table, despite not reporting on their line profile fits in this paper,
because we rederive their mass-loss rates and discuss the results for
those two early O supergiants in conjunction with the results for the
newly analysed stars in \S\ref{sec:results}.

\section{Modeling and Data Analysis Methodology}
\label{sec:modeling}

\subsection{X-ray emission line profile model}
\label{subsec:profile_model}

We use the model of X-ray emission and absorption introduced by
\citet{oc2001}. This model has the benefit of describing a general
X-ray production scenario, making few assumptions about the details of
the physical mechanism that leads to the production of shock-heated
plasma in the wind. The model does assume that the cold, absorbing
material in the wind and the hot, X-ray-emitting material both follow
a $\beta$-velocity law of the form

\begin{equation}
v = \vinf(1 - \Rstar/r)^{\beta},
\label{eqn:beta_law}
\end{equation}

\noindent
where \vinf, the terminal velocity of the wind, usually has a value
between 1500 and 3500 \kms. The $\beta$ parameter, derived from \Ha\/
and UV lines, typically has a value close to unity.  The model also
assumes that the filling factor of X-ray emitting plasma is zero below
some onset radius, \Ro, and is constant above \Ro. Such
emission-measure models with constant filling factor reproduce
observed line profiles quite well \citep{kco2003,Cohen2006}. As
recently discussed by \citet{Owocki2013} (see their fig.\ 3 and
section 4), in analogous models that explicitly account for the
expected radiative nature of embedded shocks in the relatively dense
winds of O-type stars, such fitting of the observed profiles requires
a shock heating rate that declines with radius, roughly as $1/r^2$.
With this adjustment, the form of the emission integral becomes quite
similar to that in the constant filling-factor model. To preserve
continuity with previous analyses
\citep{oc2001,kco2003,Cohen2006,Cohen2010,Cohen2011}, we retain the
latter model here, deferring to future work examination of the (likely
minor) effects of detailed differences from a model that accounts
explicitly for radiative cooling.

Our implementation of the X-ray line profile model\footnote{The {\sc
    xspec} custom model, {\it windprofile}, is publicly available at
  heasarc.gsfc.nasa.gov/docs/xanadu/xspec/models/windprof.html.}
optionally includes the effects of porosity
\citep{oc2006,Sundqvist2012} and of resonance scattering
\citep{Leutenegger2007} on the individual line profile shapes. We
explore the effects of resonance scattering for a subset of stars in
our sample, but because porosity has been shown to have a negligible
effect on observed X-ray profiles and derived mass-loss rates
\citep{Herve2013,Leutenegger2013}, we do not include its effects in
the profile modeling.

The adjustable free parameters of the profile model are generally just
the normalization, which is the photon flux, $F_{\rm line}$, the
parameter that describes the onset radius of X-ray production, \Ro,
and a fiducial optical depth parameter, \taustar, which we describe
below.  For a few high signal-to-noise lines, we allow \vinf, the wind
terminal velocity, to be a free parameter of the fit as well.
Otherwise, we fix this parameter at the literature value listed for
the star in Table \ref{tab:sample}.  The parameter \Ro\/ controls the
widths of the line via the assumed wind kinematics represented by
equation (\ref{eqn:beta_law}). Small values of \Ro\/ correspond to
more X-ray production close to the star where the wind has a small
Doppler shift, while large values of \Ro\/ indicate that most of the
X-rays come from high Doppler shift regions in the outer wind.
Hydrodynamic models show shocks developing about half a stellar radius
above the surface of the star -- albeit with some variation based on
treatments of the line force parameters and of the lower boundary
conditions in numerical simulations
\citep{fpp1997,ro2002,Sundqvist2013}.

The optical depth of the wind affects the blue-shift and asymmetry of
the line profile.  The optical depth at a given location in the wind,
and thus at a given wavelength, is given by 

\begin{equation}
  \tau(p,z) = \int^{\infty}_{z} \kappa(r')\rho(r'){\rm d}z' =  \frac{\Mdot}{4{\pi}\Rstar\vinf} \int^{\infty}_{z} \frac{\kappa(r')\Rstar {\rm d}z'}{r'^2(1-\Rstar/r')^{\beta}}, 
\label{eqn:basic_tau}
\end{equation} 

\noindent
\noindent
where $p,z$ are the usual cylindrical coordinates: the impact
parameter, $p$, is the projected distance from the z-axis centred on
the star and pointing toward the observer, and $r \equiv
\sqrt{p^2+z^2}$.The second equality arises from substituting the
$\beta$-velocity law into the general equation for the optical depth
and employing the mass continuity equation. The profile is calculated
from

\begin{equation}
  L_{\lambda} \propto  \int^{\infty}_{\Ro} {\eta}e^{-\tau} {\rm d}V, 
\label{eqn:basic_luminosity}
\end{equation} 

\noindent
where $\eta$ is the X-ray emissivity, $\tau$ is calculated using
equation (\ref{eqn:basic_tau}), and the volume integral is performed
over the entire wind above $r = \Ro$.


\begin{figure*}
\includegraphics[angle=0,angle=90,width=55mm]{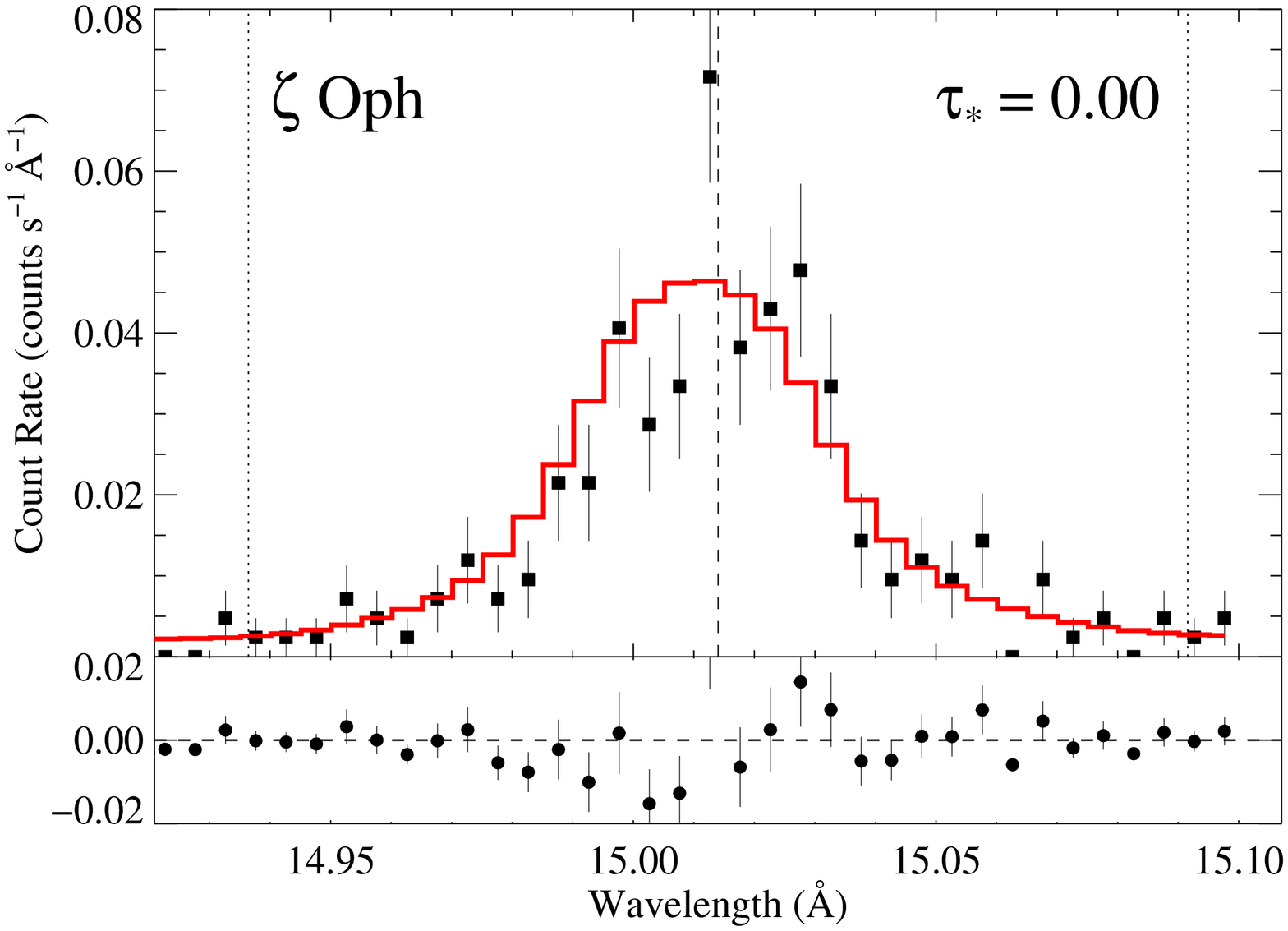}
\includegraphics[angle=0,angle=90,width=55mm]{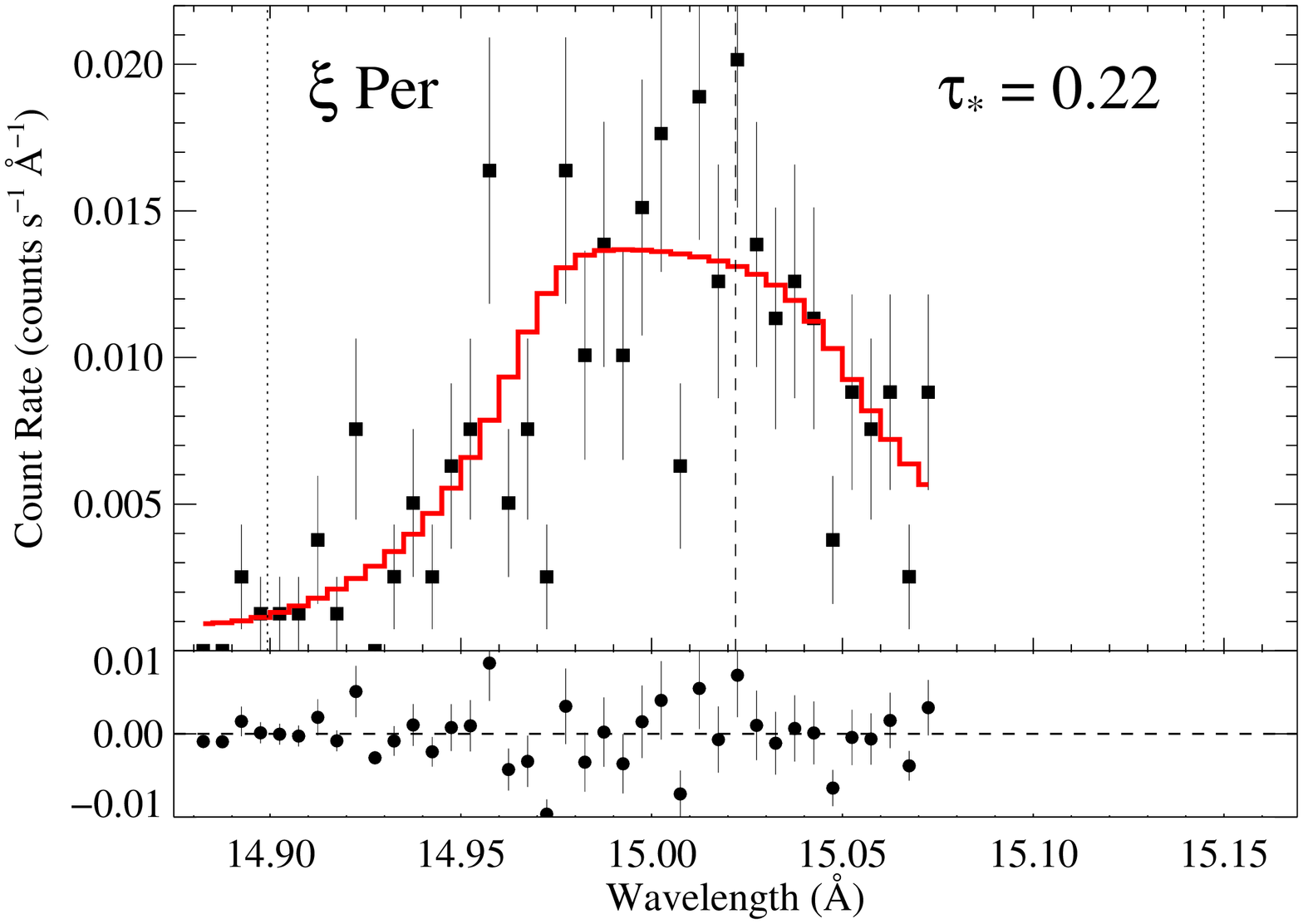}
\includegraphics[angle=0,angle=90,width=55mm]{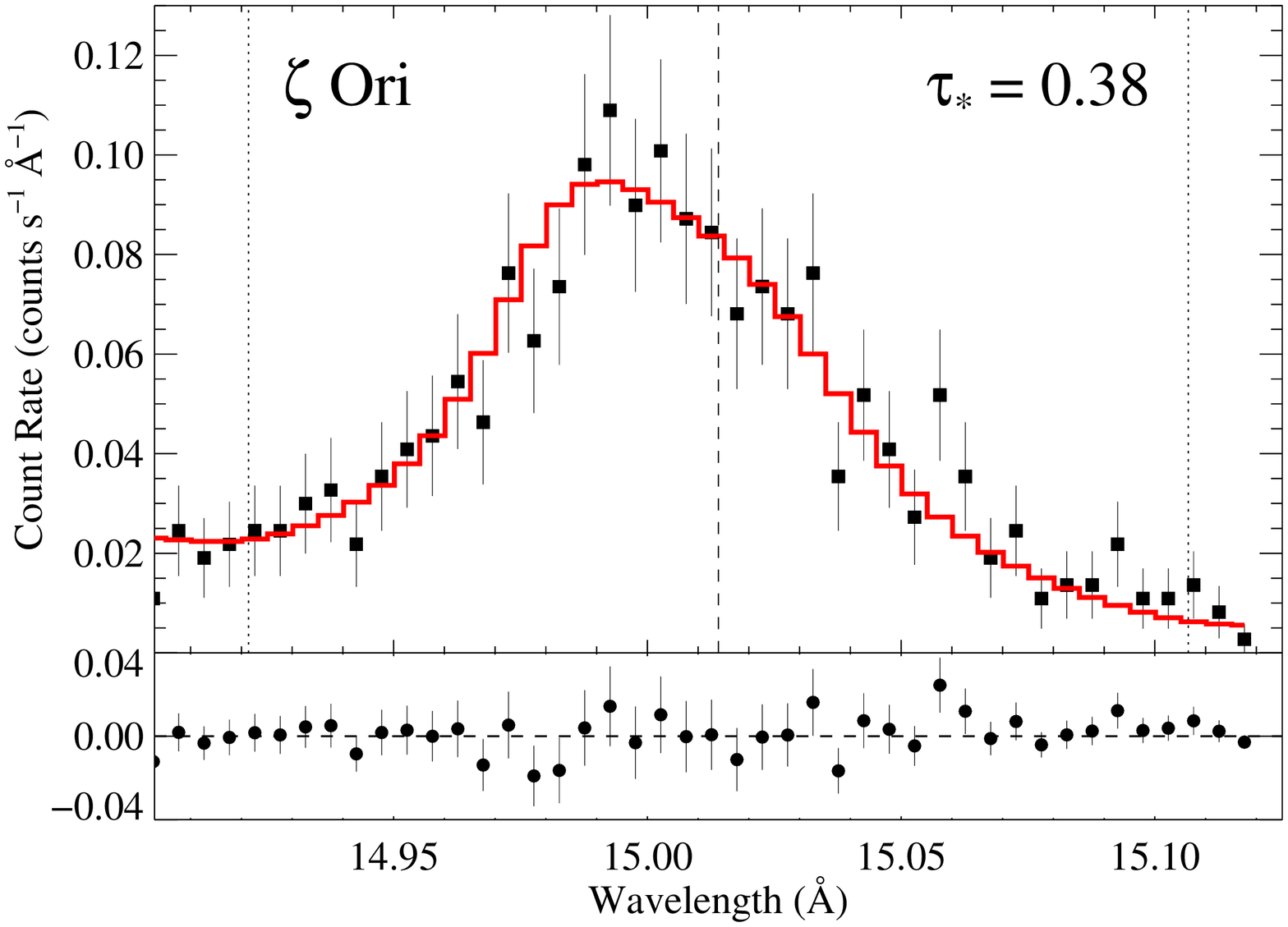}
\includegraphics[angle=0,angle=90,width=55mm]{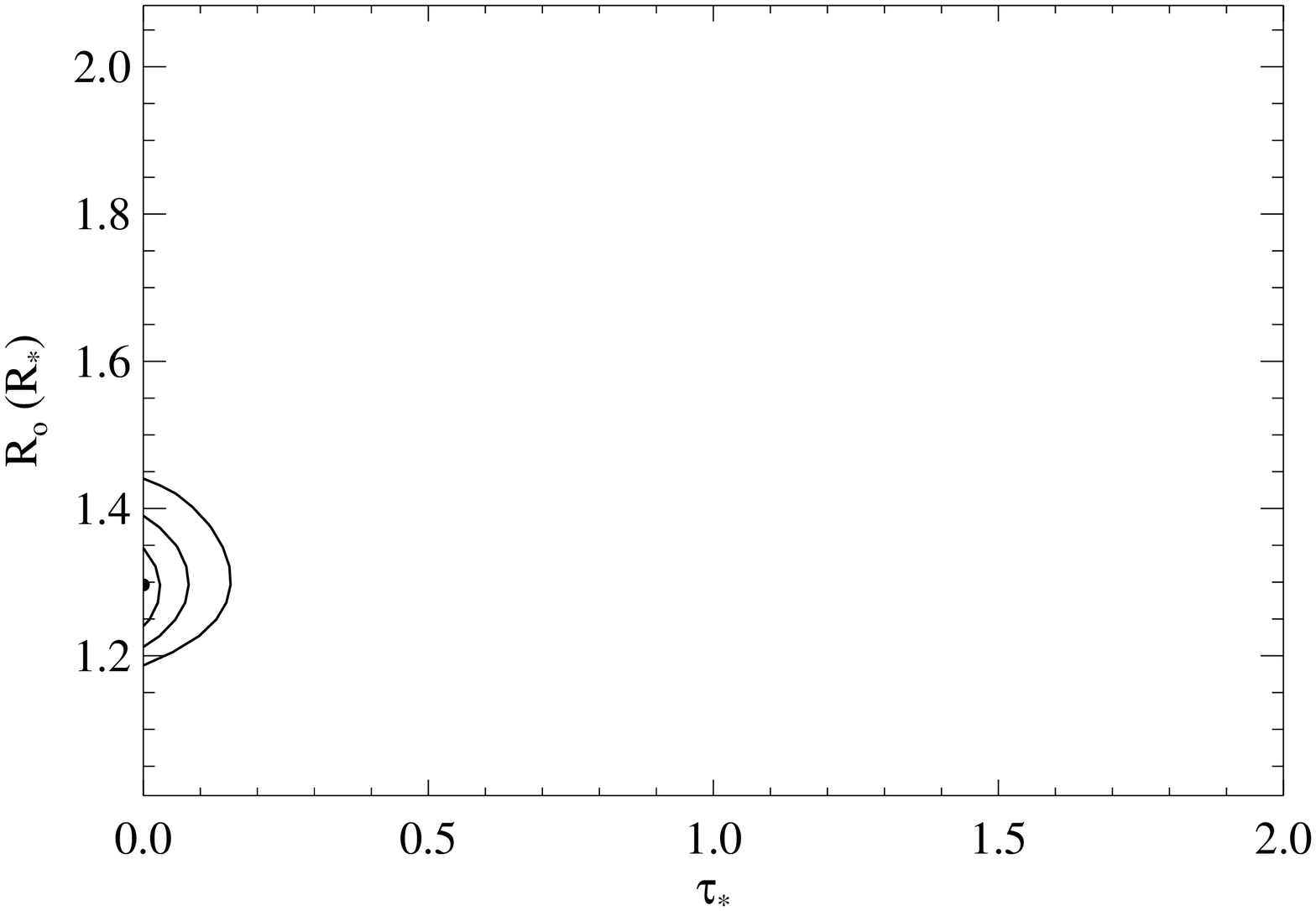}
\includegraphics[angle=0,angle=90,width=55mm]{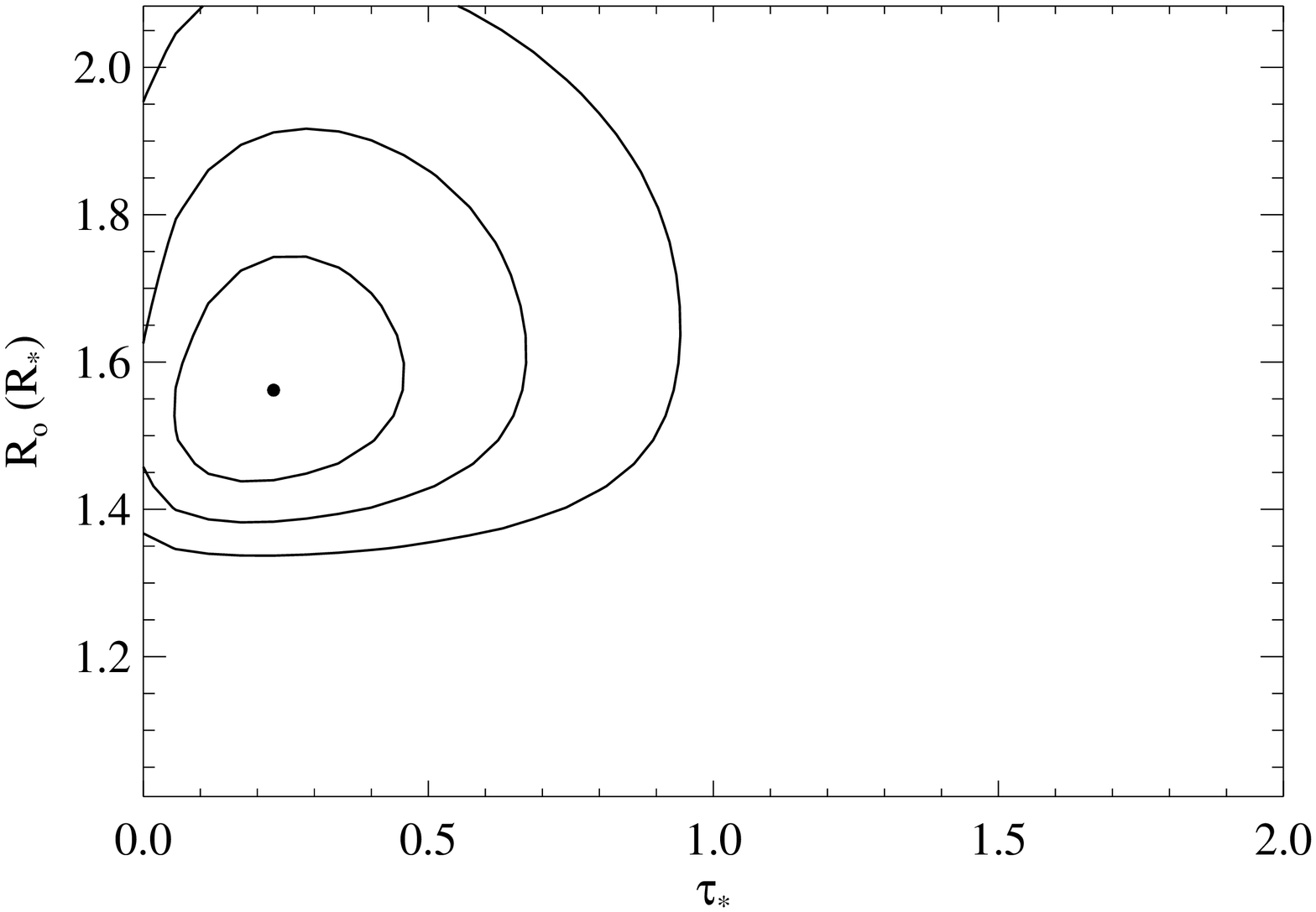}
\includegraphics[angle=0,angle=90,width=55mm]{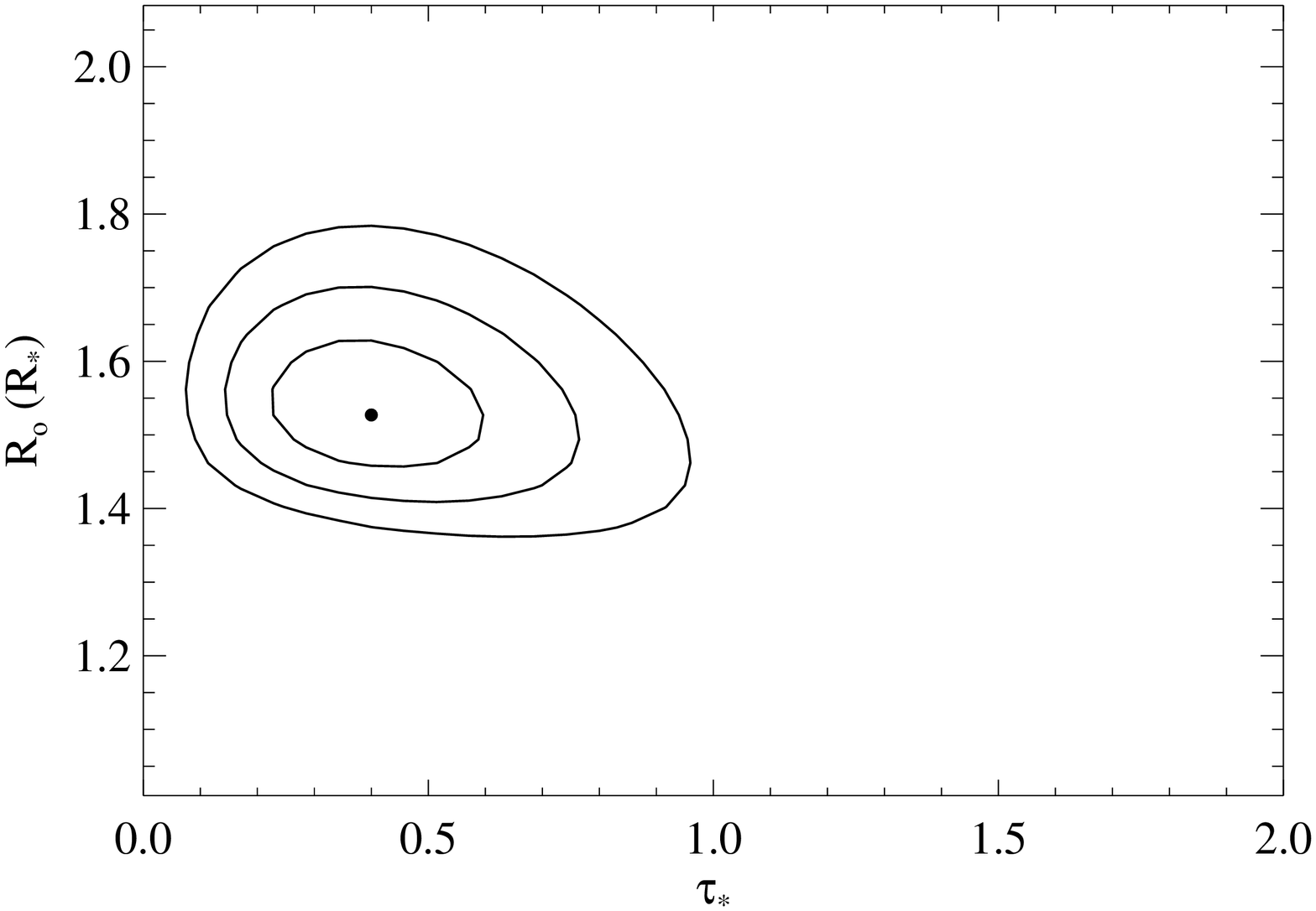}

\caption{The Fe\, {\sc xvii} line at 15.014 \AA\/ with best-fitting
  model (top row) for three of the sample stars [$\zeta$ Oph,
  $\taustar = 0.00^{+.01}_{-.00}$ (left), $\xi$ Per, $\taustar =
  0.22^{+.14}_{-.12}$ (middle), and $\zeta$ Ori, $\taustar =
  0.38^{+.13}_{-.11}$ (right)] showing various degrees of asymmetry.
  The vertical dashed lines on the profile plots represent the
  laboratory line rest wavelength and the wavelengths corresponding to
  the terminal velocity of the wind. Note that the x-axis in each
  figure in the top row encompasses the same velocity range in units
  of the wind terminal velocity, but different absolute velocity and
  wavelength ranges, due to the different terminal velocities of the
  three stars' winds. The star with the highest wind velocity, $\xi$
  Per, is subject to more blending on its red wing than are the other
  two stars.  The contours in the lower panels give the 68, 95, and
  99.7 per cent two-dimensional joint confidence limits on \taustar\/
  and \Ro, while the best-fitting models are indicated by the filled
  circles.  }
\label{fig:iron_lines}
\end{figure*}

We make an important, simplifying assumption at this point, which is
that the continuum opacity, due to photoionization in the cold wind
component, $\kappa(r)$, does not vary substantially with radius in the
wind, and can be replaced with the spatially uniform average opacity,
$\bar{\kappa}$, which we will henceforth write as $\kappa$ for
simplicity. This enables us to pull the opacity out of the spatial
optical depth integral in equation (\ref{eqn:basic_tau}), leading to

\begin{equation}
  \tau(p,z) =  \taustar \int^{\infty}_{z} \frac{\Rstar {\rm d}z'}{r'^2(1-\Rstar/r')^{\beta}}, 
\label{eqn:compact_tau}
\end{equation} 

\noindent
where the constant \taustar, given by

\begin{equation}
\taustar = \frac{\kappa\Mdot}{4\pi\Rstar\vinf},
\label{eqn:taustar}
\end{equation}

\noindent
is the single parameter that characterizes the effect of wind
absorption on the line profile shape. And, along with the
normalization, $F_{\rm line}$, and the $\Ro$ parameter described
above, \taustar\/ is the third free parameter of the profile model we
fit to the data. We note that \taustar\/ is an explicit analytic
expression for the fiducial optical depth parameter, $\tau_{\rm o}$,
first identified by \citet{MacFarlane1991} as the key parametrization
of X-ray line profile shift and asymmetry in the shock-heated winds of
OB stars.

It is key for using X-ray profile fitting to measure mass-loss rates
that \taustar\/ scales with $\Mdot$, but it should be kept in mind
that \taustar\/ also is wavelength dependent, via its dependence on
the opacity, $\kappa$. We hasten to point out, though, that while the
continuum opacity varies from line to line, it does not vary
significantly across a given line. We discuss the wind opacity, and
especially its radial dependence and the effect of our taking it to be
radially uniform, further in \S3.3.

\subsection{Fitting procedure}
\label{subsec:fitting_procedure}
 
All model fitting was done in {\sc xspec} (v12.3 to 12.6). We fit the
positive and negative first order spectra simultaneously, but not
co-added.  Co-added spectra are shown in the figures for display
purposes, however.  When there were a significant number of counts in
the HEG measurements of a given line, we included those data in the
simultaneous fit. In most cases there were negligible counts in the
HEG data and we fit only the MEG data.  Because Poisson noise
dominates these low-count \chandra\/ data, we could not use $\chi^2$
as the fit statistic, and instead used the C statistic
\citep{Cash1979}.  As with $\chi^2$, a lower $C$ value indicates a
better fit, given the same number of degrees of freedom. For placing
confidence limits on model parameters, ${\Delta}C$ is equivalent to
${\Delta}\chi^2$ with a ${\Delta}C$ value of 1 corresponding to a 68
per cent confidence bound in one dimension \citep{Press2007}.  We
establish confidence bounds on the model parameters of interest one at
a time, allowing other parameters to vary while establishing these
bounds.  There is generally a mild anti-correlation between \Ro\/ and
\taustar, so we also examined the joint constraints on two parameters,
adjusting the corresponding value of ${\Delta}C$ accordingly.  Joint
confidence limits are shown in Fig.\ \ref{fig:iron_lines}, along with
the best-fitting models, for the Fe\, {\sc xvii} line at 15.014 \AA\/
for several stars with varying degrees of wind signature strength.

To account for the weak continuum under each emission line, we first
fit a region around the line with a continuum model having a constant
flux per unit wavelength.  This continuum model was added to the line
profile model when fitting the line itself.  The fitting was generally
then done with three free parameters: \taustar, \Ro, and the
normalization, $F_{\rm line}$.  We fixed $\beta$ at 1, and \vinf\/ at
the value given in Table \ref{tab:sample}. A discussion of the effects
of changing $\beta$ and \vinf\/ as well as sensitivity to continuum
placement, treatment of blends, and other aspects of our analysis can
be found in \citet{Cohen2010}. For example, it is found that changing
the wind velocity law exponent, $\beta$, from 1.0 to 0.8 generally
leads to a change in the best-fitting \taustar\/ and \Ro\/ values of
between 10 and 20 per cent. One additional effect we account for is
the radial velocity of each star. This effect was only significant for
$\xi$ Per, which has $v_r = 57$ \kms\/ \citep{Hoogerwerf2001}; no
other star in the sample had a geocentric radial velocity during its
\chandra\/ observation that was this large.


\begin{figure}
\includegraphics[angle=0,angle=90,width=80mm]{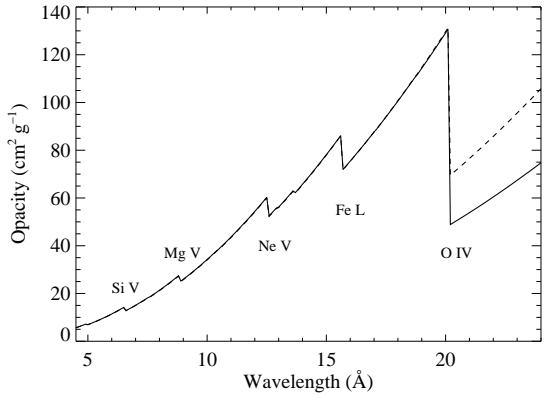}
\caption{Two different models for the wavelength-dependent opacity of
  the bulk wind, with the same simplified ionization balance assumed
  in each case, but altered C, N, and O abundances for the model shown
  as a dashed line. The solar abundance opacity model (solid) line is
  the one we use to derive mass-loss rates. Prominent ionization edges
  are labeled. Note the similarity of the two models shortward of the
  O K-shell edge, which is due to fact that despite the non-solar C,
  N, and O abundances, the metallicity, and thus the sum of the C, N,
  and O abundances, is solar for both models. }
\label{fig:opacity}
\end{figure}

The hydrogen-like \Lya\/ lines in the spectra consist of two blended
lines with wavelength separations that are much smaller than the
resolution of the \chandra\/ gratings. We fit these lines with a
single model centred at the emissivity-weighted average of the two
wavelengths.  In some cases, the lines we wish to analyse are blended.
If the blending is too severe to be modeled, as it is for the O\, {\sc
  viii} \Lyb\/ line at 16.006 \AA, we excluded the line from our
analysis entirely.  If the blended portion of the line could be
omitted from the fit range without producing
unconstrained\footnote{Unconstrained in the sense that the ${\Delta}C$
  criterion does not rule out significant portions of model parameter
  space.} results, we simply fit the model over a restricted
wavelength range.  The Ne\, {\sc x} \Lya\/ line at 12.134 \AA, for
example, produces well-constrained results, even when its red wing is
omitted due to blending with longer-wavelength iron lines.  If lines
from the same ion are blended, such as the Fe\, {\sc xvii} lines at
16.780, 17.051, and 17.096 \AA, we fit three models to the data
simultaneously, constraining the \taustar\/ and \Ro\/ values to be the
same for all the lines in the blended feature.  In the case of the
aforementioned iron complex, we also constrained the ratio of the
normalizations of the two lines at 17.096 and 17.051 \AA, which share
a common lower level, to the theoretically predicted value 
\citep{mlf2001} because the blending is too severe to be constrained
empirically.

The helium-like complexes are among the strongest lines in many of the
sample stars' spectra, but they are generally heavily blended.  The
forbidden-to-intercombination line intensity ratios are a function of
the local mean intensity of the UV radiation at the location of the
X-ray emitting plasma \citep{gj1969,bdt1972}. And so the spatial (and
thus velocity) distribution of the shock-heated plasma affects both
the line intensity ratios and the line profile shapes.  We model these
effects in tandem and fit all three line profiles, including the
relative line intensities, simultaneously, as described in
\citet{Leutenegger2006}.  In order to do this, we use UV fluxes taken
from {\sc tlusty} \citep{lh2003} model atmospheres appropriate for each
star's effective temperature and log $g$ values, as listed in Table
\ref{tab:sample}. This procedure generates a single \taustar\/ value
and a single \Ro\/ value for the entire complex; where \Ro\/ affects
both the line shapes and the $f/i$ ratios, as described above. We
generally had to exclude the results for Ne\, {\sc ix} due to blending
with numerous iron lines.

\subsection{Analysing the ensemble of line fits from each star}
\label{subsec:analysis}

To extract the mass-loss rate from a single derived \taustar\/
parameter value, a model of the opacity of the cold, unshocked
component of the wind is needed.  Then, along with values for the wind
terminal velocity and stellar radius, equation (\ref{eqn:taustar}) can
be used to derive a mass-loss rate for a given line by fitting the
ensemble of $\taustar(\lambda)$ values with \Mdot\/ as the only free
parameter. It has recently been shown for the high signal-to-noise
spectrum of \zpup\/ that if all lines in the spectrum are considered
-- but blends that cannot be modeled are excluded -- and a realistic
model of the wavelength-dependent wind opacity is used, then the
wavelength trend in the ensemble of \taustar\/ values is consistent
with the atomic opacity \citep{Cohen2010}. For other stars, the
wavelength trend of \taustar\/ expected from $\kappa(\lambda)$ may not
be evident, but may still be consistent with it, as has been shown,
recently, for HD 93129A \citep{Cohen2011}.

The opacity of the bulk, unshocked wind is due to bound-free
absorption (inner shell photoionization), and the contributions from
N, O, and Fe are dominant, with contributions from Ne and Mg at
wavelengths below about 12 \AA\/ and some contribution from C and
possibly He at long wavelengths, above the O K-shell edge near 20 \AA\
(see Fig.\ \ref{fig:opacity}; described in more detail below). Each
element has non-zero bound-free cross section only at wavelengths
shortwards of the threshold corresponding to the ionization potential.
The cross-section is always largest at threshold and decreases roughly
as $\lambda^{-3}$ below that\footnote{Near threshold resonances are
  ignored.}. The combined contributions from each abundant element
give the overall wind opacity a characteristic saw-tooth form, with
overall opacity generally being higher at longer wavelengths, but also
dependent on contributions from a smaller number of (low atomic
number) elements at those long wavelengths. For a given element,
higher ionization states have cross-section thresholds at modestly
shorter wavelengths, but very similar cross-sections at all
wavelengths below that. Thus, changes to the bulk wind ionization have
only minor effects on the overall wind opacity.

The actual wind abundances -- and uncertainties in and updates to
their values -- can affect the wind opacity, and thus the
determination of a mass-loss rate from the ensemble of fitted
\taustar\/ values. However, as shown by \citet{Cohen2010,Cohen2011},
the details of any non-solar abundances matter very little, although
the overall opacity does scale as the metallicity and so derived
mass-loss rates will be uncertain to the extent that overall
metallicity is uncertain. However, future adjustments to metallicity
determinations can be easily applied to the derived mass-loss rates,
which we determine here assuming solar metallicity
\citep{Asplund2009}. We make such a correction for \zpup\/ below.

The main reason why the detailed abundances, and specifically CNO
processing, matter very little is that the sum of the absolute
abundances of these three elements should remain the same even if
their relative concentrations are significantly altered. And at
wavelengths below the O K-shell edge, all three elements contribute to
the wind opacity and their cross-sections are very similar.  Depleted
O and enhanced N do in fact have an effect on the cross-section
longward of the O K-shell edge where only N, C, and possibly He
contribute to the wind opacity.  So, enhanced nitrogen will increase
the wind opacity longward of about 20 \AA.  However, partly because
there are few strong lines in the \chandra\/ bandpass at those long
wavelengths and partly because the ISM is generally quite optically
thick at long wavelengths, very few of our programme stars have any
measured lines in the wavelength regime that would be affected by CNO
processing and associated wind opacity modeling uncertainties.


\begin{figure*}
\includegraphics[angle=90,width=57mm]{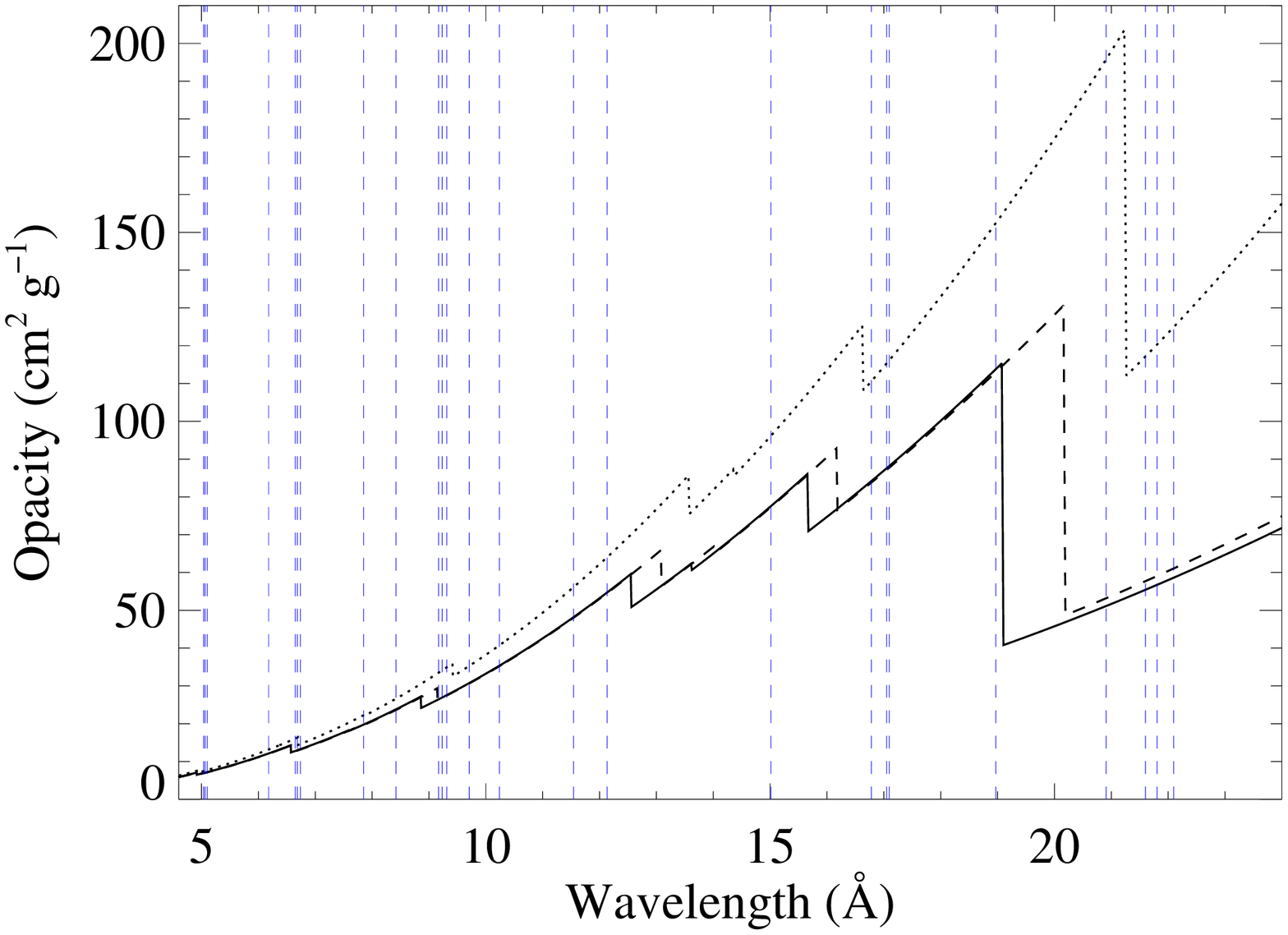}
\includegraphics[angle=90,width=57mm]{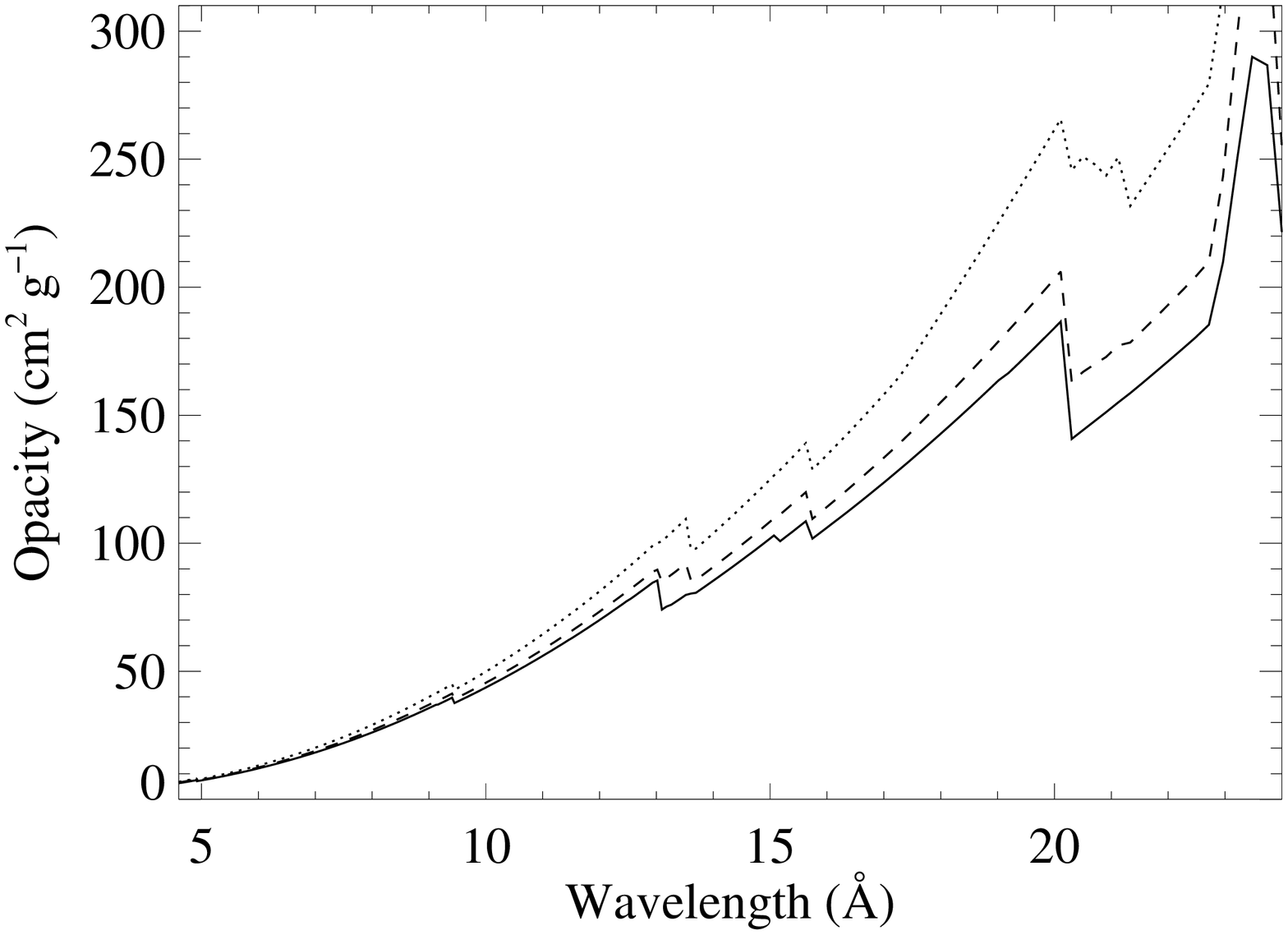}
\includegraphics[angle=90,width=55mm]{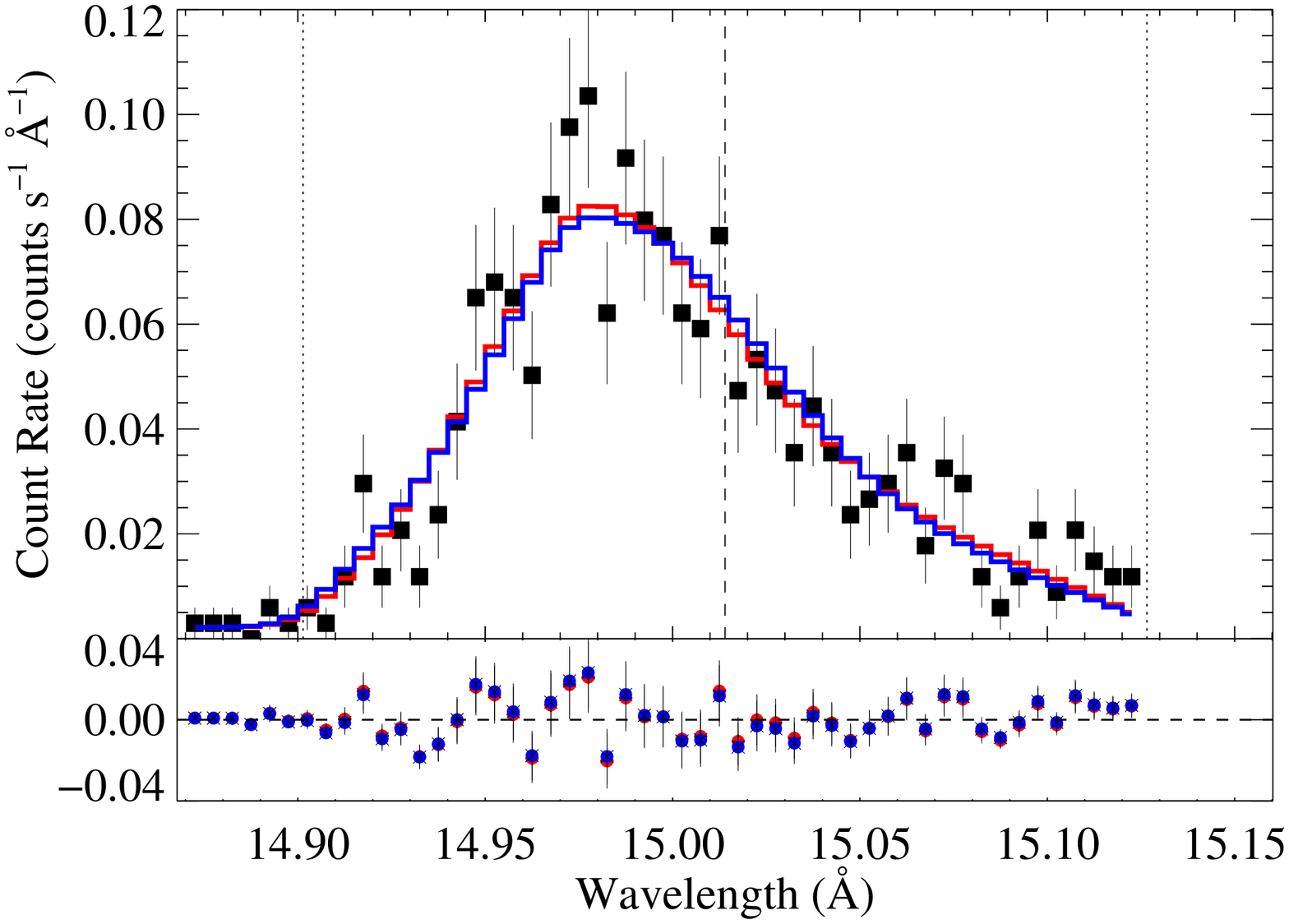}
\caption{{\bf Left:} Opacity models that demonstrate the greatest
  possible radial variation within a single, solar abundance, O star
  wind. Wavelengths of lines measured with the \chandra\/ gratings are
  indicated by the vertical dashed (blue) lines, while the three
  different opacity models assume high ionization (metals in $+4$ and
  He fully ionized; solid), medium ionization (metals in $+3$ and He
  fully ionized; dashed) and low ionization (metals in $+2$ and He
  fully recombined to He\, {\sc ii}; dotted). Clearly, the metal
  ionization differences are a small effect, and the He recombination
  is the dominant effect, but is significant only longward of the
  oxygen K-shell edge near 20 \AA. {\bf Middle:} Opacities specific to
  \zpup, computed via detailed wind modeling using {\sc cmfgen}, at
  three different radii in the wind (1.4, 3.9, 9.6 \Rstar\/ from solid
  to dashed, to dotted). (Note that the overall opacity, especially at
  long wavelengths, is somewhat higher than solar abundance models
  because of the helium abundance enhancement in \zpup\/ and also
  somewhat higher than solar metallicity in the {\sc cmfgen} model.)
  As expected, the opacity variation is small below the O K-shell
  edge, and larger above it, although not as large as the maximal
  scenario presented in the left panel. {\bf Right:} A profile fit to
  the Fe\, {\sc xvii} line at 15.014 \AA\/ of \zpup\/ (blue histogram)
  from a model where the wind opacity triples beyond $r \sim 5$
  \Rstar\/ compared to a model with constant opacity (red histogram).
  The fit with the outer-wind opacity increase demonstrates a high
  degree of degeneracy with constant opacity models, but even in this
  extreme case, the decrease in \taustar\/ is only 30 per cent, as can
  be seen by comparing the first and last rows of Table
  \ref{tab:krat}. }
\label{fig:threeopacities}
\end{figure*}


\begin{table*}
\begin{center}
  \caption{Effect of outer wind opacity increase in \zpup}

\begin{tabular}{cccccccc}
  \hline
  \hline
     &  & Fe\, {\sc xvii} at 15.014 \AA\ &  & &  & O\, {\sc viii} at 18.969 \AA\ & \\
\hline
   Extra opacity  & $\tau_{\ast}$ & R$_{\rm o}$ & C-stat & & $\tau_{\ast}$ & R$_{\rm o}$ & C-stat \\
  \hline
    0   & 1.92 & 1.56 & 280.79 & & 2.99 & 1.22 & 150.89  \\
    0.5 & 1.78 & 1.58 & 282.17 & & 2.82 & 1.24 & 150.68  \\
    1   & 1.66 & 1.60 & 283.45 & & 2.66 & 1.26 & 150.68  \\
    2   & 1.48 & 1.62 & 285.71 & & 2.31 & 1.34 & 150.98  \\
    3   & 1.33 & 1.63 & 287.66 & & 1.86 & 1.53 & 151.22   \\
  \hline
    \end{tabular}

\label{tab:krat}
\end{center}
\end{table*}  

Returning to ionization, the largest effect on the opacity due to
differences or uncertainties in the ionization comes from
recombination of He$^{++}$ to He$^{+}$ in the outer wind. Fully
ionized helium has no bound-free opacity but singly ionized helium has
significant opacity at long wavelengths (the ionization edge is at
${\rm h}{\nu} = 54.4$ eV; $\lambda = 228$ \AA), and can have an effect
on the total wind opacity longward of the O K-shell edge near 20 \AA.
This, and other secondary ionization effects, can lead to some
differences in the wind opacity as a function of radius (see, e.g.,
figs 1 and 2 of \citealt{Herve2013}). However, the significant changes
are almost entirely at the long-wavelength end of the \chandra\/
bandpass, where helium has a disproportionate effect, and where there
are very few, if any, emission lines in our programme stars' spectra.
Furthermore, although the opacity may change by roughly a factor of 2
in the outer wind, the density is so much lower there that the
contribution of the outer wind to the column density -- and thus the
optical depth -- along a typical sight line is negligible.  For
example, doubling the wind opacity beyond 5 \Rstar\/ increases the
optical depth typically by only 10 per cent along sight-lines that
pass through the densest parts of the wind.


\begin{figure*}
\includegraphics[angle=0,angle=90,width=58mm]{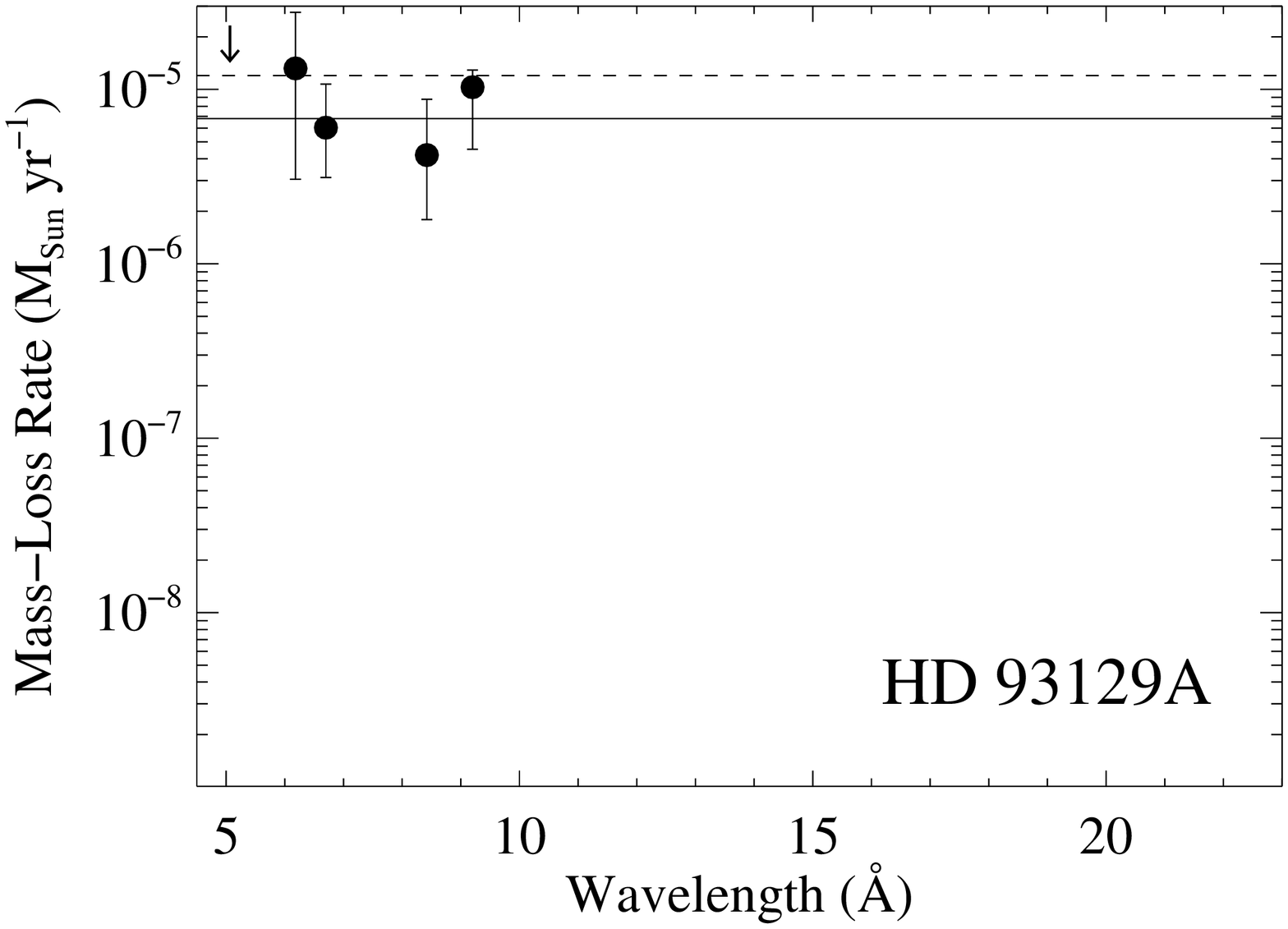}
\includegraphics[angle=0,angle=90,width=58mm]{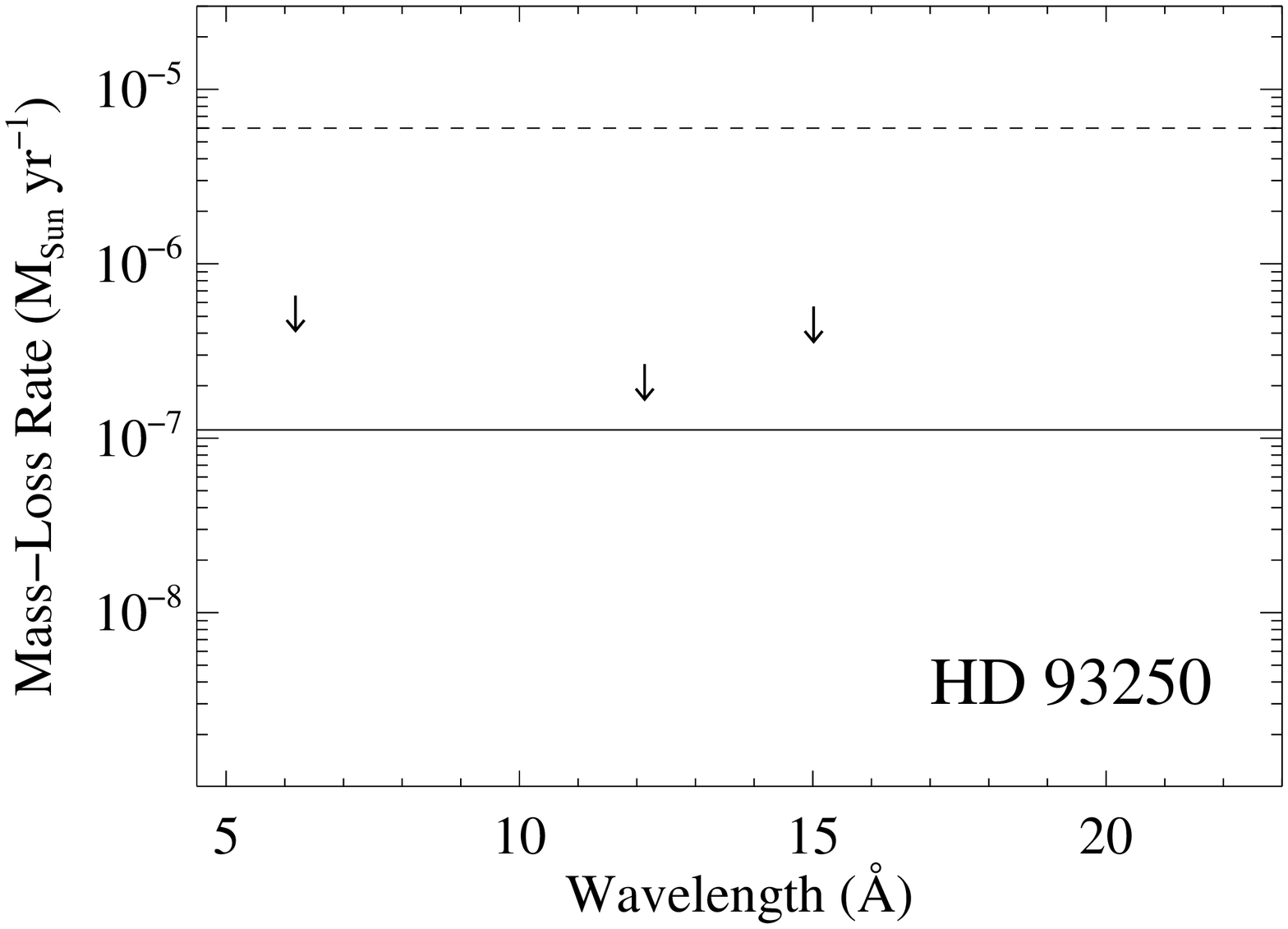}
\includegraphics[angle=0,angle=90,width=58mm]{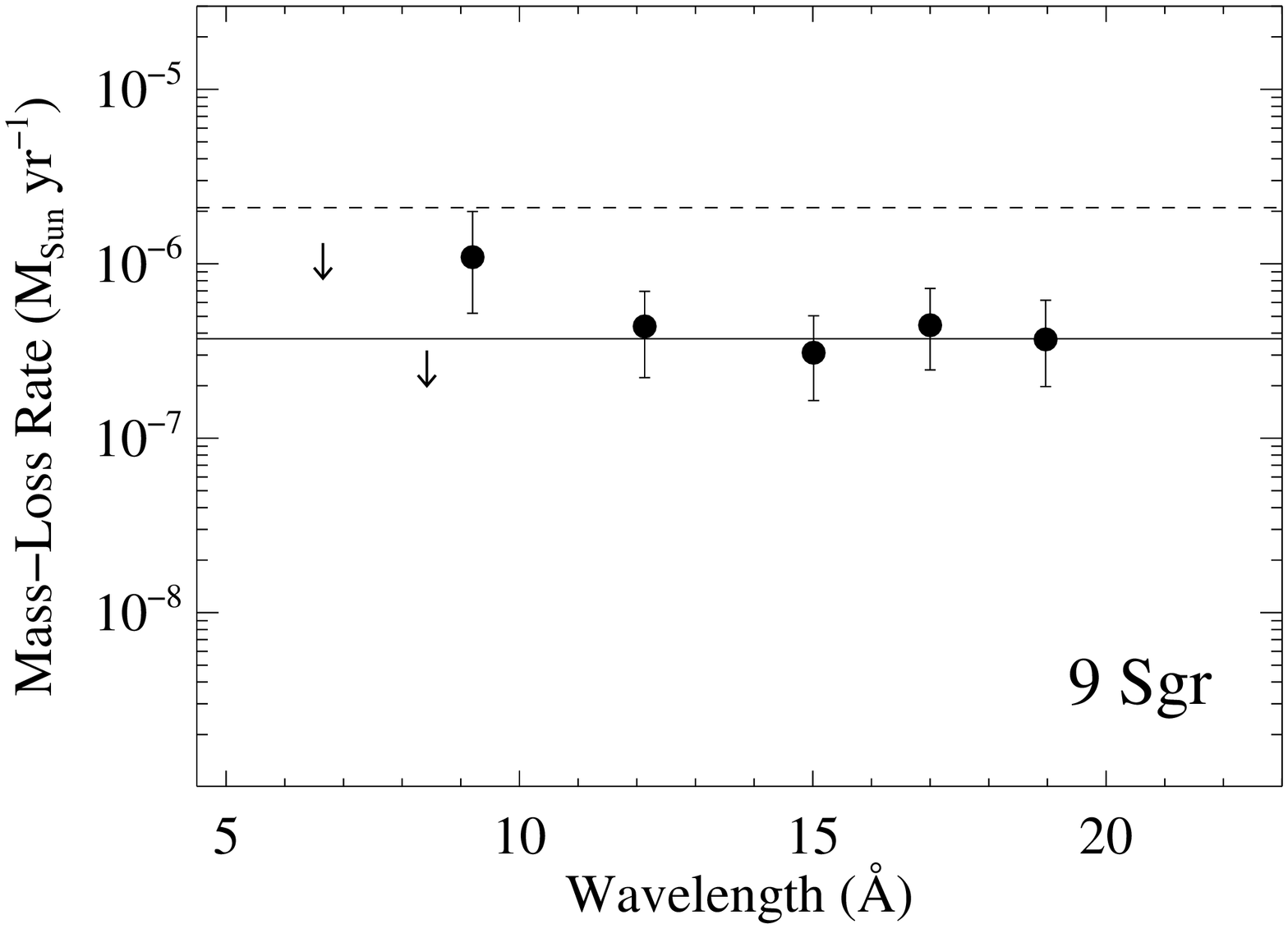}
\includegraphics[angle=0,angle=90,width=58mm]{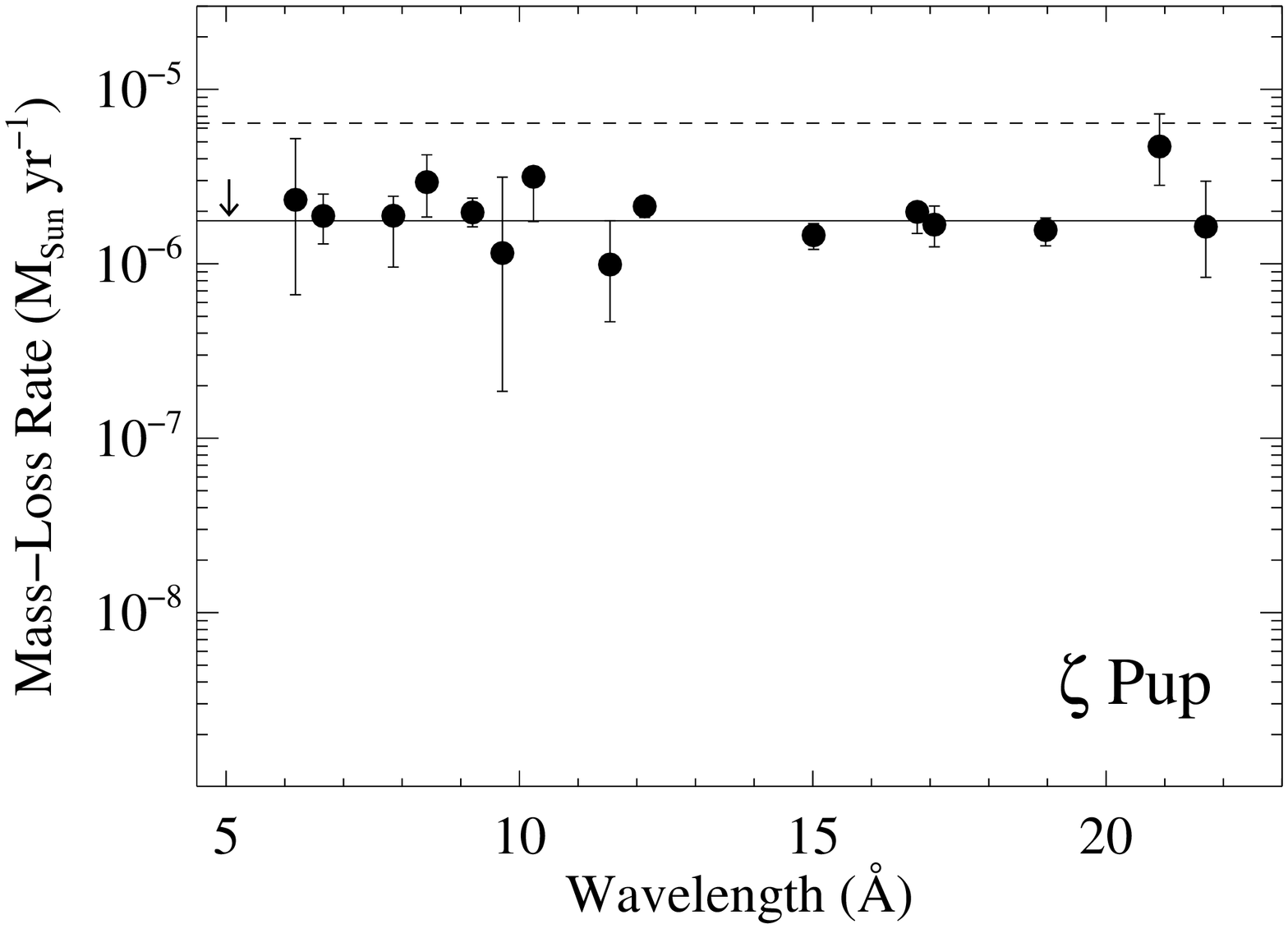}
\includegraphics[angle=0,angle=90,width=58mm]{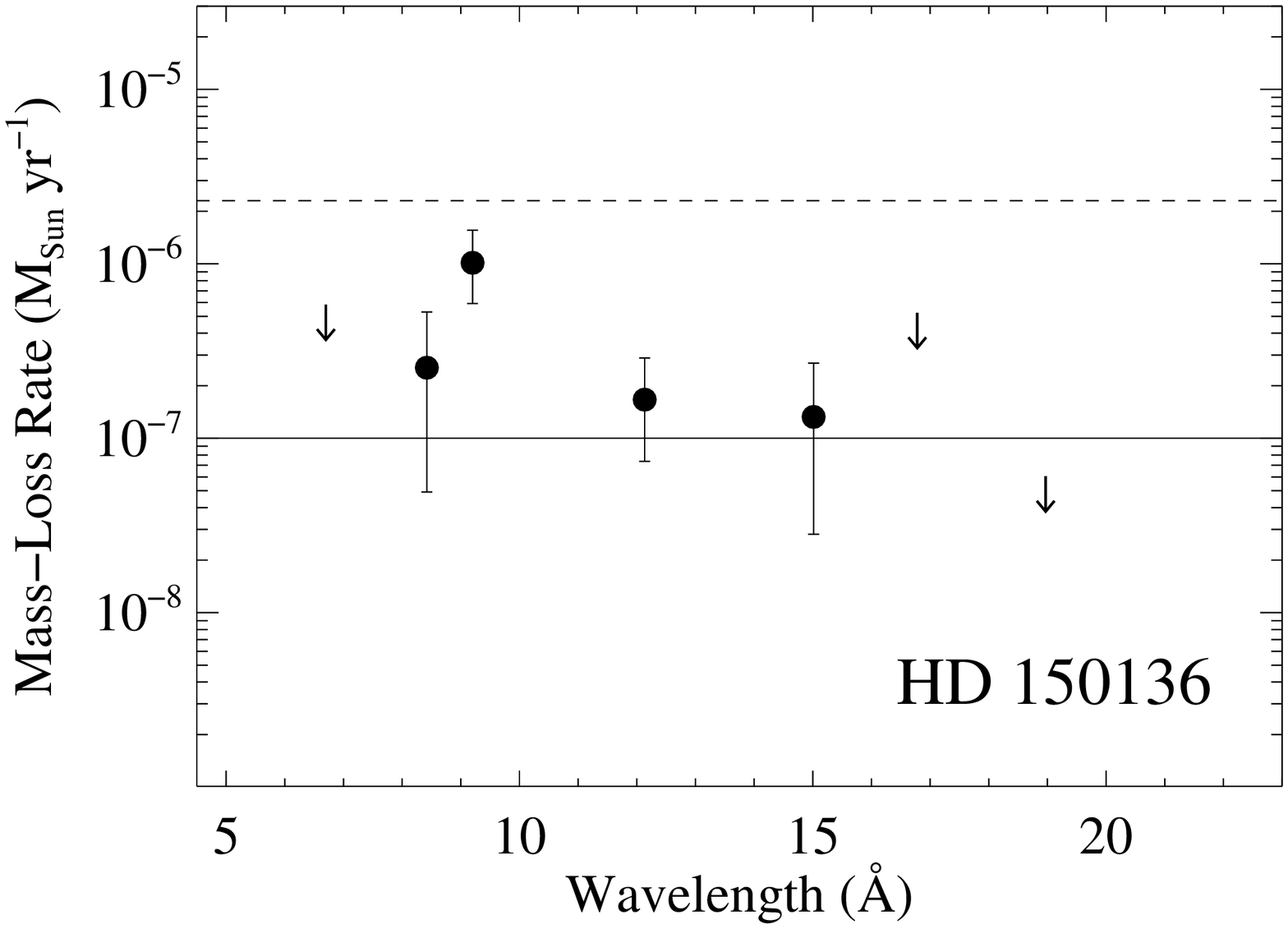}
\includegraphics[angle=0,angle=90,width=58mm]{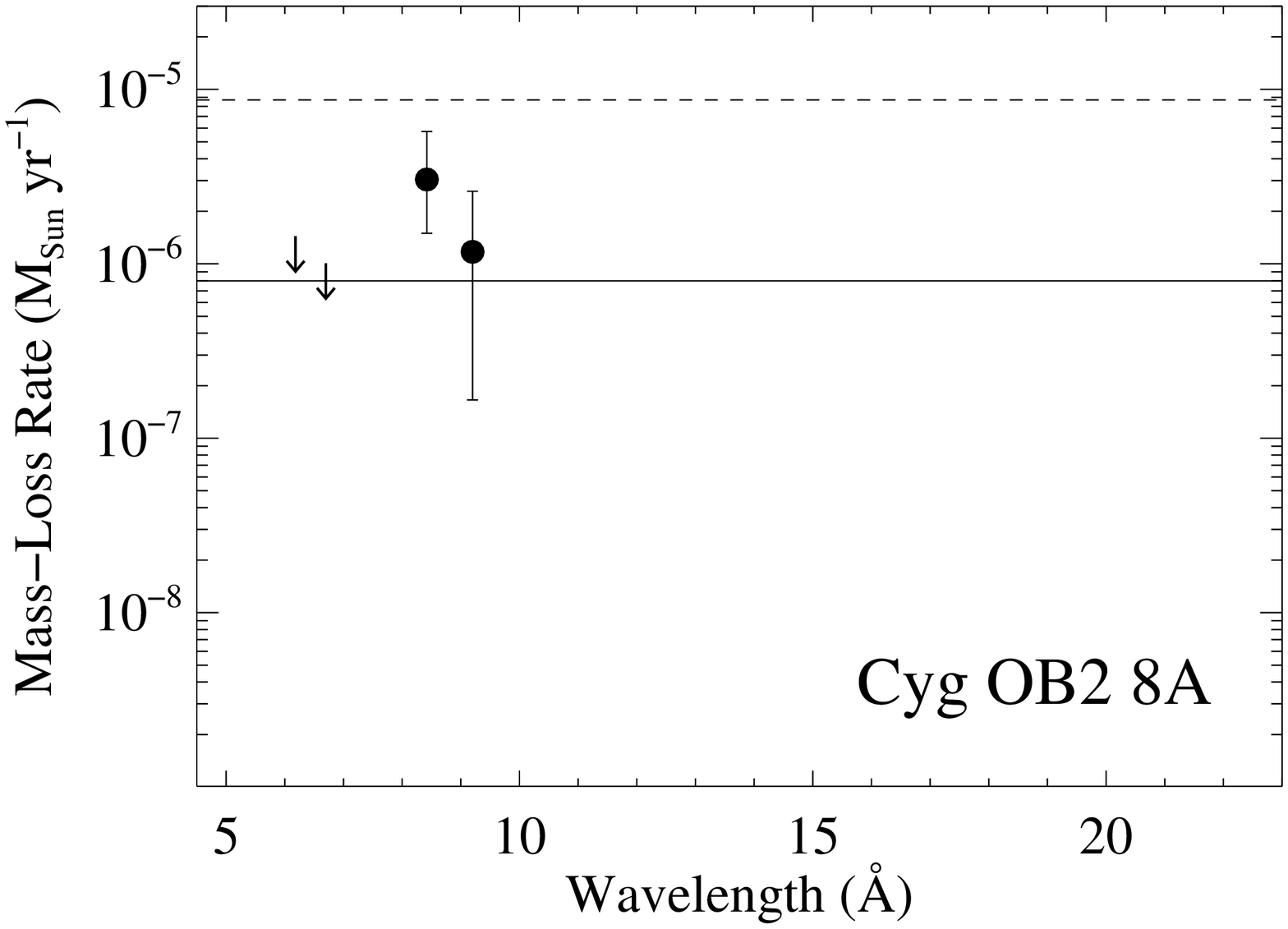}
\includegraphics[angle=0,angle=90,width=58mm]{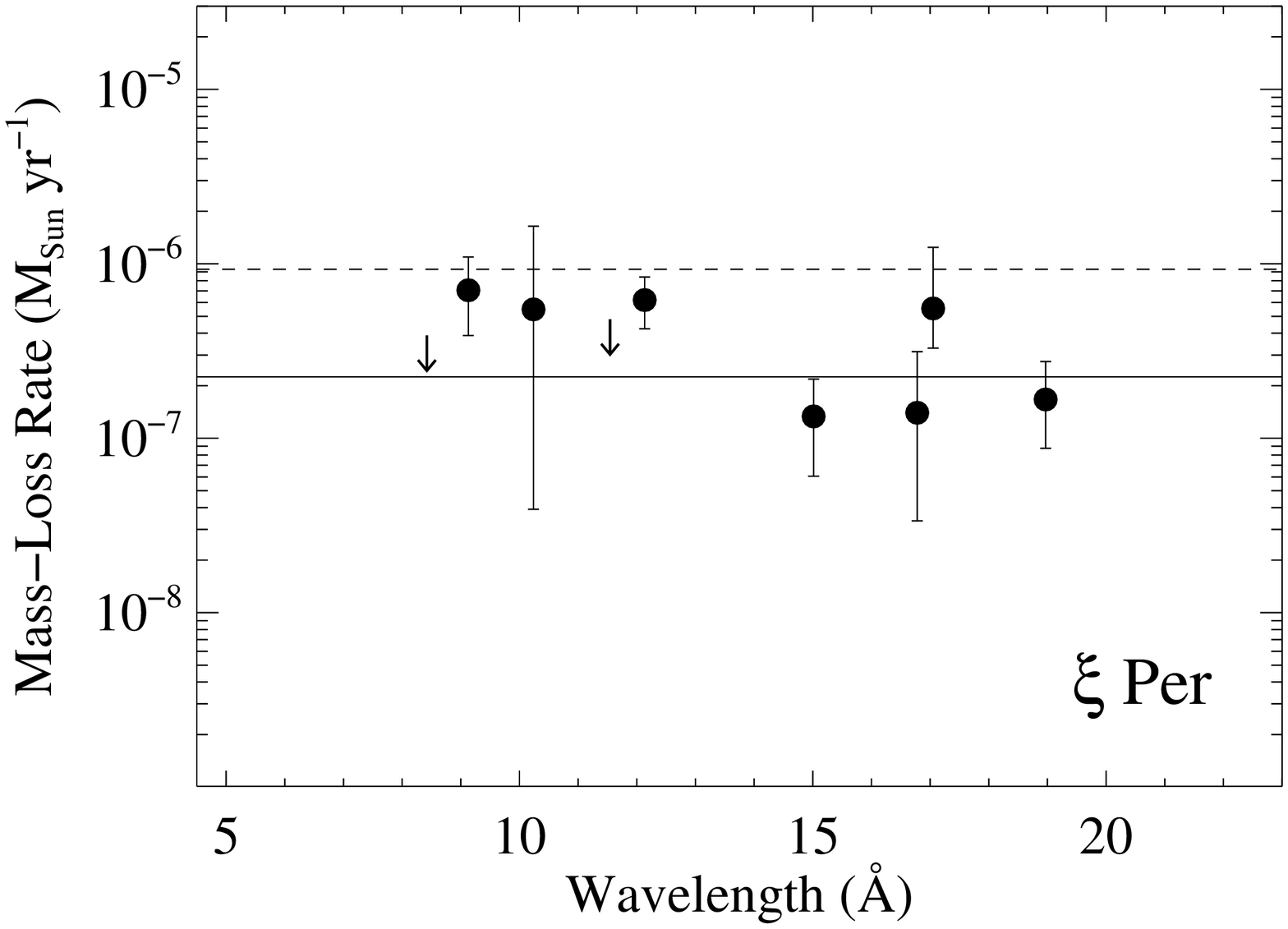}
\includegraphics[angle=0,angle=90,width=58mm]{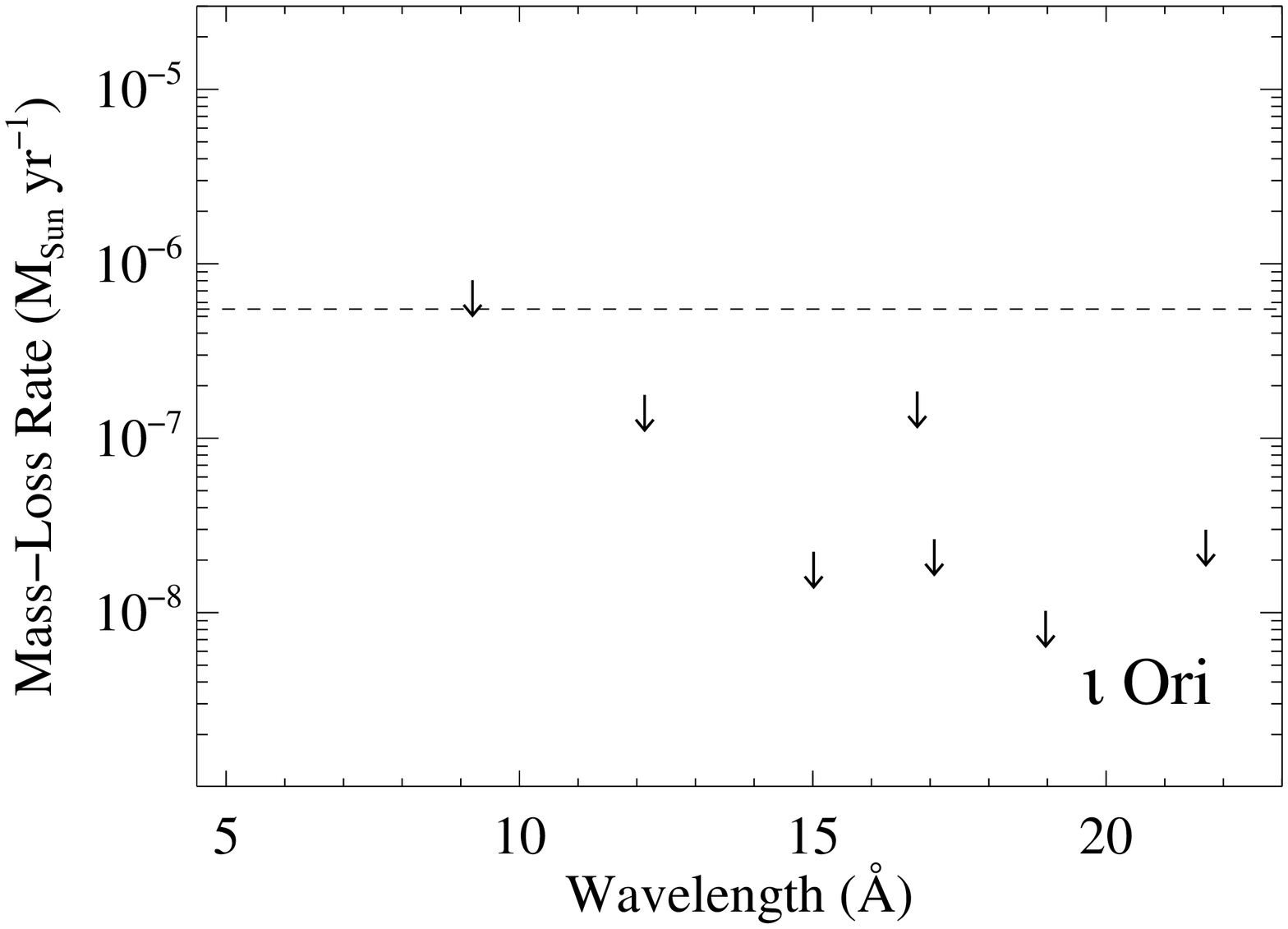}
\includegraphics[angle=0,angle=90,width=58mm]{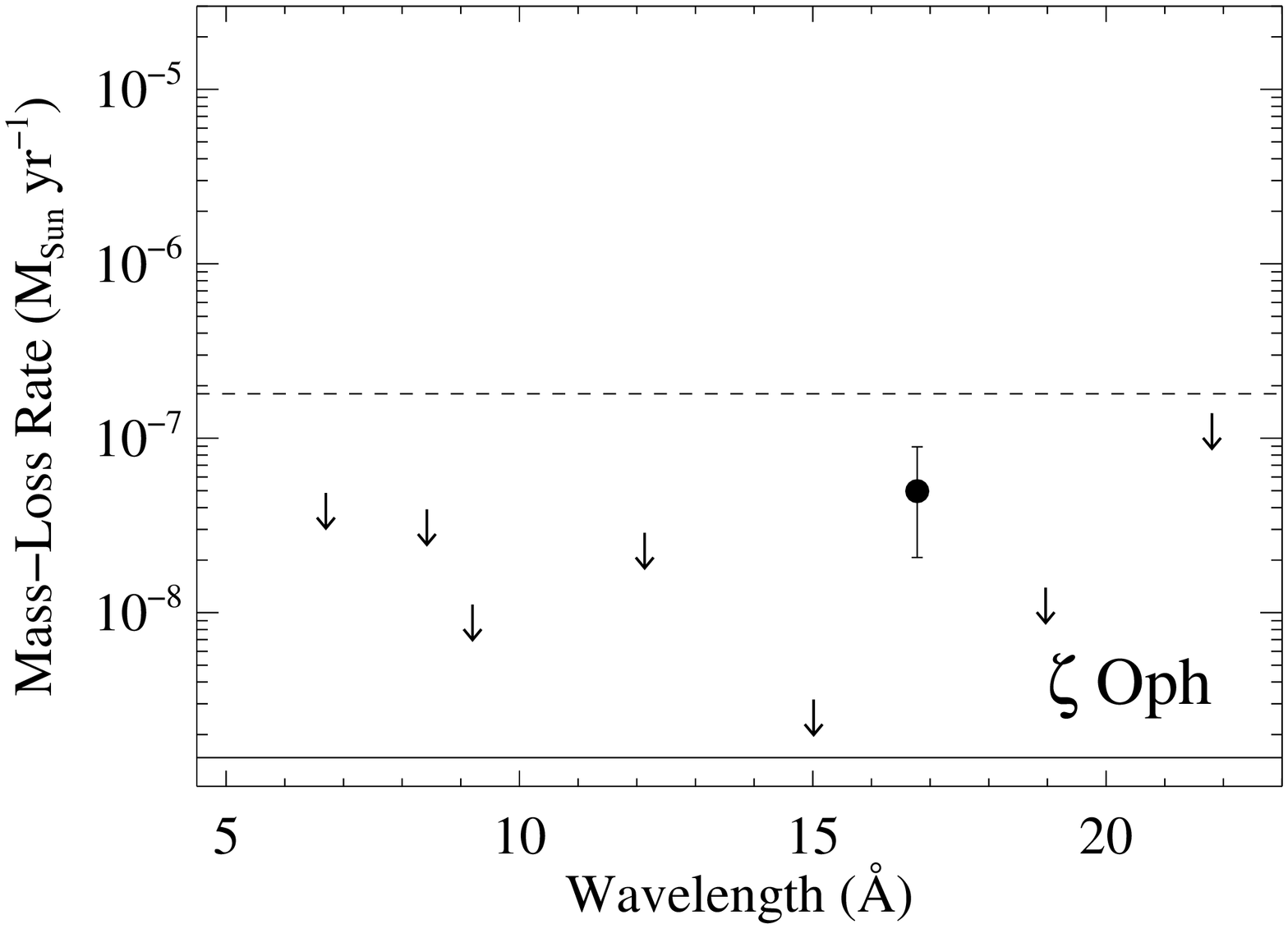}
\includegraphics[angle=0,angle=90,width=58mm]{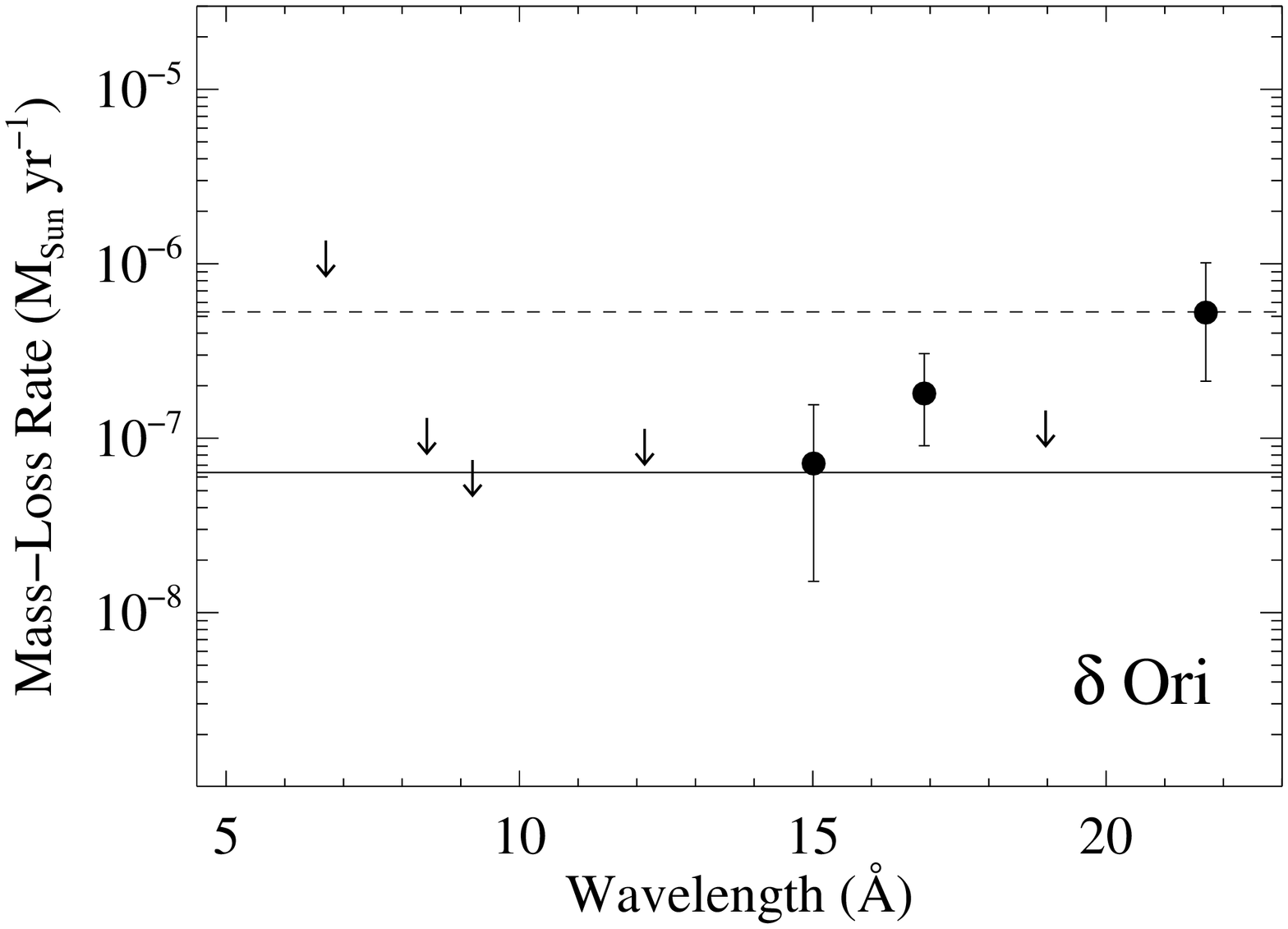}
\includegraphics[angle=0,angle=90,width=58mm]{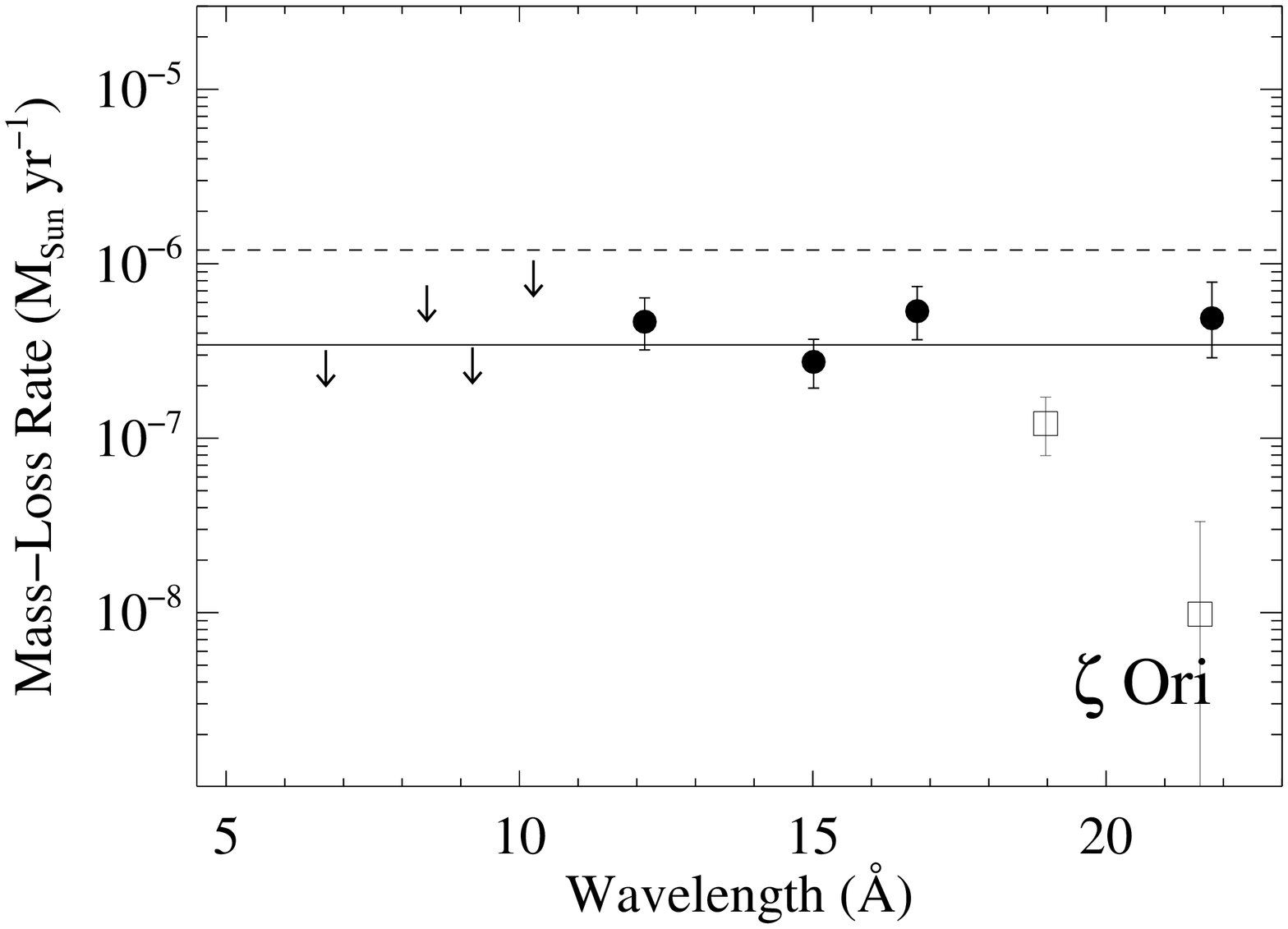}
\includegraphics[angle=0,angle=90,width=58mm]{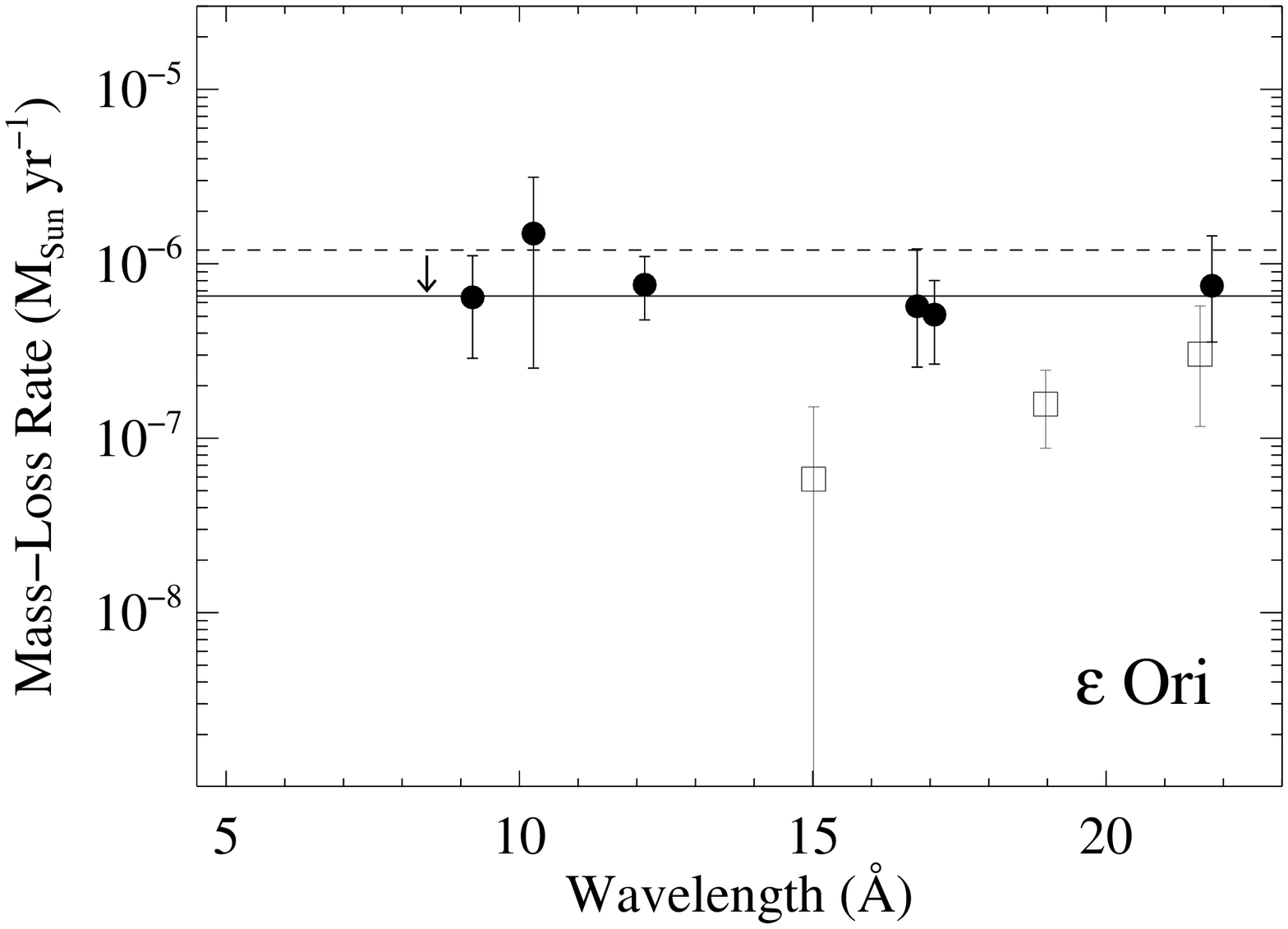}
\caption{The fitted \taustar\/ values (points), along with the 68 per
  cent confidence limits (error bars), converted to a mass-loss rate,
  are shown for each line in each star. Lines for which the fitted
  \taustar\/ values are within 1 $\sigma$ of zero are indicated as
  upper limits (at the best-fitting plus 1 $\sigma$ values).  The
  fitted mass-loss rates for each star are indicated by the solid
  lines, while the dashed line in each panel represents the
  theoretical mass-loss rate listed in Table \ref{tab:results}. Note
  that the y-axis range is the same in each panel, and that the
  best-fitting mass-loss rate for $\iota$ Ori is so low that it is off the
  bottom of that panel. For $\zeta$ Ori and $\epsilon$ Ori, we show the
  two and three points, respectively, omitted from the mass-loss rate
  fits because of resonance scattering (open squares).
}
\label{fig:taustar}
\end{figure*}


\begin{figure*}
\includegraphics[angle=0,angle=90,width=58mm]{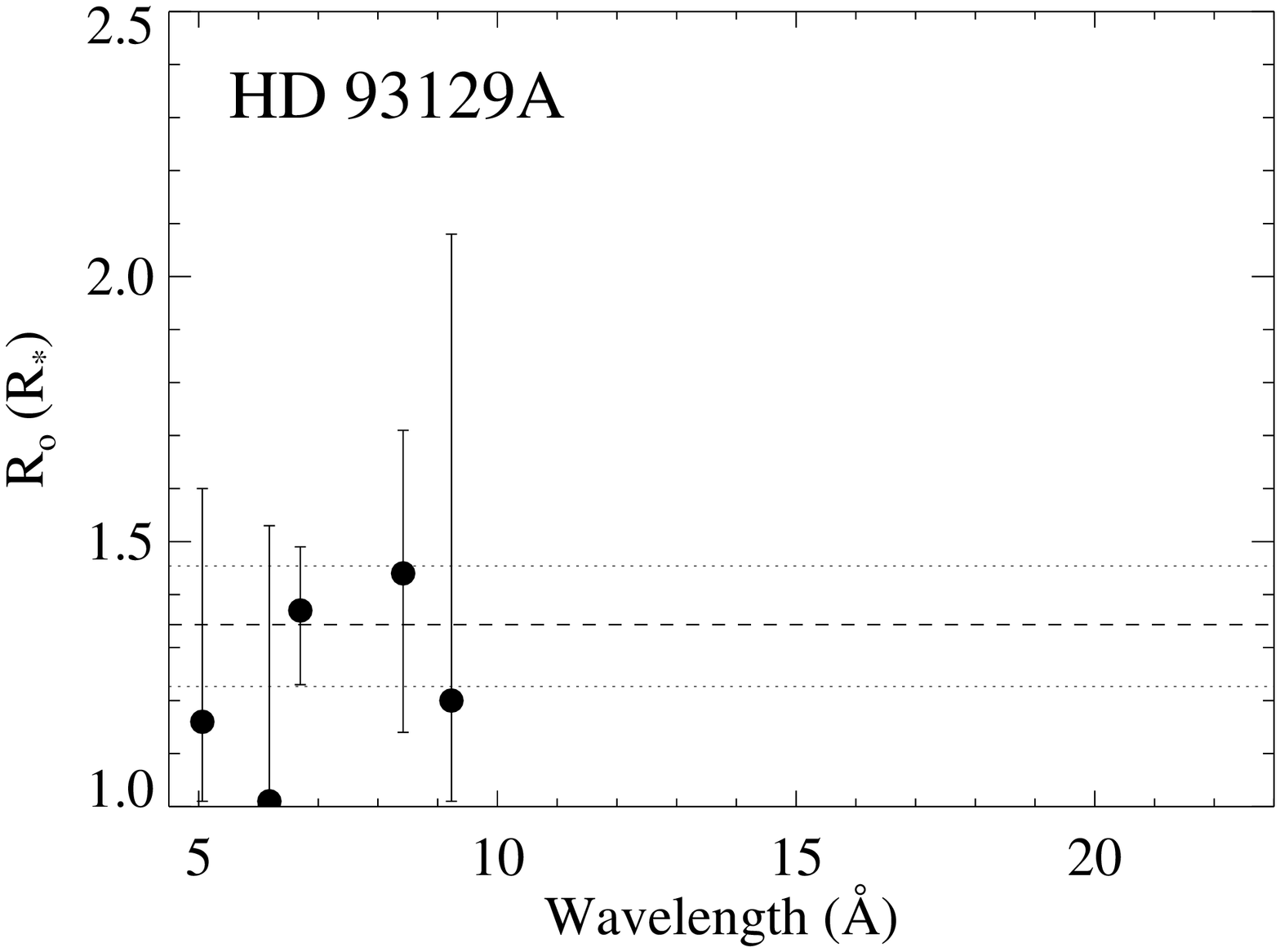}
\includegraphics[angle=0,angle=90,width=58mm]{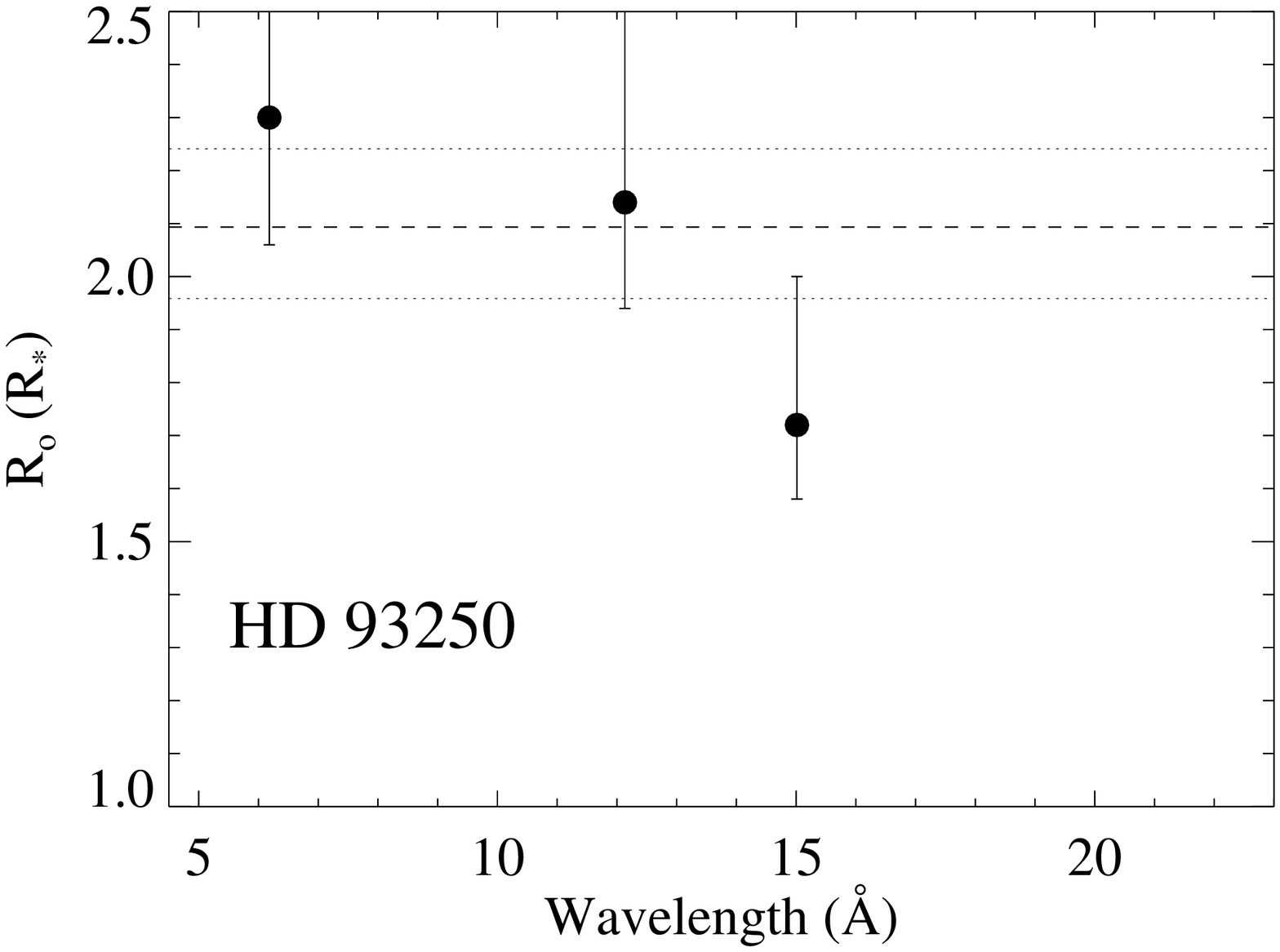}
\includegraphics[angle=0,angle=90,width=58mm]{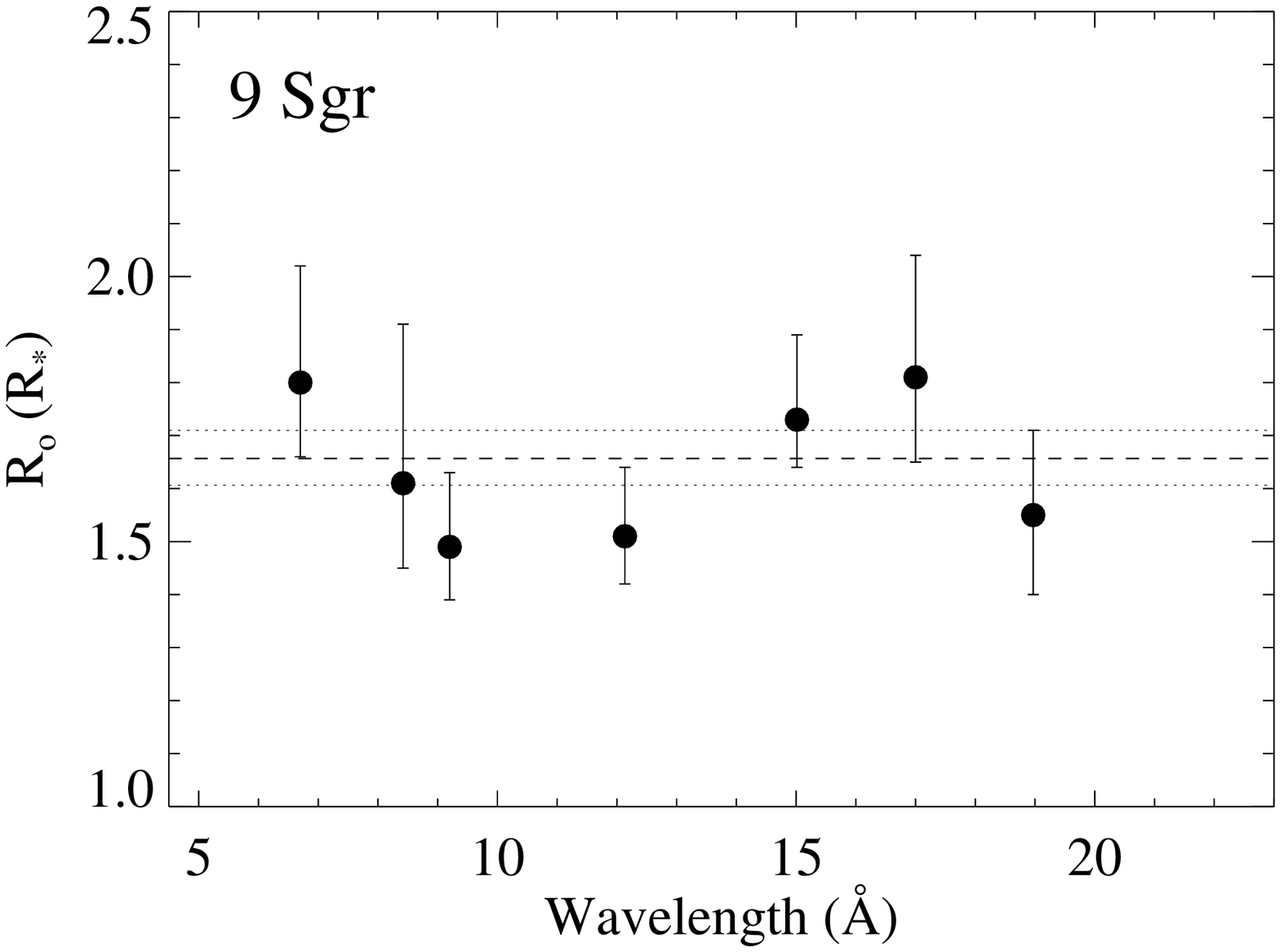}
\includegraphics[angle=0,angle=90,width=58mm]{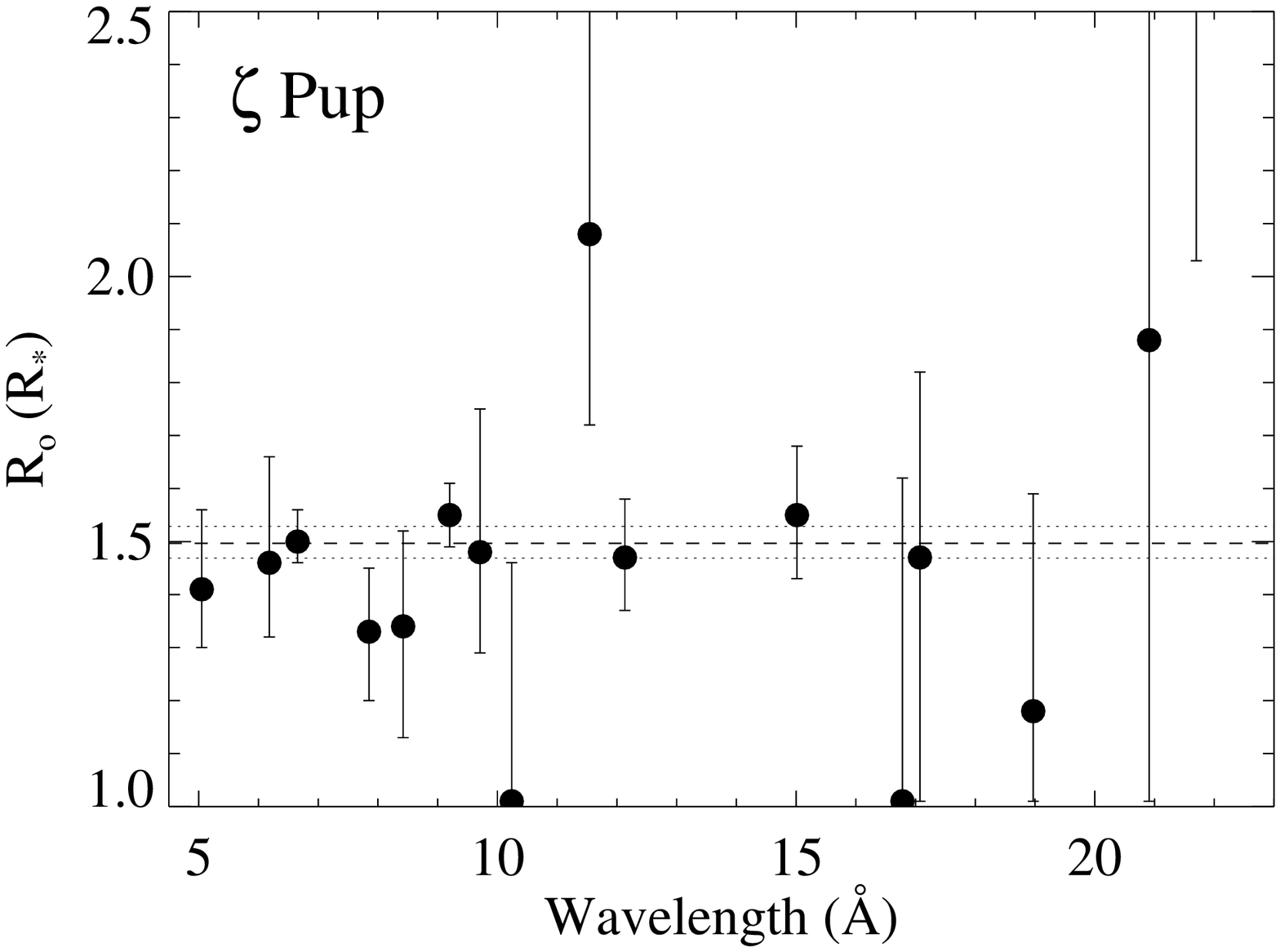}
\includegraphics[angle=0,angle=90,width=58mm]{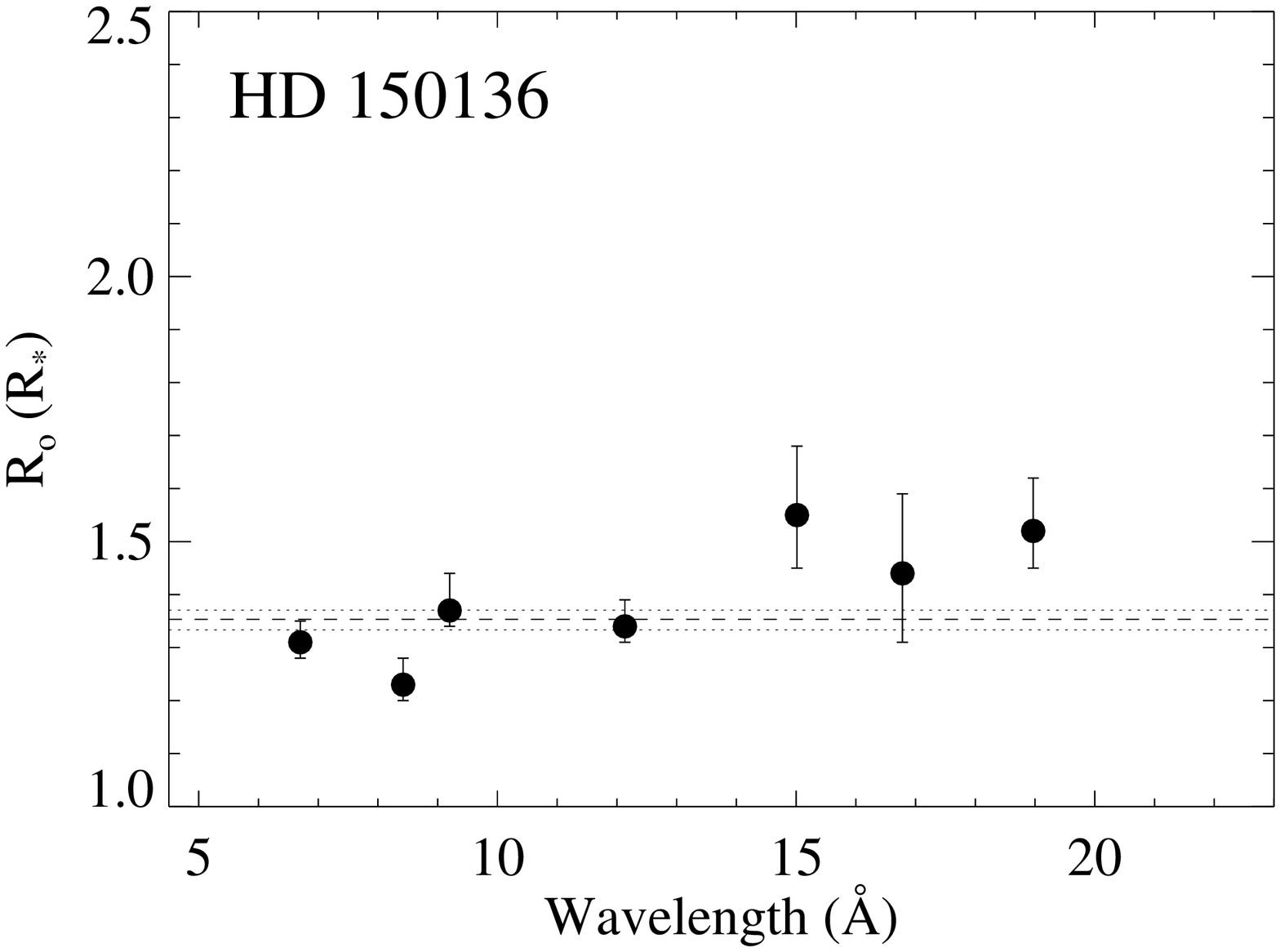}
\includegraphics[angle=0,angle=90,width=58mm]{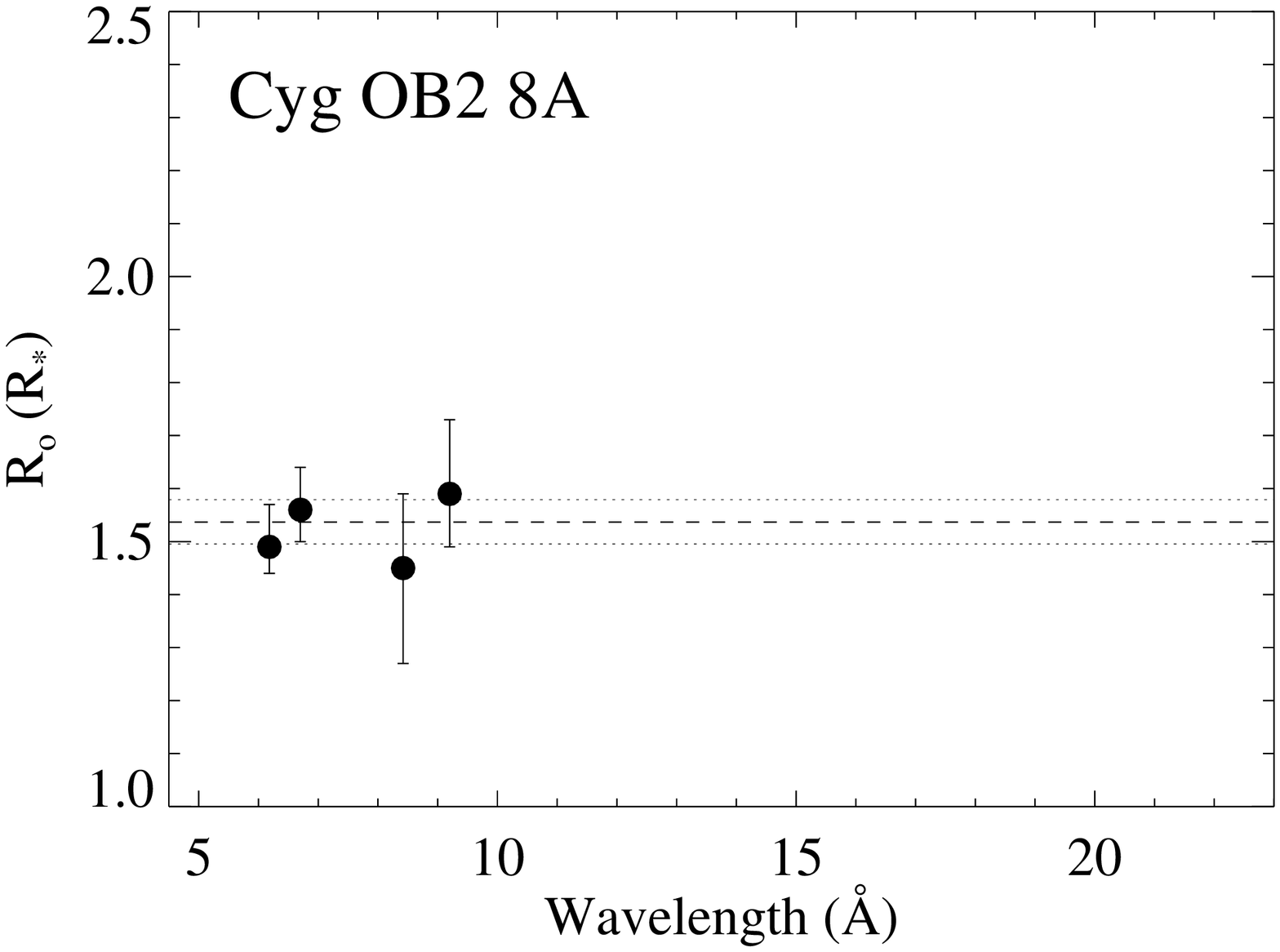}
\includegraphics[angle=0,angle=90,width=58mm]{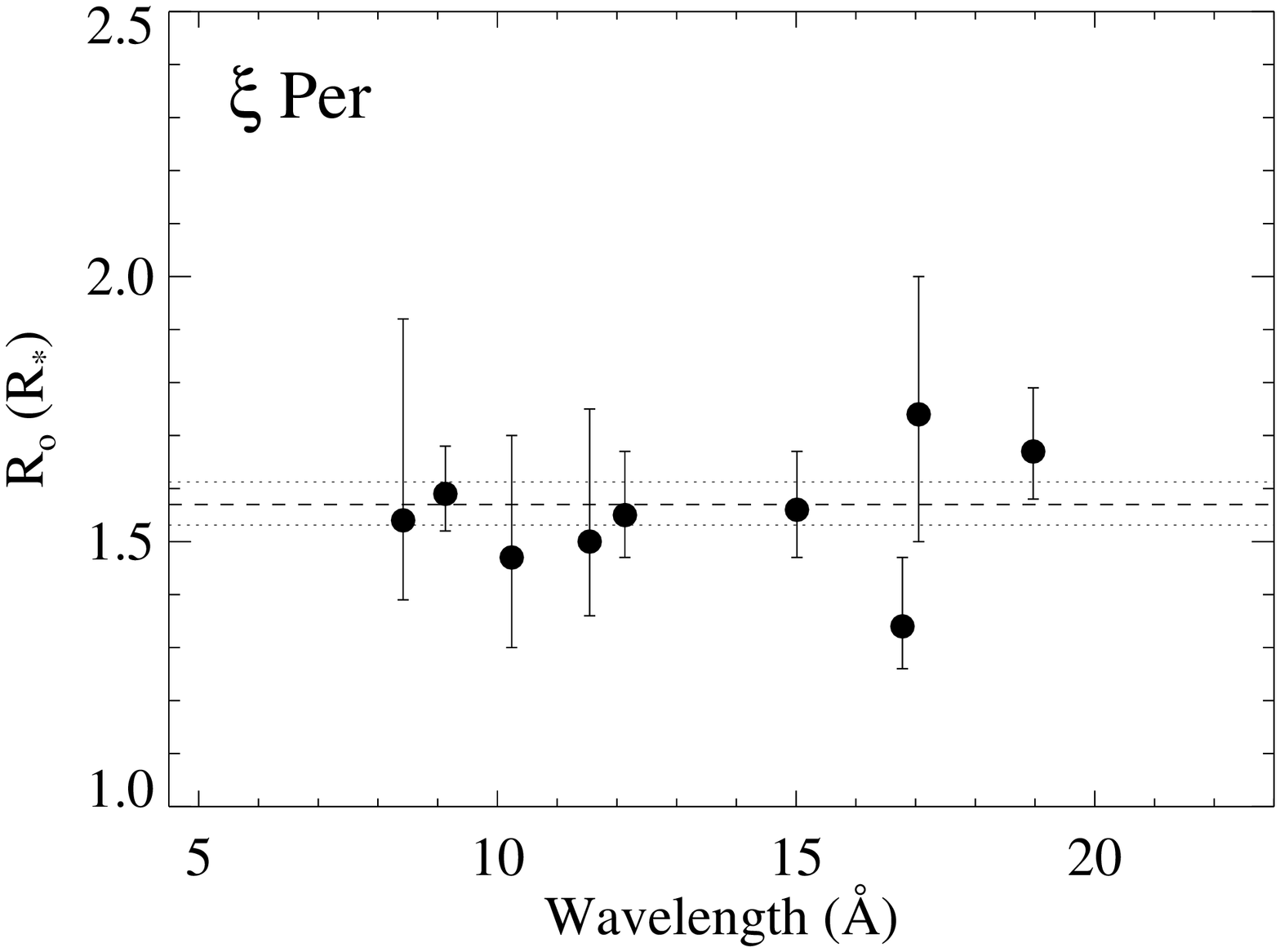}
\includegraphics[angle=0,angle=90,width=58mm]{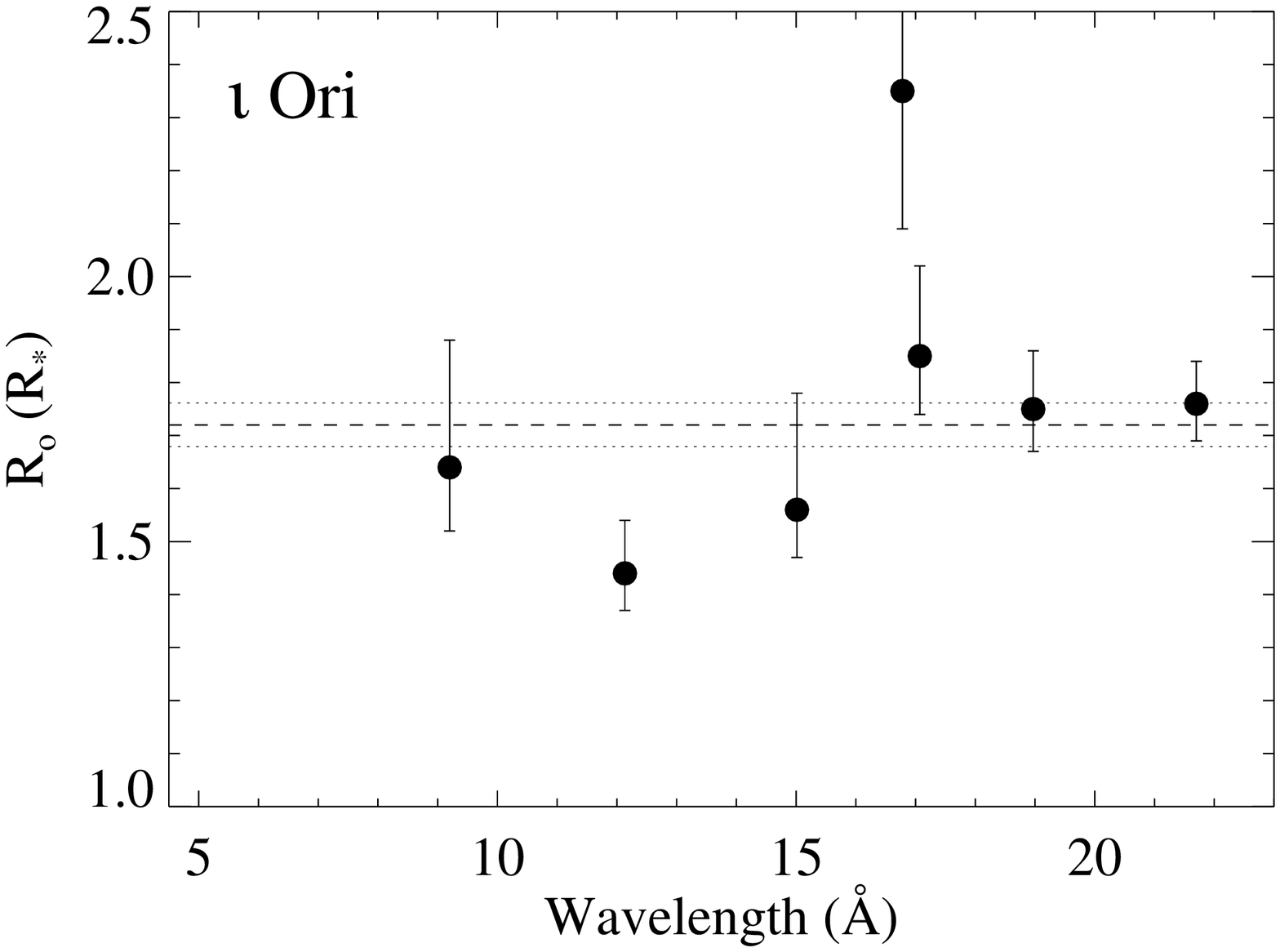}
\includegraphics[angle=0,angle=90,width=58mm]{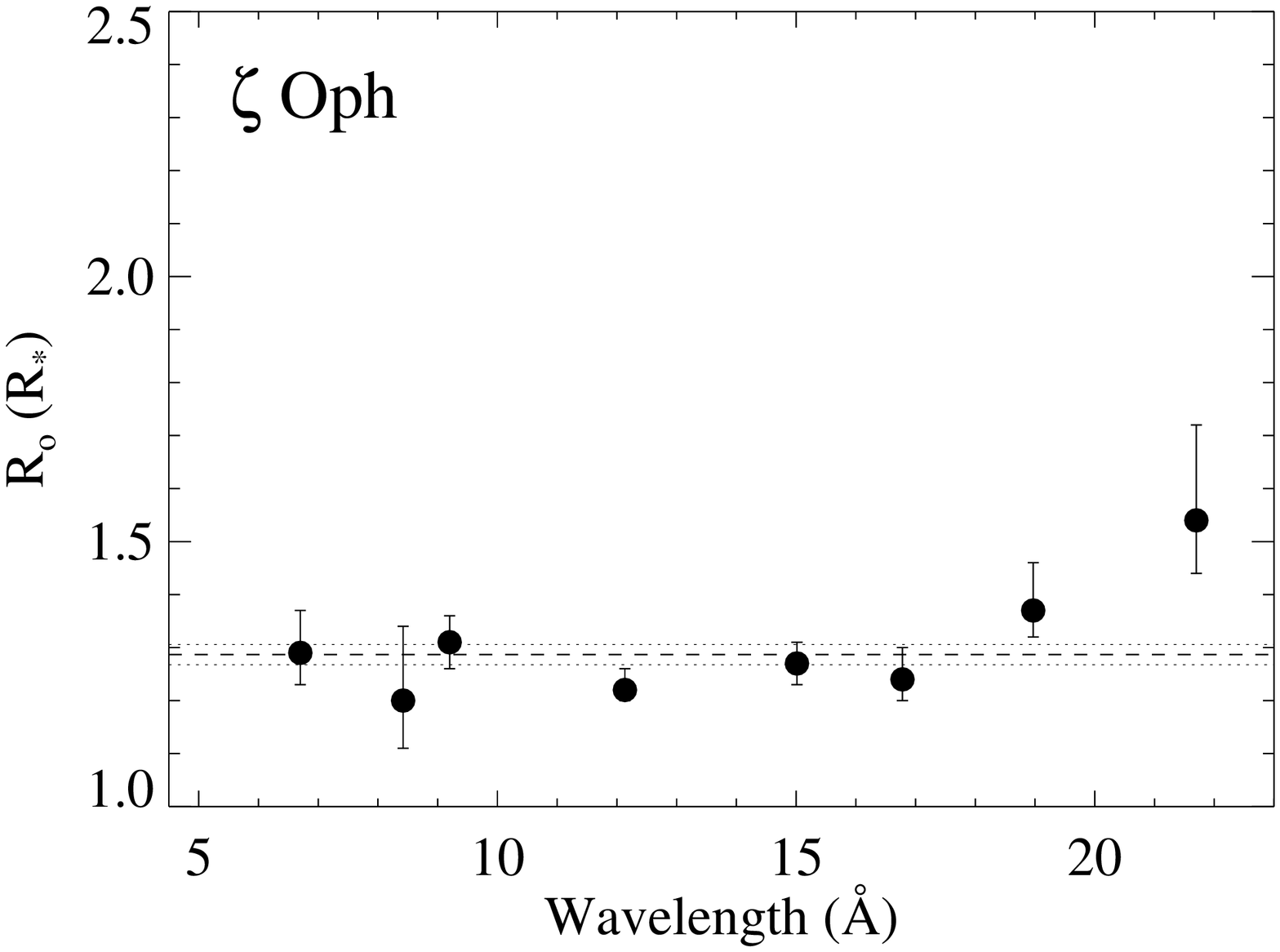}
\includegraphics[angle=0,angle=90,width=58mm]{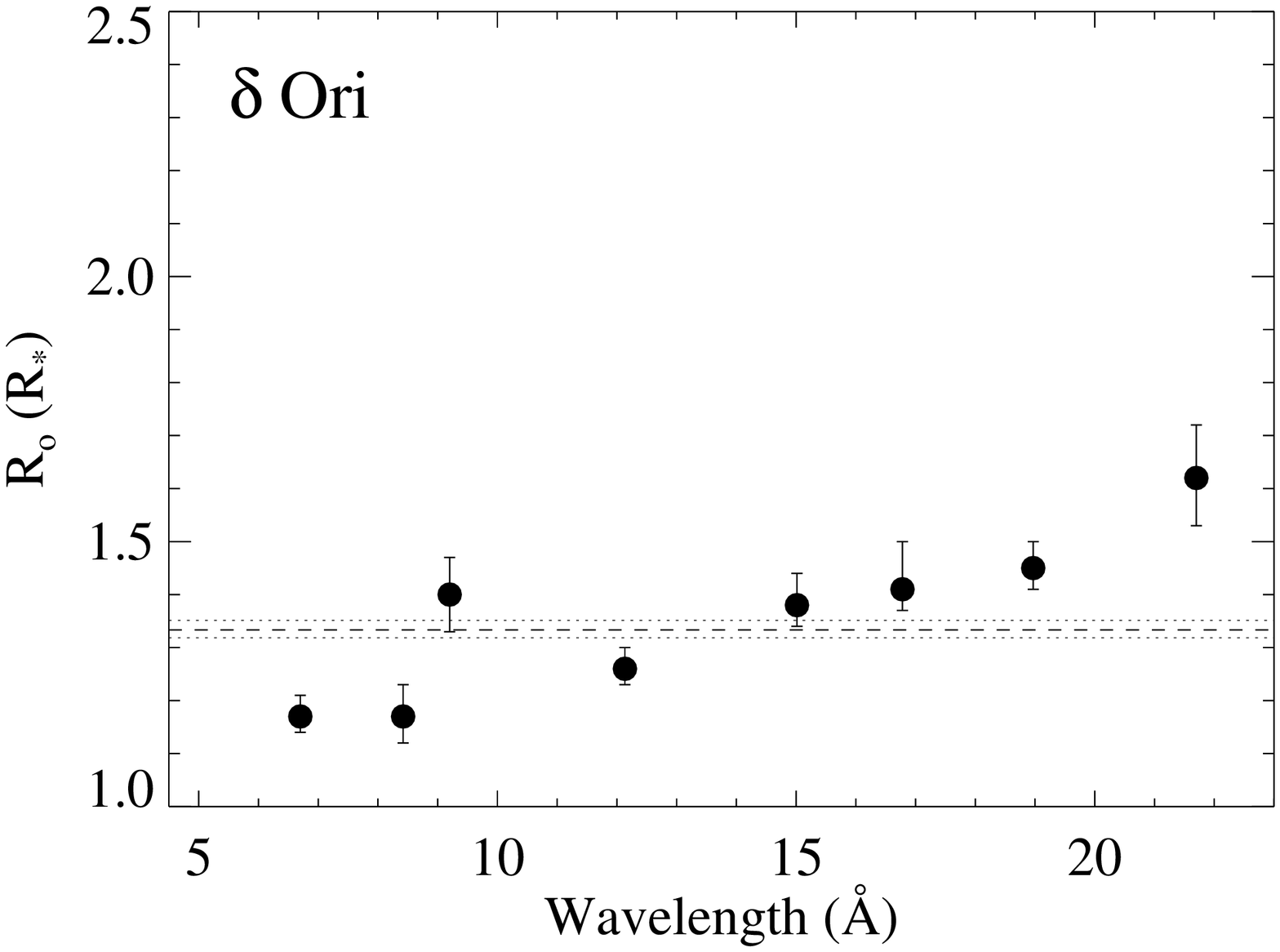}
\includegraphics[angle=0,angle=90,width=58mm]{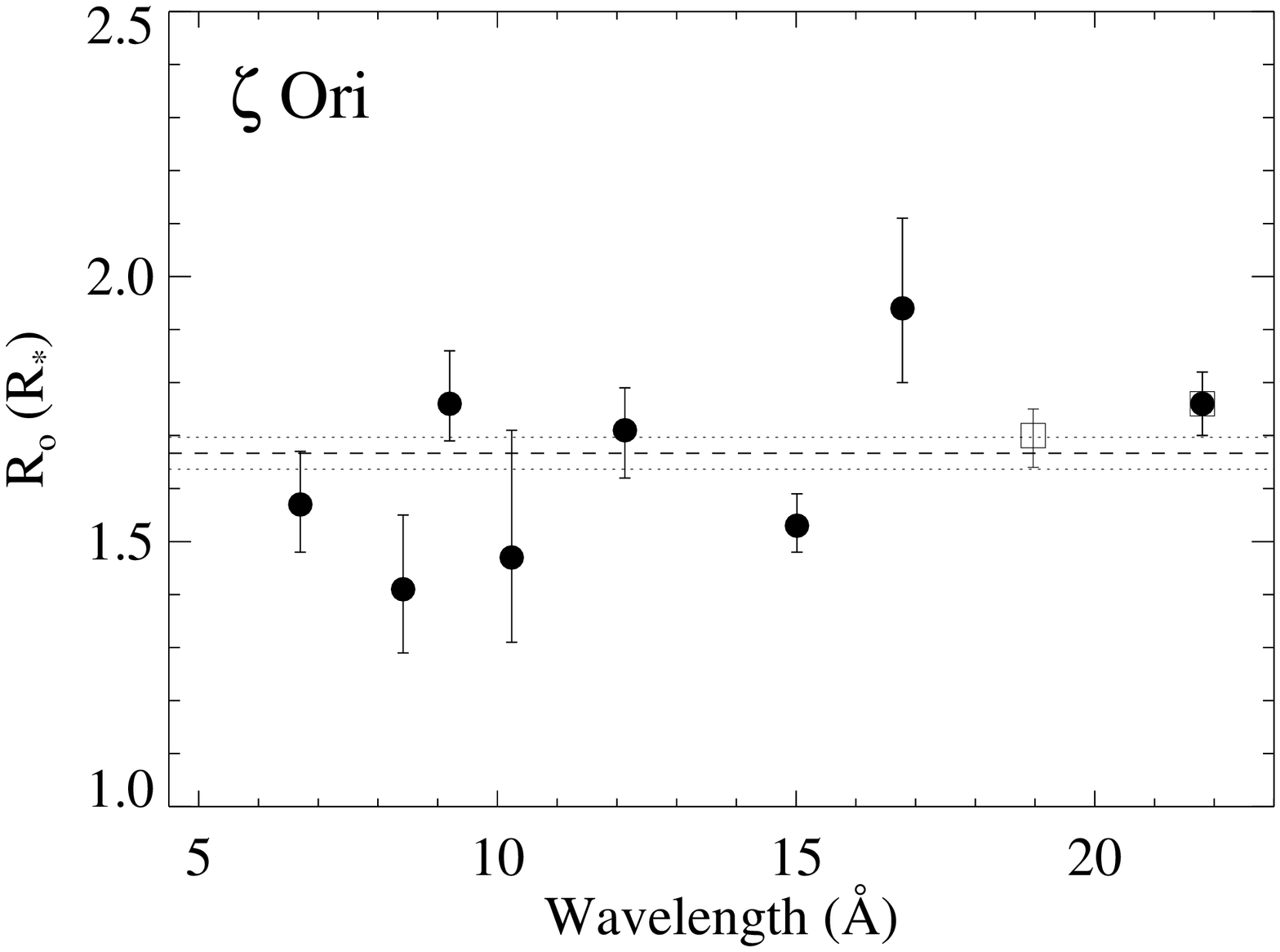}
\includegraphics[angle=0,angle=90,width=58mm]{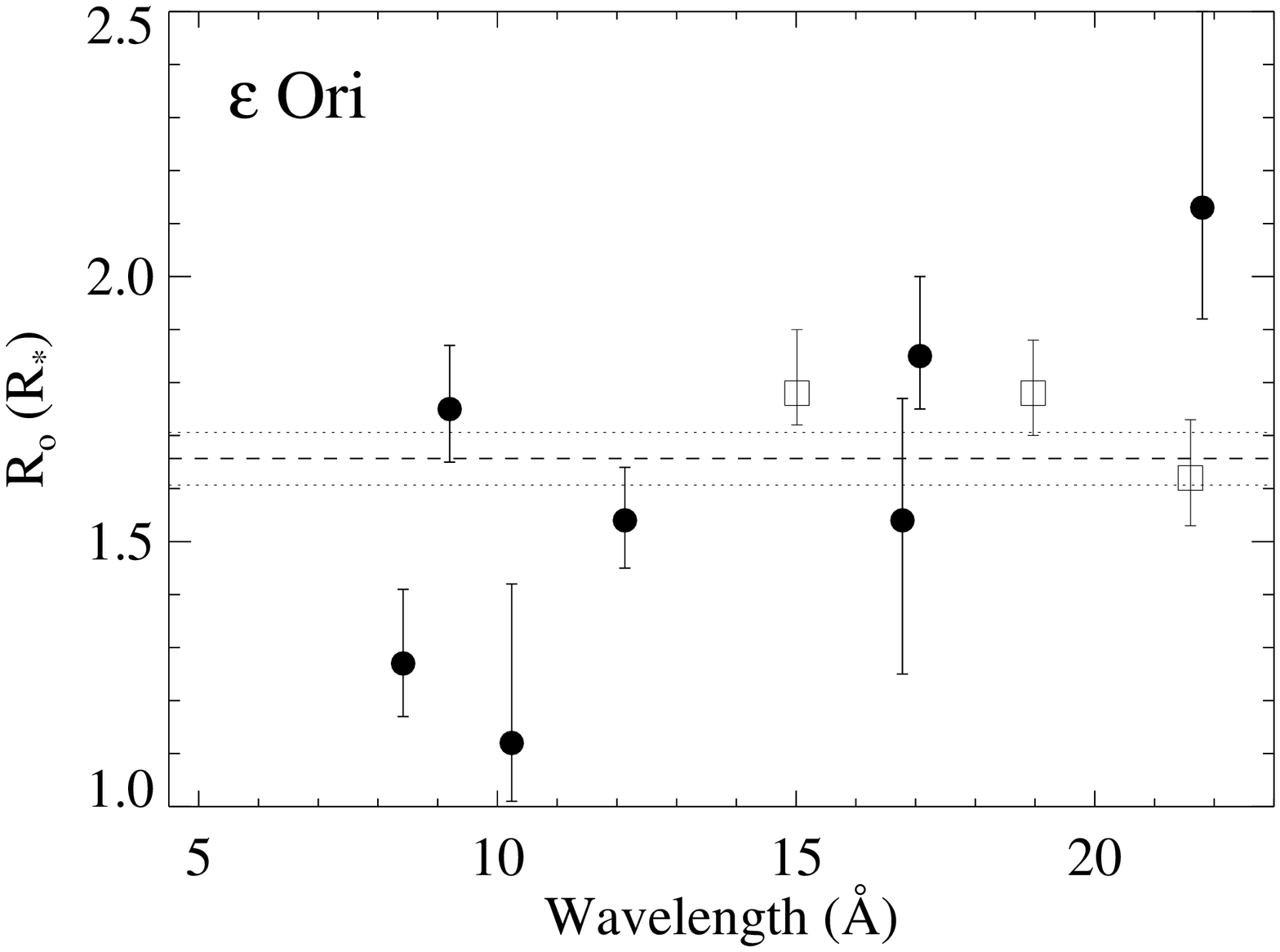}
\caption{The fitted \Ro\/ values for each line in each sample star
  (filled circles), along with the 68 per cent confidence limits
  (error bars).  The best-fitting global \Ro\/ value for each star is
  indicated in each panel by the dashed line, while the dotted lines
  indicate the extent of the 68 per cent confidence limits. The
  excluded lines for $\zeta$ Ori and $\epsilon$ Ori are shown as open
  squares.  }
\label{fig:Ro}
\end{figure*}

To further explore the uncertainty in the fitted \taustar\/ values and
ultimately the wind mass-loss rate, we tabulated three representative
opacity models that span the widest possible ranges of ionization
balance, and thus wind opacity. In Fig.\ \ref{fig:threeopacities} we
show these models, and demonstrate that below the O K-shell edge, the
wind can vary in opacity by only tens of per cent. And even above the
edge the maximum variation is no more than a factor of 2. The key
difference between the opacity models is the extent of helium
recombination in the outer wind, which, at its most extreme, can
double the opacity in the long-wavelength end of the \chandra\/
bandpass. More modest opacity variations are seen in detailed models
of the wind ionization computed with {\sc cmfgen} \citep{hm1998}, also
shown in Fig.\ \ref{fig:threeopacities}. To test the effect of such a
radial opacity increase on the derived \taustar\/ values, we modified
the line profile model to include a boost of the wind opacity above $r
= 5 \Rstar$, and fit this model to two strong lines: Fe\, {\sc xvii}
at 15.014 \AA\/ and O\, {\sc viii} at 18.969 \AA. For each line we fit
a sequence of models with different amounts of ``extra'' opacity in
the far wind; with a radial profile given by $y = 1 -
(1+(r/5\Rstar)^6)^{-0.5}$, where $y$ is the extra opacity, which can
be scaled by any desired factor. The results of these experiments --
the best-fitting \taustar\/ and \Ro\/ and the value of the fit
statistic for each value of $y$ that we tested -- are reported in
Table \ref{tab:krat}, where it can be seen, for example, that doubling
the outer wind opacity decreases the line optical depth parameter,
\taustar, but only by about 15 per cent.


\begin{table*}
\begin{center}
  \caption{X-ray derived results for each star}
\begin{tabular}{ccccccccc}
  \hline
  Star & Spectral type & $\Mdot_{\rm theory}$ & \Mdot\/ & $\chi^2$ & ${\rm N}_{lines}$ & \Ro\/ & $\chi^2$ & Primarily EWS? \\
  & & (\Msunyr) & (\Msunyr) & & & (\Rstar) & & \\
  \hline
  HD 93129A & O2 If* & $1.2 \times 10^{-5}$ & $6.8^{+2.8}_{-2.4} \times 10^{-6}$ & 1.1 & 5 & $1.34^{+.10}_{-.11}$ & 0.8 & Yes \\
  HD 93250 & O3.5 V & $6.0 \times 10^{-6}$ & $1.2^{+1.5}_{-1.2} \times 10^{-7}$ & 0.3 & 3 & $2.09^{+.15}_{-.13}$ & 2.6 & No \\
  9 Sgr & O4 V & $2.1 \times 10^{-6}$ & $3.7^{+1.0}_{-0.9} \times 10^{-7}$ & 3.3 & 7 & $1.66^{+.05}_{-.05}$ & 5.8 & Yes \\
  \zpup\ & O4 If & $6.4 \times 10^{-6}$ & $1.76^{+0.13}_{-0.12} \times 10^{-6}$ & 10.6 & 16 & $1.50^{+.03}_{-.03}$ & 13.6 & Yes \\
  HD 150136 & O5 III & $2.3 \times 10^{-6}$ & $9.4^{+4.0}_{-4.1} \times 10^{-8}$ & 8.8 & 7 & $1.35^{+.02}_{-.02}$ & 17.6 & No \\
  Cyg OB2 8A & O5.5 I & $8.7 \times 10^{-6}$ & $8.0^{+5.1}_{-5.1} \times 10^{-7}$ & 3.0 & 4 & $1.54^{+.04}_{-.04}$ & 1.2 & No \\
  $\xi$ Per & O7.5 III & $9.3 \times 10^{-7}$ & $2.2^{+0.6}_{-0.5} \times 10^{-7}$ & 11.0 & 9 & $1.57^{+.05}_{-.04}$ & 5.3 & Yes \\
  $\iota$ Ori & O9 III & $5.5 \times 10^{-7}$ & $3.2^{+84.}_{-3.2} \times 10^{-10}$ & 1.0 & 7 & $1.72^{+.04}_{-.04}$ & 16.2 & No \\
  $\zeta$ Oph & O9 V & $1.8 \times 10^{-7}$ & $1.5^{+2.8}_{-1.5} \times 10^{-9}$ & 4.7 &  8 & $1.29^{+.02}_{-.02}$ & 13.4 & Yes \\
  $\delta$ Ori & O9.5 II & $5.3 \times 10^{-7}$ & $6.4^{+3.4}_{-3.1} \times 10^{-8}$ & 5.0 & 8 & $1.33^{+.02}_{-.01}$ & 52 & Maybe \\
  $\zeta$ Ori & O9.7 Ib & $1.2 \times 10^{-6}$ & $3.4^{+0.6}_{-0.6} \times 10^{-7}$ & 5.5 & 8 & $1.67^{+.03}_{-.03}$ & 18.4 & Yes \\
  $\epsilon$ Ori & B0 Ia & $1.2 \times 10^{-6}$ & $6.5^{+1.1}_{-1.5} \times 10^{-7}$ & 1.2 & 7 & $1.66^{+.05}_{-.05}$ & 22.1 & Yes \\
  \hline
\end{tabular}
\label{tab:results}
\end{center}
\end{table*}  

In principle, empirical ionization balance and abundance determinations
for individual stars could be used to build a customized opacity model
for each star in our sample. However, abundance determinations are
sparse for O stars and also prone to systematic errors [for example,
there is a factor of $\sim 15$ range of nitrogen abundances for
\zpup\/ in the recent literature \citep{zp2007,Bouret2012}].
Similarly, ionization determinations are highly model dependent.
Although there is undoubtedly some variation in the bulk wind
ionization among our sample stars, and although some stars in the
sample certainly do have nitrogen enhancement and associated carbon
and oxygen depletion, neither of these effects will have a major
impact on the bulk wind opacity at the wavelengths with strong line
emission and therefore they will not affect the mass-loss rate
determinations. In summary, the errors in the derived mass-loss rates
due to variations and uncertainties in the wind opacity, including
those due to radial variations of the opacity in a given star's wind,
are no bigger than those due to the statistical quality of the data,
the assumptions about the wind velocity law, and the overall
metallicity of the sample stars, which we estimate to be several tens
of per cent. 

Finally, the goal of this paper is to present a homogeneously obtained
set of X-ray mass-loss rate measurements, and so we have taken a
straightforward approach to deriving the mass-loss rate from each
star's ensemble of fitted \taustar\/ values. That is, we use a single,
universal wind opacity model, which assumes solar photospheric
abundances \citep{Asplund2009} and a generic O star wind ionization
balance for each star \citep{mcw1994}. If new and reliable
determinations of programme stars' metallicities are made in the
future, our derived results can be scaled by the reciprocal of the
metallicity. We show our generic, solar abundance wind opacity model
in Fig.\ \ref{fig:opacity}, along with a model that has altered CNO
abundances, such that N is three times solar, O is 0.5 solar, and C is
0.25 solar.  Note that the sum of the absolute C, N, and O abundances
are, in this case, solar, even though the individual elemental
abundances are not.  As can be seen in the figure, the identical
metallicity of the models makes the opacity shortward of the oxygen
edge nearly the same in both models. And although there is a modest,
factor of $\sim 50$ per cent difference in the opacity longward of the
O edge, the only line that we are able to model in that part of the
spectrum is the O\, {\sc vii} line complex near 21.7
\AA\footnote{\zpup\/ also has a weak N\, {\sc vii} line at 20.91
  \AA.}.  This complex is not very strong in any of our sources, but
with higher signal-to-noise data, and when we used the
nitrogen-enhanced opacity model to derive mass-loss rates for several
of our programme stars we found the effect to be less than 10 per cent.

\section{Results}
\label{sec:results}

For each star in our sample, the simple line profile model provides
good fits to most of the emission lines and line complexes from which
we are able to derive values for \taustar\/ and \Ro, using the
formalism described in the previous section. In itself, this does not
confirm the EWS scenario of X-ray production for each of the sample
stars, as a few stars in the sample have $\taustar \approx 0$ for all
lines and profile models with $\taustar \approx 0$ are basically
indistinguishable from a generic Gaussian at the signal-to-noise and
resolution of the data.  For the stars in our sample that have
uniformly small \taustar\/ values, we therefore have to determine
whether their mass-loss rates are very low or whether some other
physical effect, such as binarity, may be producing symmetric
profiles.  However, for quite a few stars in the sample, reasonable
values of \taustar\/ and \Ro, and consistency between the \taustar\/
values and the wavelength dependence of the atomic opacity of the wind
are strong indicators that the EWS mechanism is operating and that we
can interpret the ensemble \taustar\/ values in the context of a
mass-loss rate measurement.


\begin{figure*}
\begin{center}
\includegraphics[angle=0,angle=90,width=58mm]{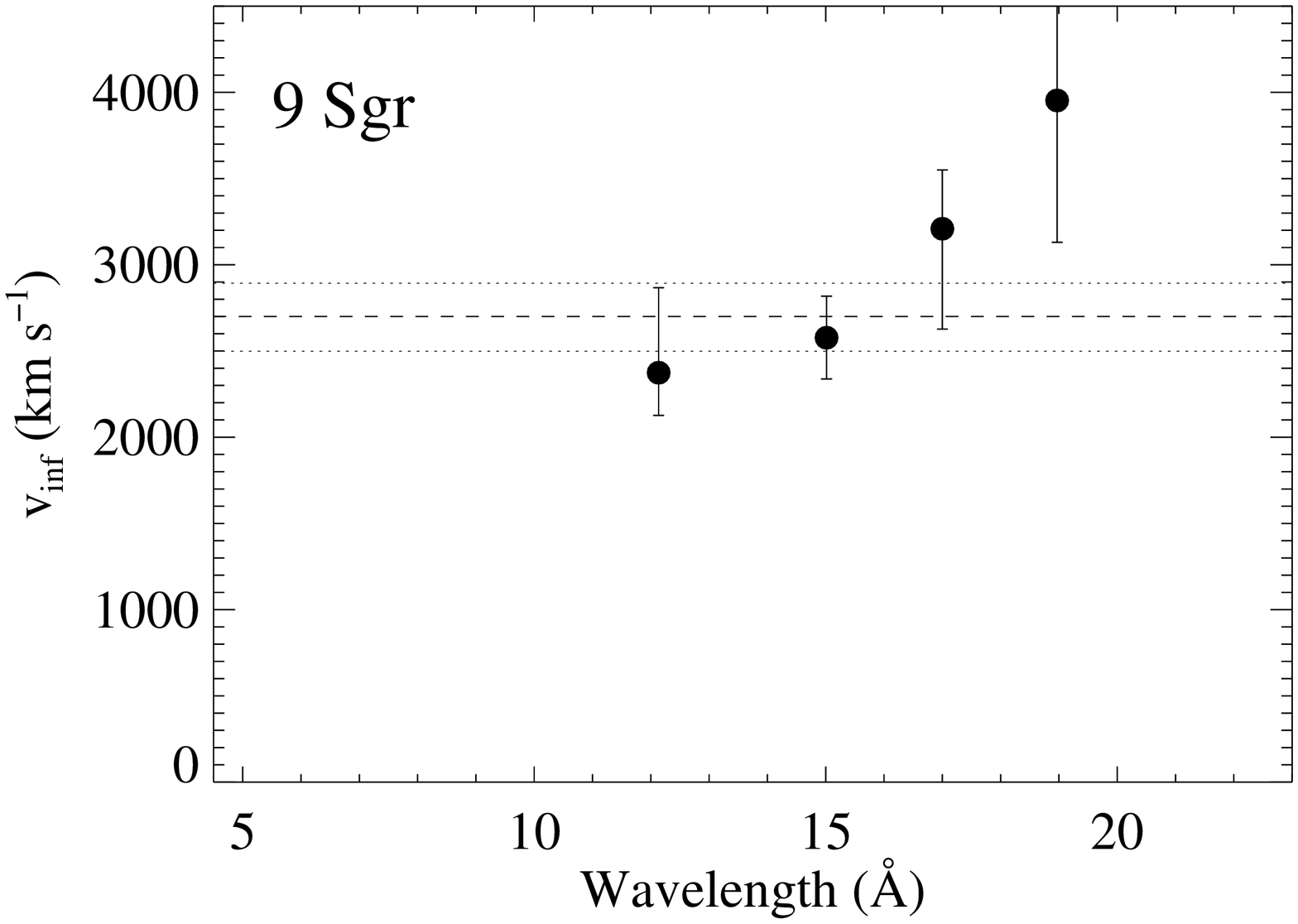}
\includegraphics[angle=0,angle=90,width=58mm]{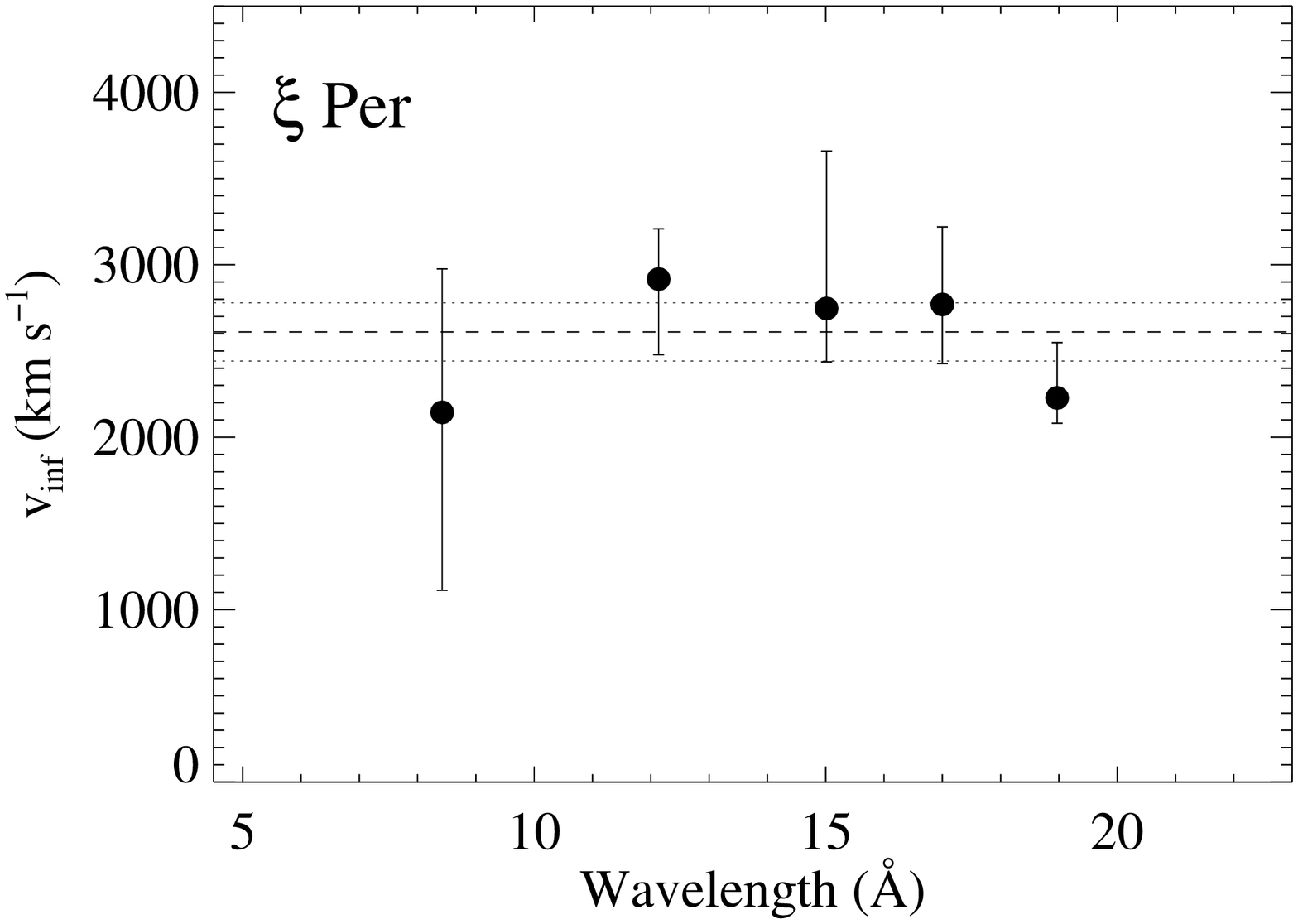}
\includegraphics[angle=0,angle=90,width=58mm]{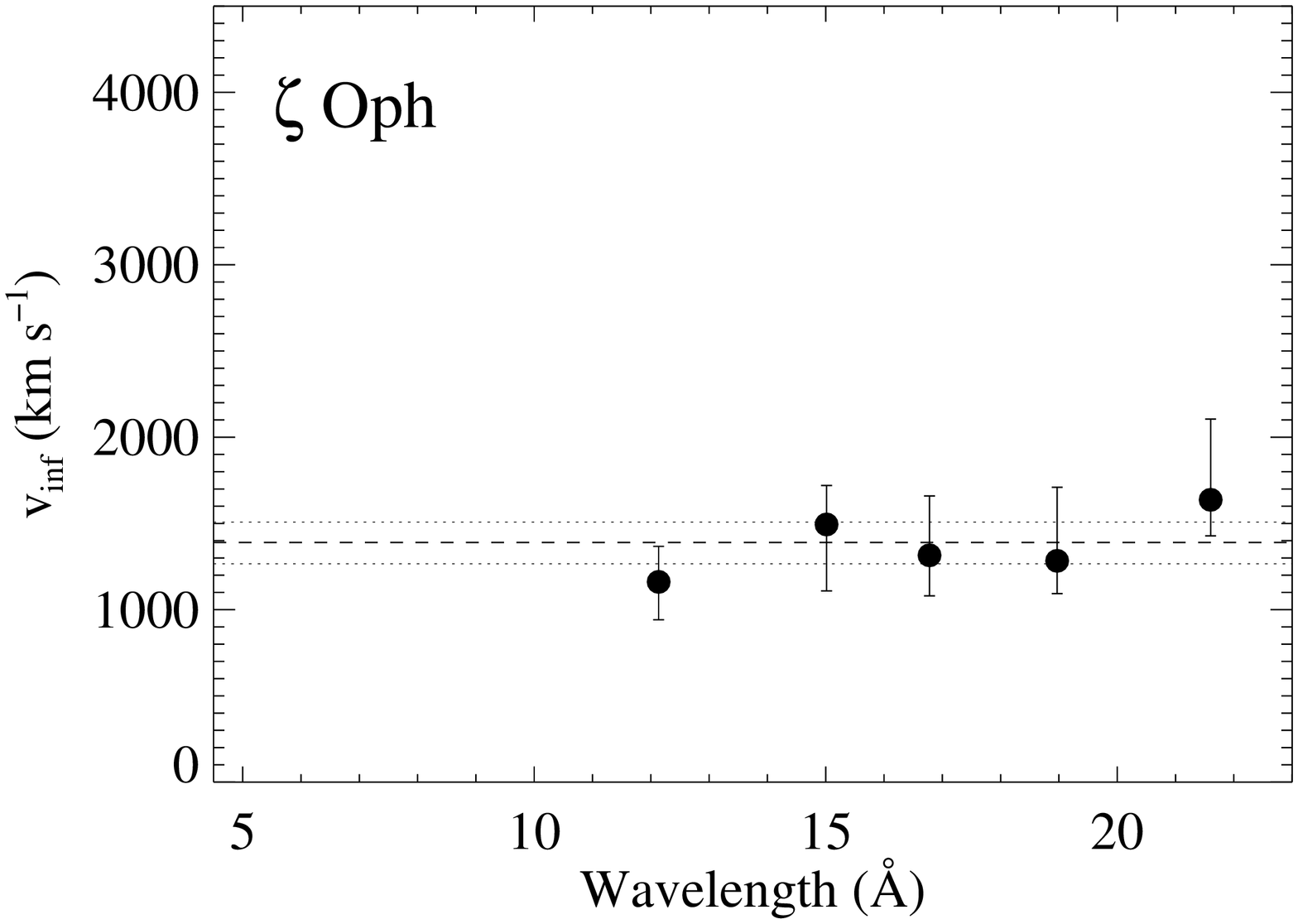}
\includegraphics[angle=0,angle=90,width=58mm]{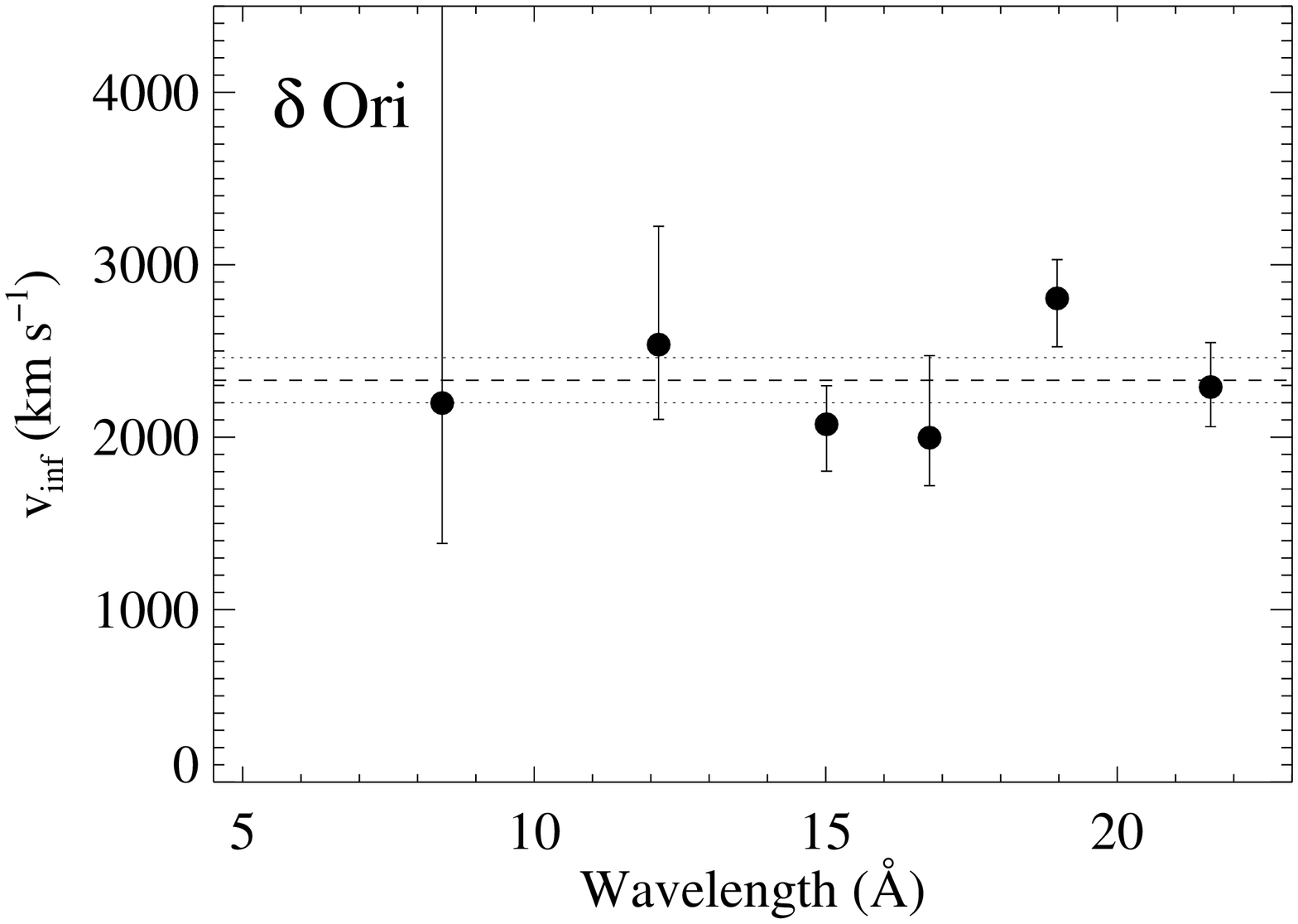}
\includegraphics[angle=0,angle=90,width=58mm]{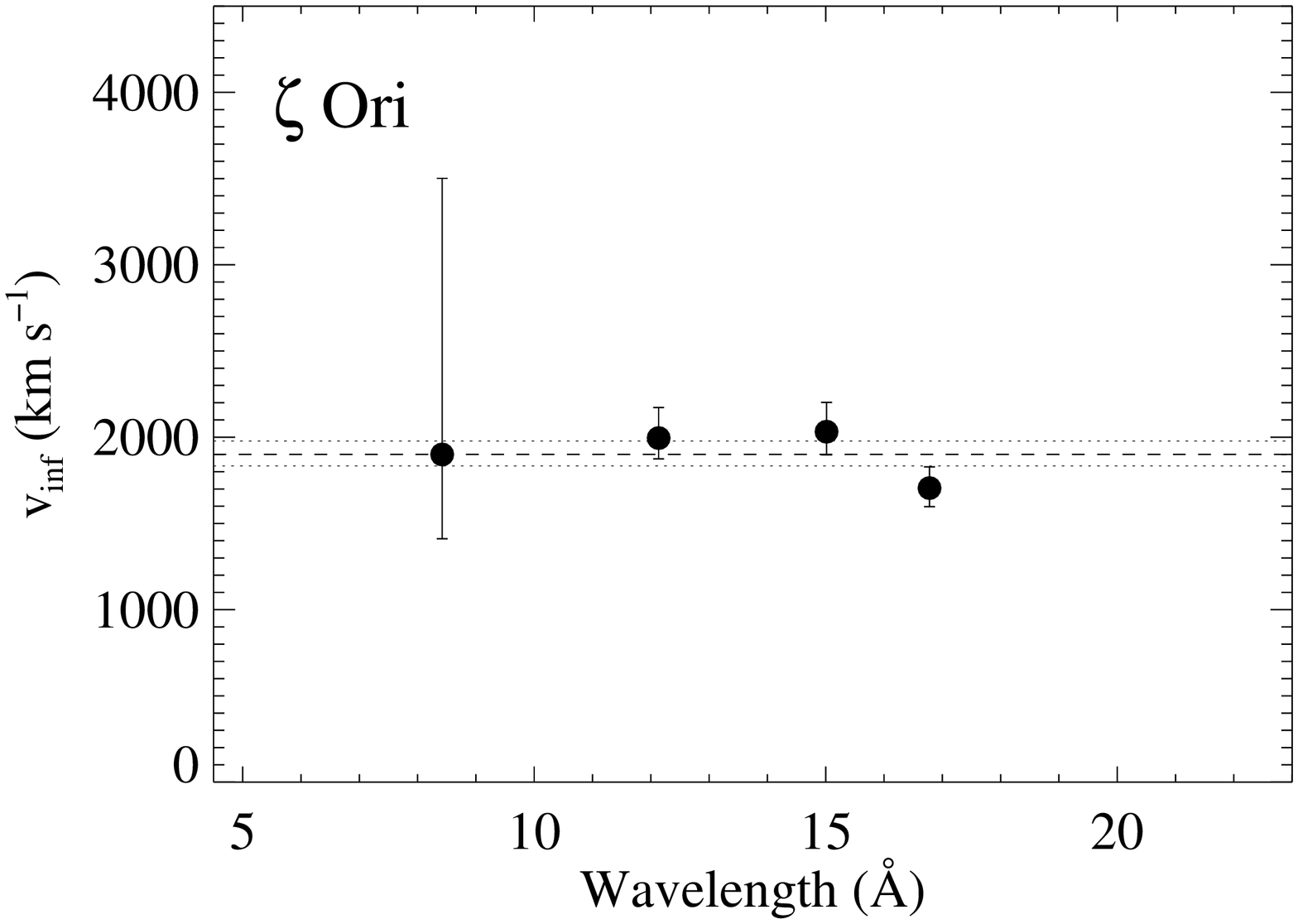}
\includegraphics[angle=0,angle=90,width=58mm]{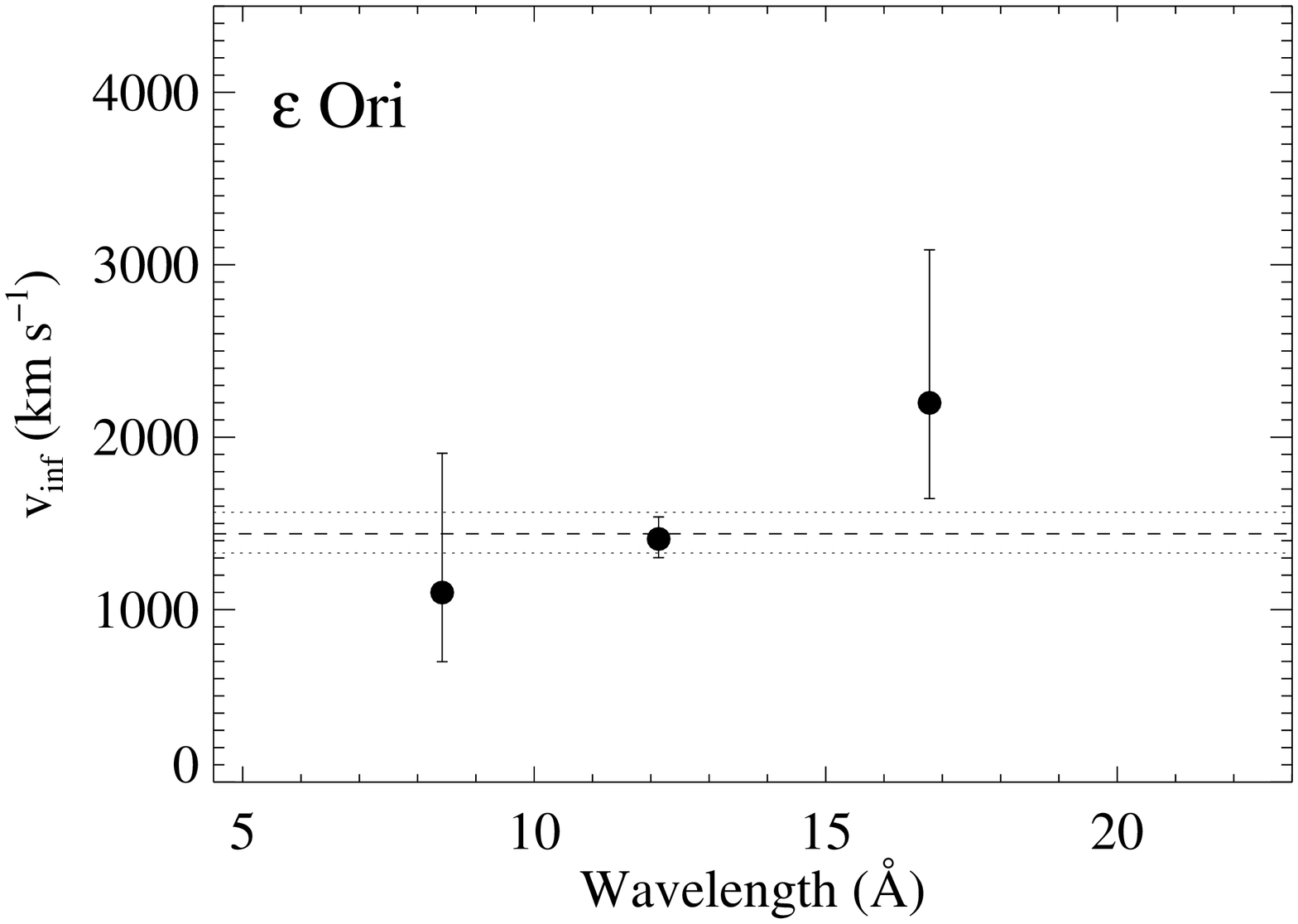}
\caption{The fitted \vinf\/ values, along with the best-fitting global
  \vinf\/ (dashed line) and its 68 per cent confidence limits (dotted
  lines).  }
\end{center}
\label{fig:vinf}
\end{figure*}


\begin{table}
\begin{center}
  \caption{Terminal velocity fit results}
\begin{tabular}{cccc}
  \hline
  Star & Spectral Type & UV \vinf\/ & X-ray \vinf\/   \\
 & & (\kms) & (\kms)  \\
  \hline
9 Sgr & O4 V & 3100 & $2700^{+193}_{-201}$ \\
$\xi$ Per & O7.5 III  & 2450 & $2610^{+169}_{-168}$ \\
$\zeta$ Oph & O9 V & 1550 & $1390^{+118}_{-124}$ \\
$\delta$ Ori & O9.5 II  & 2100 & $2330^{+132}_{-130}$ \\
$\zeta$ Ori & O9.7 Ib & 1850 & $1900^{+77}_{-67}$ \\
$\epsilon$ Ori & B0 Ia & 1600 & $1440^{+125}_{-112}$ \\
  \hline
\end{tabular}
\\ 
\label{tab:vinf}
\end{center}
\end{table}  

There are three stars in the sample for which the data quality is not
good enough to draw any meaningful conclusions: HD 206267, 15 Mon, and
$\tau$ CMa.  These are the three data sets with fewer than 2500 total
MEG + HEG counts, and for none of these stars are there more than
three emission lines for which profile fits with even marginal
constraints can be determined (and for none of the stars is there more
than one weak line that is not potentially subject to resonance
scattering and the associated ambiguity of model interpretation -- see
the resonance scattering discussion later in this section). We will
not discuss these stars further in this paper. A fourth star, HD
93250, has only three usable lines, although it has a significantly
larger number of counts in its spectrum than the three stars we are
excluding. The small number of strong lines, despite the higher
signal-to-noise spectrum, can be understood in the context of the high
plasma temperature and correspondingly strong bremsstrahlung continuum
and relatively weak lines. As we discuss in the next section, this is
a strong indication that the X-ray spectrum of HD 93250 is dominated
by hard X-ray emission from colliding wind shocks in the context of
the binary wind-wind interaction mechanism.

We summarize the fitted \taustar\/ and \Ro\/ values, and their
uncertainties, in Figs.\ \ref{fig:taustar} and \ref{fig:Ro},
respectively, with overall results for each star presented in Table
\ref{tab:results}. In the two figures, each point represents the fit
to a single line or blended line complex. In Fig.\ \ref{fig:taustar}
we convert each line's fitted \taustar\/ value to a mass-loss rate
using equation (\ref{eqn:taustar}) and the wavelength-dependent
opacity (the standard, solar-abundance-based model) shown in Fig.\
\ref{fig:opacity}. In Fig.\ \ref{fig:taustar} we also show the
best-fitting mass-loss rate we derive from fitting the ensemble of
\taustar\/ values along with the theoretical mass-loss rate
\citep{Vink2000} listed in Table \ref{tab:results}.  We show all 12
sample stars (excluding the three low-count stars mentioned above but
including HD 93129A and \zpup) in these figures, even though, as we
will discuss in the next section, we discount the interpretation of
these results in terms of a wind mass-loss rate for some of the stars.
All 12 of the mass-loss rate fits are formally good, with $\xi$ Per
showing the most scatter and largest reduced $\chi^2$, but not large
enough for the mass-loss rate fit to be formally rejected.

Among the complications of the line profile fitting is the effect of
resonance scattering in optically thick X-ray lines.
\citet{Leutenegger2007} showed that this effect is significant for
oxygen and nitrogen lines in the \xmm\/ spectrum of \zpup. And those
authors presented a ranking of the Sobolev optical depths expected for
many strong lines in the \chandra\/ bandpass. In our sample stars, the
lines most likely to be affected by resonance scattering are Fe\, {\sc
  xvii} at 15.014 \AA, O\, {\sc viii} \Lya\/ at 18.969 \AA, and the
resonance line at 21.602 \AA\/ in the O\, {\sc xvii} \Hea\/ complex.
For the spectrum of $\epsilon$ Ori, where resonance scattering seems
to be important (see \S5.1.10), we refit several of the lines,
including these three, allowing the Sobolev optical depth to be a free
parameter and the velocity law parameter $\beta$ of the hot plasma to
be either $\beta_{\rm Sob} = 0$ or 1 \citep{Leutenegger2007}.
Unfortunately, with those additional free parameters of the model, the
values of the parameters we are interested in -- \taustar\/ and \Ro\/
-- were nearly unconstrained. To account for the possible effects of
resonance scattering, then, we eliminated the affected lines from the
mass-loss rate determination.
These include all three lines mentioned above for $\epsilon$ Ori and
also the O\, {\sc viii} \Lya\/ line and the O\, {\sc vii} \Hea\/
resonance line for $\zeta$ Ori.  Note that in both cases, we were able
to include the O\, {\sc vii} intercombination line at 21.804 \AA,
which is not optically thick to resonance scattering, while excluding
the nearby resonance line\footnote{Note that the resonance lines are
  more symmetric and have lower best-fitting \taustar\/ values than do
  the intercombination lines, which is consistent with the effect of
  resonance scattering being significant.}. Excluding these lines from
the mass-loss rate fits for these two stars led to higher mass-loss
rates of a factor of 3 for $\epsilon$ Ori and 50 percent for $\zeta$
Ori. For no other star did accounting for resonance scattering make a
significant difference for the mass-loss rate determination.

There are a small number of lines for which the fits are only of
marginal quality or which provide suspect results. These include the
Si\, {\sc xiii} complex in $\zeta$ Ori, for which the fit is not
formally good, the line shapes look unusually peaked, and the formal
upper limit on \taustar\/ is remarkably small. Other suspect fits
include a few of the Ne\, {\sc ix} complexes, which are probably
affected by blending with numerous iron lines.  For $\delta$ Ori,
there is some indication that the lines are mildly red-shifted (rather
than showing the expected net blue shift due to wind absorption). This
is likely due to binary orbital motion of the primary. The results we
show in Figs.\ \ref{fig:taustar} and \ref{fig:Ro} include a redshift
(the magnitude of which was allowed to be a free parameter) in the two
longest-wavelength lines for this star. We discuss this result for
$\delta$ Ori, and the interpretation of the results for each
individual star, in the following section.

We fit an average \Ro\/ value for each star based on the ensemble of
line-fit results, and we show that average, and its 68 per cent
confidence limits, in Fig.\ \ref{fig:Ro}. For many of the stars, a
single value provides a good fit, but for HD 150136, $\iota$ Ori,
$\delta$ Ori, $\zeta$ Ori, and $\epsilon$ Ori the fits are marginal
(rejected at $\approx~95$ per cent confidence). For the latter two
stars, at least, there is a modest correlation between \Ro\/ and
wavelength. These overall results, of a basically uniform onset radius
of $\Ro \approx 1.5$ \Rstar, with possibly somewhat higher values for
the longest wavelength lines, are, we note, true for the He-like
complexes as well as the other lines, which do not have any particular
radial line ratio sensitivity. This is in contrast to Gaussian line
profile fits to the same helium-like complexes in many of these same
stars which assume a single formation radius for each line complex,
and which show a much wider variation in X-ray source location based
on the forbidden-to-intercombination line ratio values \citep{wc2007}.
Those results seem to be an artefact of the overly simplistic
assumption that all the X-rays form at a single radius.

Finally, for a few lines in some of the sample stars' spectra, we
treat the wind terminal velocity, \vinf, as a free parameter (as
described in \S\ref{subsec:fitting_procedure}).  These results are
shown in Fig.\ 6 and listed in Table \ref{tab:vinf}. For all the stars
with EWS emission, we fit a single \vinf\/ value to the ensemble of
line results, and in each case the fit is formally good and consistent
with the bulk wind terminal velocity at the 95 per cent (2 $\sigma$)
confidence level.

\section{Discussion and Conclusions}
\label{sec:discussion}

While the empirical line profile model provides good fits to nearly
all the lines in all the sample stars, one of the primary results of
this study is the overall weakness -- or even absence -- of wind
absorption signatures in the \chandra\/ grating spectra of O stars.
This has been noted before by various authors examining individual
objects, generally via fitting Gaussian profile models (e.g.\
\citealt{Miller2002}), but here we have systematically quantified this
result using a more physically meaningful line profile model.  There
are three classes of explanations for the weak wind-absorption
signatures we measure, and the associated low mass-loss rates: (1) the
line profile model is missing some crucial physics; (2) processes
other than embedded wind shocks are contributing to the X-ray line
emission and thereby diluting the characteristic shifted and skewed
profiles that are the signature of wind absorption; and (3) the actual
mass-loss rates of these stars are lower than expected from theory and
from older empirical determinations made from \Ha, UV, or radio/IR
data that ignore wind clumping.

Examining the trends shown in Figs.\ \ref{fig:taustar} and
\ref{fig:Ro}, we can identify several stars with extremely low wind
optical depths and/or shock onset radii that deviate significantly
from the expectations of the EWS scenario.  These include HD 93250, HD
150136, $\iota$ Ori, $\zeta$ Oph, and $\delta$ Ori.  As we show below,
it is likely that most of these stars, and also Cyg OB2 8A, have a
significant contribution from CWS in their observed X-ray line
profiles.  The other stars in the sample: 9 Sgr, $\xi$ Per, $\zeta$
Ori, and $\epsilon$ Ori (as well as HD 93129A and \zpup) have line
profiles that are consistent with the expectations of the EWS
scenario, with \taustar\/ values that, while low, are well within an
order of magnitude of the theoretically expected values and are
consistent with the expected wavelength trend of the atomic opacity of
their winds. We note that $\delta$ Ori is a borderline case.

The mass-loss rates we derive for these stars from their ensembles of
\taustar\/ values are listed in Table \ref{tab:results} and are all
lower than the theoretical values computed by \citet{Vink2000}. We
summarize the X-ray-derived mass-loss rates for all the stars in the
sample (even those for which the derived values cannot be trusted) in
Fig.\ \ref{fig:xray_vs_vink}, and compare these mass-loss rates to the
theoretical values. We will discuss the results shown in this figure
further, but first let us consider each star in our sample with an eye
towards differentiating among the three scenarios outlined above for
explaining the weaker-than-expected line profile wind absorption
signatures.

\subsection{Individual stars}

\subsubsection{HD 93129A}

Fits to the small number of lines in this very early O supergiant's
\chandra\/ grating spectrum have already been presented
\citep{Cohen2011}, and here we rederive the mass-loss rate from the
previously fitted \taustar\/ values using the standard, solar
abundance wind opacity model we described in \S3.3. We find the same
mass-loss rate reported by \citet{Cohen2011}, who used a wind opacity
model with altered CNO abundances. As noted in that paper, this star
has an early-type visual companion at a separation of roughly 50
mas detected with the {\it Hubble Space Telescope} Fine Guidance
Sensor \citep{Nelan2004,Nelan2010}. But at that separation any
colliding wind X-ray emission is negligible compared to the observed
EWS X-ray emission \citep{Cohen2011}.

\subsubsection{HD 93250}

The \chandra\/ grating spectrum of this early O main sequence star is
quite hard and bremsstrahlung dominated, indicating that the spectral
hardness is due to high plasma temperatures rather than being a
by-product of wind and/or ISM absorption. HD 93250 was identified as
being anomalous in X-rays in the recent Chandra Carina Complex Project
\citep{Townsley2011}, with an X-ray luminosity even higher than that
of HD 93129A, and a high X-ray temperature derived from
low-spectral-resolution \chandra\/ ACIS data \citep{Gagne2011}.  Those
authors suggest that the X-rays in HD 93250 are dominated by colliding
wind shocks from interactions with an assumed binary companion having
an orbital period greater than 30 d. Soon after the publication of
that paper, \citet{Sana2011} announced an interferometric detection of
a binary companion at a separation of 1.5 mas, corresponding to 3.5
au.  Thus it seems that the hard and strong X-ray spectrum and the
symmetric and unshifted X-ray emission lines can be readily explained
in the context of CWS X-ray emission.

\subsubsection{9 Sgr}

This star is known to be a spectroscopic binary with a massive
companion in an 8 or 9 yr orbit \citep{Rauw2005}. The X-ray properties
of 9 Sgr were described by \citet{Rauw2002} based on \xmm\/
observations.  These authors noted blue-shifted line profiles, based
on Gaussian fits, and also a somewhat higher than the normal
$\Lx/\Lbol$ ratio and a moderate amount of hot ($T \approx 20$ MK)
plasma based on fits to the \xmm\/ European Photon Imaging Camera
(EPIC) spectrum, although only about 1 per cent of the X-ray emission
measure is due to the hot component. A simple CWS model computed by
\citet{Rauw2002} shows that the observed X-ray emission levels cannot
be explained by colliding wind shocks, and the authors conclude that
the X-ray emission is dominated by embedded wind shocks.  Presumably
the separation of the components and/or their relative wind momenta
are not optimal for producing CWS X-ray emission. It is reasonable to
assume that while there may be a small amount of contamination from
CWS X-rays, the line profiles we measure in the \chandra\/ grating
spectra are dominated by the EWS mechanism, and therefore the
mass-loss rate we derive from the profile fitting is indeed a good
approximation to the true mass-loss rate. We note, also, that
according to the radial velocity curve shown in \citet{Rauw2005} the
\chandra\/ grating spectrum we analyse in this paper was taken during
a phase of the orbit when the primary's radial velocity was close to
zero. And finally, we note that the published value of the wind
parameter $\beta = 0.7$ gives $\Ro = 1.4$ \citep{Cohen2010b}, which is
somewhat lower than the value we find here, using the standard $\beta
= 1$.

\subsubsection{\zpup}

As with HD 93129A, we refit the mass-loss rate from the previously
published ensemble of \taustar\/ values \citep{Cohen2010}. In this
case, though, we find a mass-loss rate that differs from the published
value due to our use of a solar abundance wind opacity model in this
paper. We find a mass-loss rate of $\Mdot = 1.76 \times 10^{-6}$
\Msunyr\/ (and find the same value when we used our altered CNO wind
opacity model, shown as the dashed line in Fig.\ \ref{fig:opacity}).
This is nearly a factor of two below the value found by
\citet{Cohen2010} because their wind opacity was based on empirical C,
N, and O abundance determinations that had a net metallicity of about
half solar. All of the change in our new, lower mass-loss rate is due
to our use of a wind opacity model that assumes solar metallicity.

\subsubsection{HD 150136}

A well-known spectroscopic binary, with a period of only 2.662 d
\citep{ng2005}, and a third O star in the system at a somewhat larger
separation \citep{Sana2013}, the HD 150136 system has previously been
studied in the X-ray using the same data we reanalyse here
\citep{Skinner2005}.  Those authors find a very high X-ray luminosity
but a soft spectrum with broad X-ray emission lines.  They also detect
some short period X-ray variability that they tentatively attribute to
an occultation effect. A more recent determination of the ephemeris
\citep{Mahy2012} is consistent with an occultation effect causing the
observed X-ray variability \citep{Russell2013}. And although colliding
wind binaries with strong X-ray emission are generally thought to
produce hard X-ray emission, it has recently been shown that many
massive O+O binaries have relatively soft emission and modest X-ray
luminosities, especially if their orbital periods are short
\citep{Gagne2011,Gagne2012}. We also note that this star's X-ray
emission stands out from the other giants and supergiants in the X-ray
spectral morphology study of \citet{wnw2009} by virtue of its high
H-like/He-like silicon line ratio, indicating the presence of some
hotter plasma. We conclude that although a few of the X-ray emission
lines measured in this star's spectrum have non-zero \taustar\/
values, overall the lines are too heavily contaminated by X-rays from
colliding wind shocks to be used as a reliable mass-loss rate
indicator.

\subsubsection{Cyg OB2 8A}

With phase-locked X-ray variability, a high $\Lx/\Lbol$, and a
significant amount of hot plasma with temperatures above 20 MK
\citep{deBecker2006}, Cyg OB2 8A has X-ray properties characteristic
of colliding wind shocks.  It is a spectroscopic binary with a 21 d
period in an eccentric orbit, and a semi-major axis of 0.3 au.  The
small number of short-wavelength lines we are able to fit are not
terribly inconsistent with the expectations of the embedded wind shock
scenario, although the inferred mass-loss rate is roughly an order of
magnitude lower than the theoretically expected value.  However,
because they are only present at short wavelengths, where the wind
opacity is low, they do not provide very much leverage on the
mass-loss rate, and, with their large error bars, they are also
generally consistent with $\taustar = 0$ (although the Mg\, {\sc xii}
\Lya\/ line has $\taustar = 0.75^{+.66}_{-.38}$). We included this
star in our sample because of a prior analysis of the same \chandra\/
grating data under the assumption of EWS emission from a single star
\citep{Waldron2004}, but given the thorough analysis by
\citet{deBecker2006}, we must conclude that the X-rays are dominated
by colliding wind shocks, at least to a large extent, and that the
profile fits we present here do not provide much information about
embedded wind shocks or the wind mass-loss rate.

\subsubsection{$\xi$ Per}

A runaway star without a close binary companion and with a constant
radial velocity \citep{Sota2008}, $\xi$ Per should not have any binary
colliding wind shock emission contaminating the X-ray emission lines
we analyse. It does, however, show significant UV and \Ha\/
variability, at least some of which is rotationally-modulated
\citep{deJong2001}.  Thus the assumptions of spherical symmetry and a
wind that is smooth on large scales are violated to some extent.
Still, the X-ray line profiles should provide a relatively reliable
mass-loss rate.  The \taustar\/ values we find are significantly
larger than zero and are consistent with the expected wavelength
trend. The mass-loss rate we derive is a factor of 4 or 5 below the
theoretical value from \citet{Vink2000}.

\subsubsection{$\iota$ Ori}

Of all the stars in the sample, $\iota$ Ori shows the least amount of
line asymmetry and blue shift, with all seven lines and line complexes
we analyse having \taustar\/ values consistent with zero. Taken at
face value, the derived mass-loss rate is three orders of magnitude
below the theoretical value.  The star is in a multiple system, with
the closest component a spectroscopic binary in a highly eccentric, 29
d orbit \citep{Bagnuolo2001}. The \chandra\/ observations were made at
a time when the stars' radial velocities were very close to zero.
Although there are no definitive signatures of CWS X-ray emission
(such as orbital modulation of the X-rays), it is very likely that the
quite broad but symmetric emission lines we have measured are from
colliding, rather than embedded, wind shocks.

\subsubsection{$\zeta$ Oph}

This star also has a nearly complete lack of wind absorption
signatures in its line profiles.  And its lines are narrower than
expected in the EWS scenario, as shown by the low \Ro\/ values in
Fig.\ \ref{fig:Ro}. Unlike the other stars in the sample with X-ray
profiles that are difficult to understand in the context of embedded
wind shocks, $\zeta$ Ori does not have a binary companion likely to
produce colliding wind shock X-rays.  It is, however, a very rapid
rotator ($v\sin{i} = 351$ \kms; \citealt{ce1977}), goes through \Ha\/
emission episodes that qualify it as an Oe star \citep{bb1974}, and
has an equatorially concentrated wind \citep{Massa1995}.  The wind's
deviation from spherical symmetry could explain the relatively
symmetric and narrow X-ray emission lines. The wind is likely truly
weak as well \citep{Marcolino2009}, and so our measurements can place
a 1 $\sigma$ upper limit on the mass-loss rate that is a factor of 40
below the theoretically predicted value \citep{Vink2000}, if the
wind's deviation from spherical symmetry is not important for the
X-ray emission. This low mass-loss rate is in fact consistent with
those found by \citet{Marcolino2009}. The X-rays allow for even lower
mass-loss rates, too, but not higher ones.

\subsubsection{$\delta$ Ori}

With a quite small amount of wind attenuation evident in its line
profiles and narrower than expected lines, the results from $\delta$
Ori are also suspect, although there are some emission lines with
non-zero \taustar\/ values in its \chandra\/ spectrum. This star is a
member of a multiple system that includes an eclipsing, spectroscopic
binary companion with an orbital period of 5.7 d. The companion is
an early B star, and an earlier analysis of these same \chandra\/ data
indicated that colliding wind shocks were not likely to be strong
enough to account for the X-ray luminosity of $\Lx \approx 10^{32}$
erg s$^{-1}$ \citep{Miller2002}. However, it seems likely that
between occultation effects and modest wind-wind interaction with the
known companion that there is some degree of contamination of the wind
absorption signal in the context of our basic, spherically symmetric
single-star emission line model. Preliminary analysis of a new, long,
phase-resolved \chandra\/ observation does indeed indicate some
possible effects of the companion star on the X-ray line profiles
\citep{Corcoran2013,Nichols2013}. As far as the mass-loss rate is
concerned, we can only be quantitative to the extent that we can say
that if all of the X-ray emission comes from embedded wind shocks in
the spherically symmetric wind of the primary, then the mass-loss rate
of $\delta$ Ori is an order of magnitude below the \citet{Vink2000}
mass-loss rate.

\subsubsection{$\zeta$ Ori}

Significant wind absorption signatures are seen in the X-ray profiles
of $\zeta$ Ori (as demonstrated in \citealt{Cohen2006}), which has the
highest signal-to-noise \chandra\/ spectrum of any of the stars in our
sample.  The expected wavelength trend is seen in the \taustar\/
results, especially after the O\, \Lya\/ and \Hea\/ lines are excluded
due to resonance scattering. The fitted \Ro\/ values are consistent
with $\Ro = 1.5$ \Rstar, expected in the EWS scenario.  While it is
possible that there could be some contamination from CWS X-ray
emission, the binary companion of $\zeta$ Ori is two magnitudes
fainter than the primary and is at a separation of about 100 \Rstar,
making strong CWS emission an unlikely scenario
\citep{Hummel2000,Rivinius2011}. A more distant companion is resolved
in \chandra\/ images and contaminates the \chandra\/ grating spectra
at a level of about 10 per cent.

\subsubsection{$\epsilon$ Ori}

The only B star in our sample, $\epsilon$ Ori is a B0Ia MK standard,
and given its evolved state and high luminosity, its wind is as strong
as many of the O stars in our sample.  Nearly all of the X-ray
emission lines show wind signatures with \taustar\/ values that
deviate significantly from zero. It is also the only star in our
sample for which eliminating the lines most likely subject to
resonance scattering has a very significant effect on our derived
mass-loss rate, increasing it from $2.1 \times 10^{-7}$ to $6.4 \times
10^{-7}$ \Msunyr. Eliminating those lines also significantly improves
the quality of the fit. And the low wind terminal velocity of
$\epsilon$ Ori makes resonance scattering Sobolev optical depths
larger, all things being equal, so the importance of the effect here,
but not apparently in most of the other stars, is reasonable.  Thus,
we report the higher mass-loss rate in Table \ref{tab:results} and
show the fit from which that value is derived in Fig\
\ref{fig:taustar}. There is no reason to believe that CWS X-ray
emission affects the star's \chandra\/ spectrum.  Its only known
companion is at 3 arcmin \citep{Halbedel1985} (which would be easily
resolved by \chandra) but is not seen in the \chandra\/ data, while
interferometric observations show no binary companion down to small
separations \citep{rp2002}.

\subsection{Discussion}

Before discussing the mass-loss rates and EWS properties of the sample
stars, we must note that a not insignificant fraction of the sample
seems to be contaminated by binary colliding wind X-ray emission.
Stars like Cyg OB 8A show characteristic time-variable, hard X-ray
emission. But other stars, like $\iota$ Ori and HD 150136 show X-ray
emission that is not obviously orbitally modulated or very hard (with
HD 93250 being something of an intermediate case).  All four of these
stars have known O star binary companions at relatively small
separations, and thus we can attribute the bulk of their X-ray line
emission to the CWS mechanism and therefore we cannot infer a wind
mass-loss rate nor any EWS shock properties from their X-ray line
profiles. While idealized CWS models predict distinctive X-ray
emission line profile shapes \citep{hsp2003}, such shapes are not
observed in real systems (e.g.\ \citealt{hsp2005}), perhaps because of
shock instabilities and the associated mixing and large random
velocity components of the X-ray emitting plasma \citep{pp2010}.
Therefore, when a mixture of CWS and EWS X-rays is present, the
observed, hybrid line profiles should be relatively symmetric and
moderately broad, mimicking pure EWS profiles with little or no
absorption. And as we have already mentioned, binary CWS X-ray
emission does not necessarily have to be hard or at significantly
elevated levels, depending on the binary orbital parameters
\citep{Gagne2012}.


\begin{figure}
\includegraphics[angle=90,width=90mm]{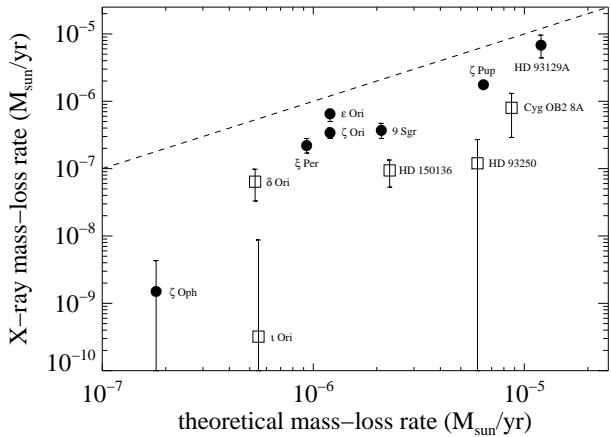}
\caption{The X-ray derived mass-loss rates for each star in our sample
  (and also \zpup\/ and HD 93129A) compared to the theoretically
  expected mass-loss rates \citep{Vink2000}. Stars dominated by EWS
  are shown as filled circles, while those where our line profile
  model breaks down, in most cases due to CWS X-rays, are shown as
  open squares.  The dashed line indicates the region where both
  mass-loss rate estimates are equal. }
\label{fig:xray_vs_vink}
\end{figure}

In addition, the X-ray line emission from the late O supergiant
$\delta$ Ori may very well be affected by the presence of an early-B
close binary companion, which at the very least should break the
spherical symmetry of the primary's wind. As we show from the profile
fitting and discuss in the last subsection, there is some evidence of
EWS signatures in the profiles of this star, and so it is most likely
a hybrid case, and thus the profile fitting provides a lower limit on
the mass-loss rate, assuming that EWS emission is the dominant
contribution, and that limit is a factor of 12 below the theoretically
expected value \citep{Vink2000}. Thus $\delta$ Ori and the four sample
stars discussed in the previous paragraph -- the five stars denoted by
open symbols in Fig.\ \ref{fig:xray_vs_vink} -- fall to one extent or
another into categories (1) and (2) discussed at the beginning of this
section; their X-ray emission is not well described by physics
assumptions such as spherical symmetry or it is not dominated by the
EWS mechanism.

For the other seven stars -- indicated by filled symbols in Fig.\
\ref{fig:xray_vs_vink} -- it is unlikely that a non-EWS mechanism is
significantly affecting the X-ray line emission and so we can
interpret their small to modest wind absorption signatures in terms of
low, but measurable, wind mass-loss rates. The systematically low
values of these mass-loss rates compared to the theoretically
predicted values is the main result of this study, but the \taustar\/
values we fit for the ensemble of X-ray emission lines from these
stars are indeed consistent with the wavelength trend expected from
the atomic opacity of their winds. And the low mass-loss rate values
we find are themselves consistent with other recent multi-wavelength
wind studies \citep{Najarro2011,Sundqvist2011,Bouret2012} that find
mass-loss rates a factor of a few lower than those predicted by
\citet{Vink2000}. The most luminous, earliest star in our sample, HD
93129A (O2 If*) has an X-ray mass-loss rate a factor of 2 below the
\citet{Vink2000} theoretical value, while \zpup, 9 Sgr, $\zeta$ Ori,
and $\xi$ Per have X-ray mass-loss rates a factor of 3 to 6 lower than
the theoretically predicted values.  The early B supergiant $\epsilon$
Ori shows similar results, but when we exclude the emission lines that
might be affected by resonance scattering the resulting higher
mass-loss rate is only a factor of 2 below the theoretical value.  The
average mass-loss rate reduction with respect to the theoretical
values is a factor of 3 for these six stars. Finally, the least
luminous star in our sample, $\zeta$ Oph, has essentially no wind
signatures in its \chandra\/ emission lines, and although to some
extent this may be due to rapid rotation and associated asphericity,
the X-ray mass-loss rate we derive of only a few $10^{-9}$ \Msunyr\/
is consistent with other recent determinations of the mass-loss rate
of this weak-wind star \citep{Marcolino2009}.

The mass-loss rate measurements we present here, based on wind
absorption, are important because they are not subject to the
density-squared clumping effects that make the traditional mass-loss
rate diagnostics problematic. But with these X-ray mass-loss rates in
hand, we can use the density-squared diagnostics to measure the
clumping factor in the diagnostic formation region, via $f_{\rm cl} =
(\Mdot_{den-sq}/{\Mdot_{\rm X-ray}})^2$, where $\Mdot_{den-sq}$ is the
mass-loss rate derived from \halpha, IR, or radio under the assumption
of a smooth wind. In practice, it is more reliable to model the
density-squared diagnostics using the X-ray derived mass-loss rate and
varying the clumping factor, $f_{\rm cl}$ and the clump onset radius,
$R_{\rm cl}$. Of course, the clumping factor may vary with location in
the wind. For the O stars in our sample the \Ha\/ is formed mainly in
the inner wind, whereas radio emission originates at much larger radii
and thus probes the conditions in the outer wind \citep{Puls2006}.


\begin{figure}
\includegraphics[angle=90,width=85mm]{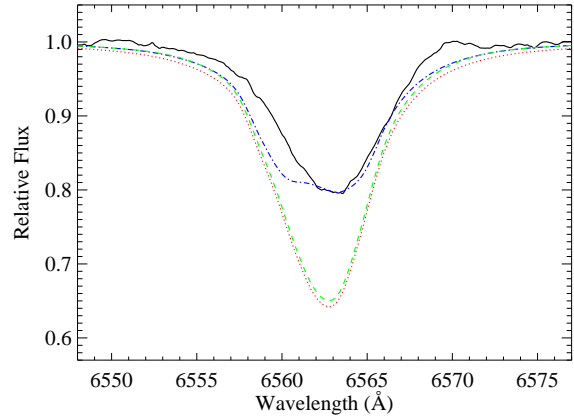}
\caption{The mean observed \halpha\/ profile of $\xi$ Per (solid,
  black) is well-fitted by a wind model that has the low, X-ray
  derived mass-loss rate, of $2.2 \times 10^{-7}$ \Msunyr, and $f_{\rm
    cl} = 20$ with a clumping onset, $R_{\rm cl}$, just above the
  photosphere (dash-dot, blue). Neither an unclumped model (dotted,
  red) nor a model with $f_{\rm cl} = 20$ but a clumping onset of
  $R_{\rm cl} = 1.3$ \Rstar\/ (dashed, green) fits the data.}
\label{fig:halpha}
\end{figure}

To demonstrate this technique, we fit the \halpha\/ line (mean
profile) measured in one sample star, $\xi$ Per, accounting for
optically thin clumping using the synthesis technique developed by
\citet{Puls2006} and \citet{Sundqvist2011}. Fig.\ \ref{fig:halpha}
shows several model profiles computed using our X-ray derived
mass-loss rate. Models without clumping or with clumping that starts
well above the photosphere fail to produce enough \halpha\/ emission,
but a model with a clumping factor of $f_{\rm cl} = 20$ and a clumping
onset radius immediately above the photosphere does reproduce the
observed \halpha\/ emission level. This clumping factor is completely
consistent with the smooth-wind \Ha\/ mass-loss rate measured by
\citet{Repolust2004} when using the scaling law in the previous
paragraph. The big difference between the dash-dotted (blue) curve and
the dashed (green) curve in Fig.\ \ref{fig:halpha} shows that the
\Ha\/ wind emission in $\xi$ Per originates almost entirely in layers
just above the photosphere.  A similar result was found for HD 93129A
\citep{Cohen2011}, where a radially constant clumping factor of
$f_{\rm cl} = 12$ with an onset just above the photosphere fits the
\halpha\/ data along with the X-ray-derived mass-loss rate.

Although the profile fitting presented here is, like any diagnostic
technique, subject to various systematic effects, we have quantified
those effects and find that they are generally on the order of a few
tens of per cent. Uncertainties in the wind opacity, which must be
modeled in order to derive a mass-loss rate from an ensemble of
\taustar\/ values, may be the biggest source of error. But although
radial variations within a given wind, and uncertainty about the
ionization state and detailed elemental abundances contribute modestly
to the systematic errors, the biggest wind opacity uncertainty arises
from uncertainty about the overall metallicity of the wind. The
opacity is directly proportional to the metallicity and, indeed, using
a solar abundance wind opacity model, as we do in this study, has led
us to reduce the mass-loss rate estimate of the canonical O
supergiant, \zpup, from $3.5 \times 10^{-6}$ \Msunyr\/
\citep{Cohen2010} to $1.8 \times 10^{-6}$ \Msunyr. This value is very
close to the newly derived values of $2.1 \times 10^{-6}$ \Msunyr\/
from the analysis of hydrogen lines in the near-IR
\citep{Najarro2011}, and of $2.0 \times 10^{-6}$ \Msunyr\/ from an
optical and UV analysis \citep{Bouret2012}, while the global X-ray
modeling of \citet{Herve2013} finds a modestly higher value of $3.5
\times 10^{-6}$ \Msunyr\/. \citet{Najarro2011} also includes our
programme star $\epsilon$ Ori, for which those authors find $\Mdot =
4.3 \times 10^{-7}$ \Msunyr. That value is bracketed by our two
values, the higher of which accounts for resonance scattering.

For the EWS sources in our sample, the fits to the X-ray emission
lines also provide information about the spatial distribution and
kinematics of the shock-heated wind plasma. In general, we find
consistency with models in which the bulk wind and the embedded X-ray
plasma have the same kinematics, described by the standard beta wind
velocity law, with terminal velocities given by optical and UV
diagnostics holding for the X-ray plasma as well as the bulk wind. The
shock onset parameter, \Ro, is consistent with $\Ro \approx 1.5$
\Rstar, or a little less, which is in line with published 1D and 2D
numerical simulations of the instability
\citep{fpp1997,ro2002,do2003,do2005}.  However, recent wind structure
simulations that account for both sound-wave driven excitation of the
wind instability and limb darkening, do show more structure near the
base of the wind, and certainly well below $r = 1.5$ \Rstar\/
\citep{Sundqvist2013}. But the role of such inner wind structure for
the onset of X-ray emission is not yet clear. To reliably predict the
X-ray emission from clump-clump collisions, which is likely to be the
dominant mode of Line Deshadowing Instability (LDI)-induced EWS X-ray
emission \citep{fpp1997}, may require fully 3-D simulations of clump
formation.

From a diagnostic perspective, the \Ro\/ parameter is governed to a
large extent by the line widths and thus the kinematics of the X-ray
plasma. If X-ray emitting plasma near the wind base actually does
exist, but is moving systematically faster than the velocity predicted
by the beta law, then our modeling technique would likely overestimate
the value of \Ro. It should be kept in mind that there is no intrinsic
limitation to the pre-shock flow speed at small radii, as the nature
of the LDI is to rapidly accelerate a small fraction of the
line-driven wind mass to higher-than-ambient velocities. Another
factor to consider is that different lines, sensitive to plasma of
different temperatures, may form in different spatial locations (e.g.\
\citealt{Herve2013}).  There is some indication from the \Ro\/ results
shown in Fig.\ \ref{fig:Ro} that longer wavelength lines, which tend
to arise in relatively cooler plasma, form farther out in the wind,
and so perhaps some of the shorter wavelength lines, indicative of
plasma with temperatures approaching or exceeding $10^7$ K, do form at
smaller radii, consistent with the base wind shocks seen in the
simulations presented in \citet{Sundqvist2013}.

Regardless of the X-ray shock onset constraints, the consistent
\halpha\/ and X-ray fitting seems to require -- now for $\xi$ Per,
too, as shown in Fig.\ \ref{fig:halpha}, in addition to HD 93129A and
\zpup\/ -- that clumping begins very close to the photosphere.
It is possible for the LDI to produce clumping without also producing
significant X-ray emission if the shocks are not strong enough to heat
the wind plasma to more than $10^6$ K. It is also quite possible that
the clumping in O stars begins already in the photosphere, perhaps due
to the radiation-driven magneto-acoustic instability \citep{fs2013}.
Future simulations will have to address these issues of clump
formation and X-ray production in the context of the LDI.

Unfortunately, it is unlikely that many more O stars will be observed
at high X-ray spectral resolution in the near future, as the X-ray
brightest O stars in the sky are all in the current sample, and as we
showed, detailed spectral analysis requires several thousand counts in
the \chandra\/ gratings. However, wind absorption of X-rays has an
effect on the broad-band X-ray emission in addition to the emission
lines, and modeling the global thermal emission spectrum in
conjunction with the broad-band wind absorption holds promise for
making mass-loss rate measurements \citep{Leutenegger2010}. In fact,
this technique has already been applied to HD 93129A and gives results
consistent with the line profile fitting approach we use in this paper
\citep{Cohen2011}.

In summary, then, the new findings presented in this paper include:
(1) mass-loss rates can be determined from X-ray line profile shapes
without having to correct for optically thin clumping; and (2) this
clumping-insensitive diagnostic finds mass-loss rates that are on
average a factor of 3 lower than the theoretical rates of
\citet{Vink2000}; but (3) in the case of $\zeta$ Oph, which is a
previously determined weak-wind star, the mass-loss rate discrepancy
is closer to two orders of magnitude; (4) the spatial distribution of
the X-ray plasma and its kinematics is roughly consistent with the
predictions of numerical simulations of these O star winds; (5)
clumping that affects \halpha\/ begins very close to the photosphere
while the X-ray emission onset is farther out in the wind; and finally
(6) a perhaps surprising number of programme stars seem subject to
contamination by CWS X-ray emission, even in some cases where the
overall X-ray emission is neither unusually strong nor unusually hard.

\section*{Acknowledgements}

Support for this work was provided by the National Aeronautics and
Space Administration through the ADAP award NNX11AD26G and \chandra\/
award numbers TM3-14001B and AR2-13001A to Swarthmore College and
award number TM6-7003X to University of Pittsburgh. EEW was supported
by a Lotte Lazarsfeld Bailyn Summer Research Fellowship from the
Provost's Office at Swarthmore College. JOS and SPO acknowledge
support from NASA award ATP NNX11AC40G to the University of Delaware
and JOS also acknowledges support from DFG grant Pu117/8-1. Special
thanks to V\'{e}ronique Petit for her careful reading of the
manuscript and her numerous helpful suggestions.


\end{document}